\documentclass[epj]{svjour}
\usepackage{graphics}
\usepackage{graphicx}
\usepackage{amsmath}
\usepackage{xcolor}
\usepackage[labelfont=footnotesize, textfont=footnotesize]{caption}

\usepackage[english]{babel}
\usepackage[utf8x]{inputenc}
\usepackage[T1]{fontenc}
\usepackage{amsfonts}
\usepackage{cite}
\usepackage{multirow}
\usepackage{booktabs}
\usepackage{subfig}

\usepackage[switch]{lineno}

\begin{document}\sloppy
\raggedbottom
\hugehead
\title{PANDA Phase One}
%
%
\author{PANDA collaboration \\[5bp]
G.~Barucca\inst{1} \and 
F.~Davì\inst{1} \and 
G.~Lancioni\inst{1} \and 
P.~Mengucci\inst{1} \and 
L.~Montalto\inst{1} \and 
P. P.~Natali\inst{1} \and 
N.~Paone\inst{1} \and 
D.~Rinaldi\inst{1} \and 
L.~Scalise\inst{1} \and 
B.~Krusche\inst{2}
% ...
}
\institute{
Università Politecnica delle Marche-Ancona,{ \bf Ancona}, Italy \and 
Universität Basel,{ \bf Basel}, Switzerland
% ...
}


\abstract{The Facility for Antiproton and Ion Research (FAIR) in Darmstadt, Germany, provides unique possibilities for a new generation of hadron-, nuclear- and atomic physics experiments. The future antiProton ANnihilations at DArmstadt (PANDA or $\overline{\rm P}$ANDA) experiment at FAIR will offer a broad physics programme, covering different aspects of the strong interaction. Understanding the latter in the non-perturbative regime remains one of the greatest challenges in contemporary physics. The antiproton-nucleon interaction studied with PANDA provides crucial tests in this area. Furthermore, the high-intensity, low-energy domain of PANDA allows for searches for physics beyond the Standard Model, \textit{e.g.} through high precision symmetry tests. This paper takes into account a staged approach for the detector setup and for the delivered luminosity from the accelerator. The available detector setup at the time of the delivery of the first antiproton beams in the HESR storage ring is referred to as the \textit{Phase One} setup. The physics programme that is achievable during Phase One is outlined in this paper.
\PACS{
       {07.05.Fb}{Design of experiments} \and
       {29.30.-h}{Spectrometers and spectroscopic techniques} \and
       {24.85.+p}{Quarks, gluons, and QCD in nuclear reactions} \and 
       {13.75.-n}{Hadron-induced low- and intermediate-energy reactions and scattering} \and
       {21.30.Fe}{Forces in hadronic systems and effective interactions} \and 
       {25.43.+t}{Antiproton-induced reactions} \and
       {13.40.Gp}{Electromagnetic form factors} \and
       {14.20.Jn}{Hyperons} \and 
       {13.75.Ev}{Hyperon-nucleon interactions} \and
       {14.40.-n}{Mesons} \and 
       {14.40.Pq}{Heavy quarkonia} \and
       {14.40.Rt}{Exotic mesons} \and
       {14.40.Lb}{Charmed mesons} \and
       {13.30.-a}{Baryon decay} \and
       {13.60.Rj}{Baryon production} \and
       {13.88.+e}{Polarization in interactions and scattering}
}}

\maketitle

\section{Introduction}
\begin{figure*}[h!]
\centering
\includegraphics[width=1.0\linewidth,trim={0cm 2cm 0cm 3cm},clip]{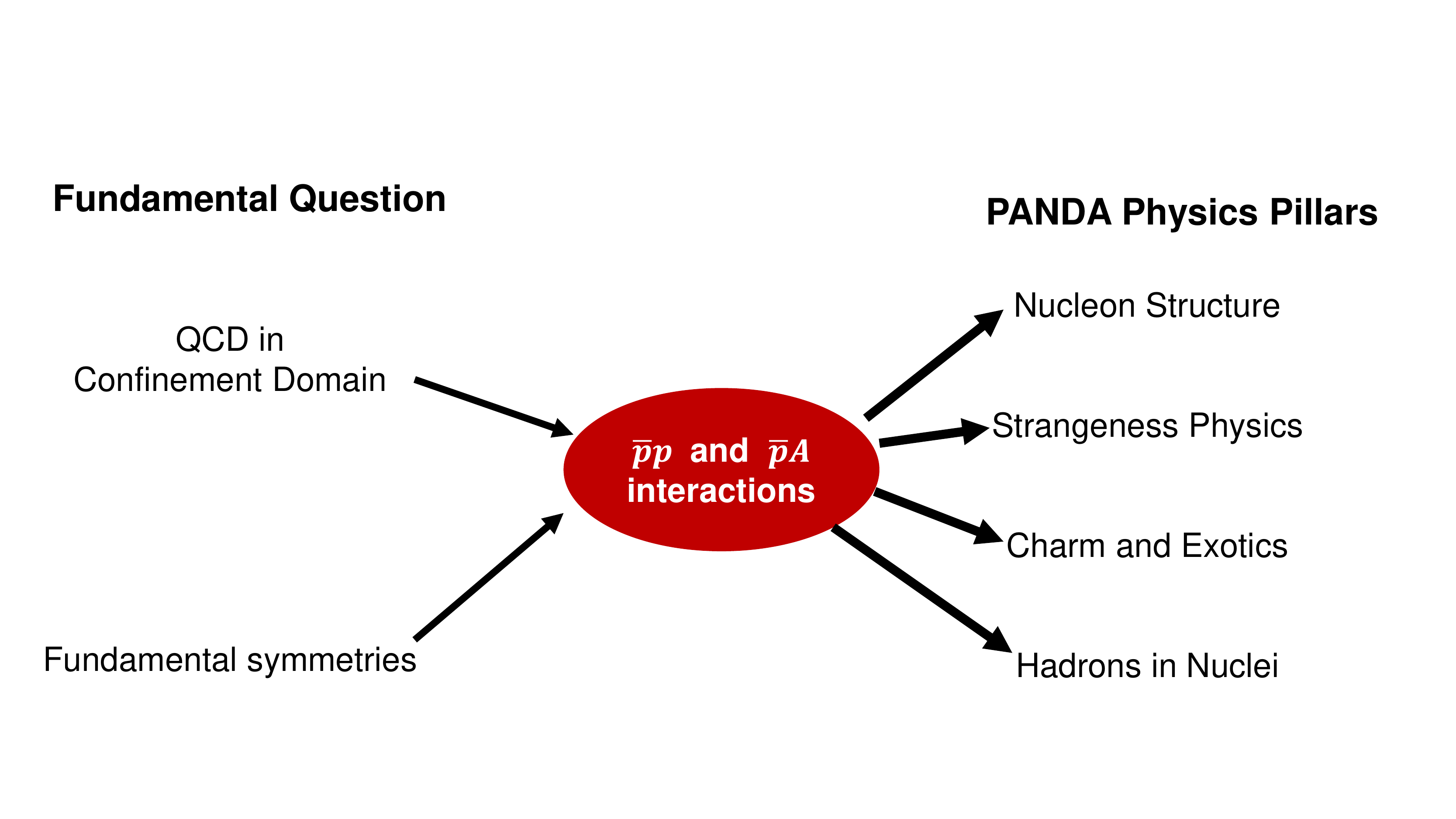}
\caption{The PANDA physics domains emerging when using antiproton interactions with nucleons and nuclei as diagnostic tools to shed light on some of the most challenging unresolved problems of contemporary physics.}
\label{fig:map}
\end{figure*}
The Standard Model (SM) of particle physics has to date successfully described elementary particles and their interactions. However, many challenging questions are yet to be resolved. Some of these are being studied at the high energy frontier at \textit{e.g.} the LHC at CERN. A different approach is the high precision/high intensity frontier provided by exclusive measurements of hadronic reactions at intermediate energies. This will be exploited in the upcoming PANDA experiment at FAIR, where antiproton-proton and antiproton-nucleus interactions serve as diagnostic tools. The PANDA physics programme consists of four main physics domains: a) Nucleon structure b) Strangeness physics c) Charm and exotics and d) Hadrons in nuclei, as illustrated in Fig.~\ref{fig:map}.


\noindent The theory describing the strongly interacting quarks and gluons is Quantum Chromodynamics (QCD)~\cite{fritsch}. At high energies, or short distances, the strong coupling $\alpha_s$ is sufficiently weak to enable a perturbative treatment \textit{i.e.} pQCD. Quarks act as free particles due to \textit{asymptotic freedom}, an inherent property of QCD \cite{wilczek}, and the predictions from pQCD have been rigorously and successfully tested in experiments~\cite{PDG}. At low and intermediate energies, $\alpha_s$ increases and pQCD breaks down. The strongly interacting quarks and gluons are confined into \textit{hadrons} within a radius of $\approx$1 fm. A quantitative description of the strong interaction at the scale where quarks and gluons form hadrons and up to the onset of pQCD, belongs to the most challenging questions in contemporary physics. 
This manifests itself in the nucleon whose inherently non-perturbative properties such as the spin \cite{pspin1,pspin2} and, partly controversial, the mass \cite{pmass} remain objects of intense discussions and research. Understanding the former requires detailed knowledge about the distribution and motion of the quarks and gluons inside the hadrons. These can be quantified by \textit{e.g.} electromagnetic structure observables such as form factors and parton distributions. 

The mass of a purely light-quark system such as the nucleon, is to a very large extent generated dynamically by the strong interaction via the QCD intrinsically generated scale $\Lambda_{\rm QCD}$, rather than the Higgs mechanism. Nature is close to the chiral-limit case of massless quarks. Explaining the mass of nucleons and other hadrons requires a detailed theoretical understanding of the low-energy aspects of QCD, which goes hand in hand with the experimental determination of the hadronic excitation spectrum.
In particular, it is illuminating to study hadrons whose building blocks have different masses, from the massless gluons on one hand, to heavy quarks, \textit{e.g.} charm, on the other.
In the latter case the charm quark(s) can be viewed as a near static color source(s) surrounded by the
strongly interacting light quarks, a scenario that leads to additional, new forms of matter.

\textit{Glueballs}, suggested by theorists since more than 40 years \cite{glueballs1}, constitute one extreme since they consist solely of massless gluons. Hence, 100\% of the glueball mass is dynamically generated by the strong interaction. However, unambiguous evidence for their existence has not yet been found.
Also the mass of \textit{hybrids}~\cite{hybrids1}, states which consist of massive quarks and massless gluons, is expected to be largely generated dynamically.
 
The other extreme are "pure" quark systems containing heavier quarks, \textit{e.g.} strange or charm. The experimentally well-established \textit{hyperons} are baryons just like the nucleons, but contain one or several heavier quarks. Strange systems provide a bridge between nucleons, composed of essentially relativistic and non-perturbative quarks, on one side, and the fairly non-relativistic systems containing heavy charm or beauty quarks on the other.
With a strong coupling at the charm scale of $\alpha_s \approx$ 0.3, the validity of perturbative QCD becomes questionable. In general, we note that
the limit of applicability of perturbative QCD is by itself a complex subject. However, it is a reasonable approximation to describe states and
processes in terms of quark and gluon degrees of freedom.
Meson-like systems with hidden charm ($c\bar{c}$) show interesting features;
in particular the experimentally discovered states above the open-charm production threshold that do not
fit into the naive conventional quark-antiquark picture and, thereby, likely must have a more
complicated structure~\cite{ccbarrev1,rev_XYZ,Brambilla:2019esw}. These states are often referred to as
the \textit{XYZ} states, whereby $Z$ refers to isospin one cases, $Y$ to those states
that are formed in $e^+e^-$ annihilations with spin-parity of $J^{PC}$=$1^{--}$,
and $X$ to the other remaining resonances.
Even for the conventional nucleon,
the existence of nonperturbative intrinsic charm, a first-principle property of QCD, has been proposed~\cite{brodsky1980,harris1996,franz2000},
but experimentally not firmly confirmed.   

At the next level of complexity, where nucleons form nuclei, a long-standing question is how the nuclear force emerges from QCD. The short-distance structure of nuclei, studied in hadronic interactions with atomic nuclei, can shed light on this issue. At high energies, the strong interaction is predicted to be reduced due to \textit{colour transparency} \cite{colortrans}. At low energies, hadrons are implanted in the nuclear environment and form bound systems with finite life-time. Those could be \textit{hypernuclei} where one (or several) nucleon(s) in a nucleus is replaced by a hyperon. Studies of hypernuclei could shed light on the long-standing hyperon puzzle of neutron stars. Here, hyperon-nucleon and hyperon-hyperon interactions give rise to hyperon pairing which can suppress the cooling of neutron stars ~\cite{bielich:2008}. 

Finally, the validity and limitations of the SM itself remain an open question at the most fundamental level. One example is the matter-antimatter asymmetry, or baryon asymmetry, of the Universe, that cannot be explained within the SM. Unless fine-tuned in the Big Bang, the baryon asymmetry should be of dynamical origin, referred to as \textit{baryogenesis}~\cite{Sakharov:1967dj}. This would however require \textit{e.g.} CP violating processes to an extent that so far have not been observed experimentally.

To summarise, despite the many successes of the SM, many unresolved puzzles remain. Various efforts from both, theoretical and experimental frontiers are in progress or planned in the near future to address these puzzles~\cite{lutz:2015}. In this paper, we highlight PANDA, a future facility that will exploit the annihilation of antiprotons with protons and nuclei to shed light on the mysteries behind the fundamental forces in nature.
PANDA has unique features associated with the usage of antiprotons and a versatile detector that provide a complementary discovery potential with respect
to other facilities with the capability to carry out precision studies in the field of particle, hadron, and nuclear physics. In this paper, we outline these
features and give an overview of the PANDA physics objectives with emphasis on the programme foreseen for the first phase of operation of PANDA,
in the following referred to as \textit{Phase One}. The structure of the paper is as follows. First, we elaborate on the advantages of antiprotons as a probe. Next, we give a detailed presentation of the PANDA experiment in general and the Phase One conditions in particular. We go through each one of the PANDA physics sections and discuss their underlying purpose and aims, the present experimental status and the potential for PANDA Phase One. Finally, we conclude each part by providing a discussion on its impact and long-term perspectives in which we also briefly outline additional follow-up aspects for the subsequent phases of PANDA. 





\section{Opportunities with antiprotons}
\label{sec:antiprotons}



\begin{figure*}[t]
\centering
\includegraphics*[clip,trim= 2 2 2 2,width=\textwidth,angle=0]{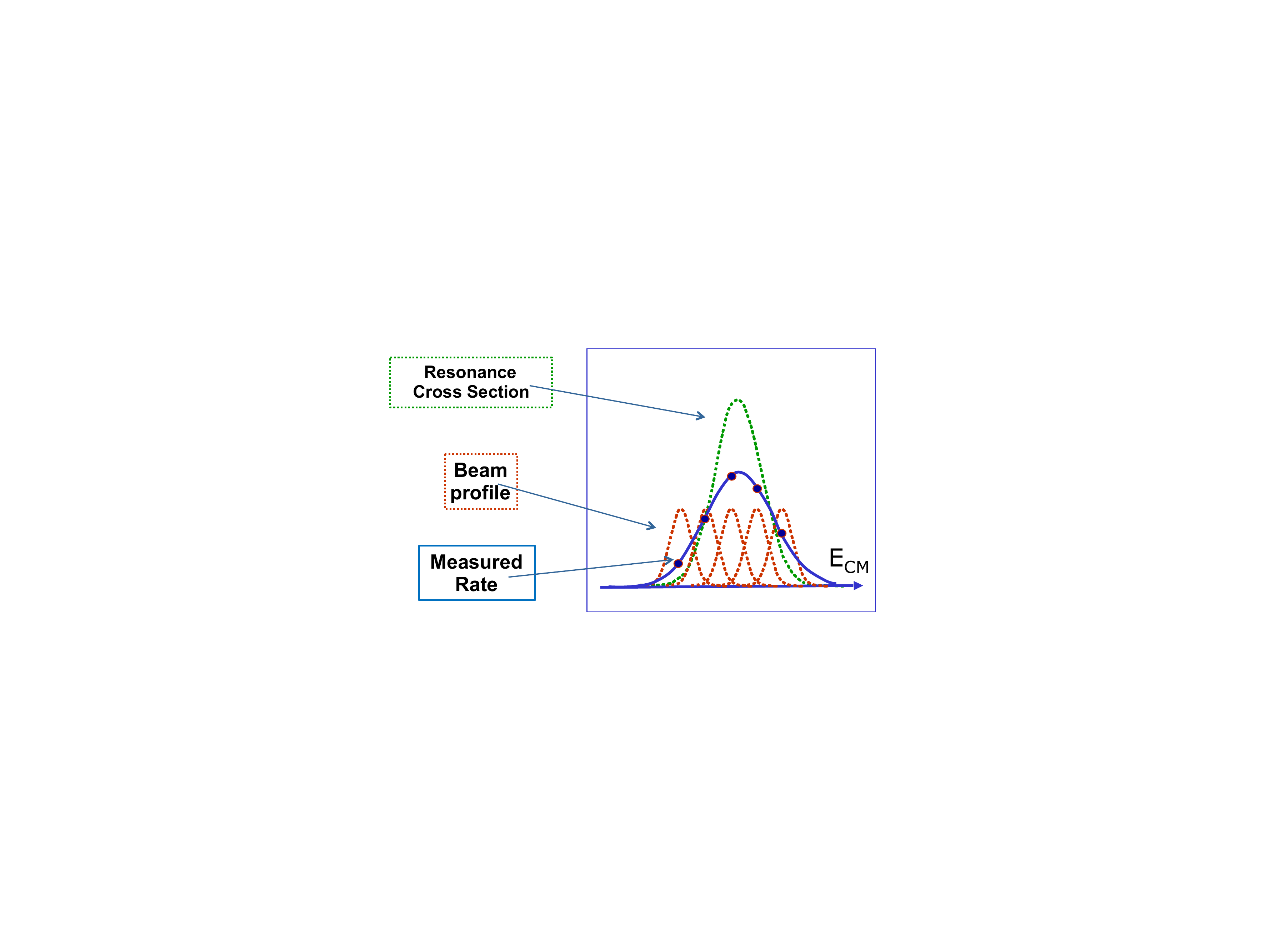}
\caption{Schematics of a resonance energy scan: The true energy dependent cross-section (dashed green line), the beam momentum spread (dashed red line), the measured yields (black markers), and the effectively measured energy dependent event rate (solid blue line) are illustrated.}
\label{fig:scanscheme1} 
\end{figure*}

\noindent The intense and precise antiproton beam foreseen in PANDA has many advantages:

\begin{itemize}
    \item The cross sections of hadronic interactions are generally large.
    \item Individual meson-like states can be produced in formation without severe limitations in spin and parity combinations.
    \item Baryons with various flavour, spin and parity can be produced in two-body reactions.
    \item The annihilation process could proceed via gluons and, in that case, providing a gluon-rich environment. 
\end{itemize}

\noindent In the following, we elaborate on these points in more detail.

The cross sections associated with antiproton-proton annihilations are generally several orders of magnitude larger than those of experiments using electromagnetic probes. This enables excellent statistical precision already at the moderate luminosities available in Phase One ($\sim$10$^{31}$~cm$^{-2}$s$^{-1}$). In particular, hadrons composed of strange quarks and hadrons with explicit gluon degrees-of-freedom are abundantly produced as demonstrated at a multitude of previous experiments at LEAR, CERN~\cite{Amsler19}. 

Hadronic reactions can be divided into two classes: \textit{formation} and \textit{production}. In formation, the initial systems fuse into one single state. The line shape of such a state can be determined from the initial system, using a technique called \textit{resonance energy scan}. The beam momentum is changed in small steps thereby varying the centre-of-mass energy in the mass region of the state of interest and the production rate is measured. Each resulting data point is a convolution of the beam profile and the resonance cross section according to Fig.~\ref{fig:scanscheme1}. The true energy-dependent cross section (green dashed line) is determined by the effectively measured cross section (solid blue line) based on the measured yields (markers) and the beam momentum spread (red dotted line).
The smaller the momentum spread of the beam, the more precise the measurement of the resonant line shape will be. In formation, the possible quantum numbers of the formed state depend on the probes. In $e^+e^-$ annihilations, processes in which the formed state has the same quantum numbers as the photon, \textit{i.e.} $J^{PC}~=~1^{--}$, are strongly favoured. States with any other quantum number are strongly suppressed and these therefore have to be produced together with a system of recoiling particles, \textit{i.e.} in \textit{production}, or from decays of the $1^{--}$ state. The disadvantage of production with recoils is that the state of interest needs to be identified by the decay products. As a consequence, the mass resolution is limited by the detector resolution, which is typically several orders of magnitude worse than the beam momentum spread. In antiproton-proton annihilations, any state with $\bar{q}q$-like, or non-exotic, quantum numbers can be created in formation. With a cooled antiproton beam, like the one foreseen for PANDA, the centre-of-mass energy resolution is excellent reaching a precision that is expected to be about 85~keV
at a centre-of-mass energy of 3.8~GeV at Phase One.
Experiments of this kind are therefore uniquely suited for precision studies of masses, widths and line-shapes of meson-like states with non-exotic quantum numbers that are different from $1^{--}$. A prominent example of this is the hidden-charm $\chi_{c1}(3872)$\footnote{This state is also known as the $X(3872)$. In this paper, we use the notation used by the Particle Data Group~\cite{PDG}.} with $J^{PC}=1^{++}$, that we will discuss further in Section \ref{sec:scan}. 
Furthermore, PANDA is unique in its capability to probe resonances with high spin. These are difficult to produce using electromagnetic probes, as well as in decays of \textit{e.g.} $B$ mesons.

Baryons and antibaryons can be produced in two-body reactions $\bar{p}p \to \bar{B}_1B_2$. The final state baryons can carry strangeness or charm provided the $\bar{B}_1B_2$ system is flavour neutral. In particular for multi-strange hyperons, this is an advantage compared to meson or photon probes, where strangeness conservation requires that the hyperon is produced with the corresponding number of associated kaons. As a result, the final state comprises at least three pseudo-stable particles, which complicates the partial-wave analysis necessary in hyperon spectroscopy. Two-body reactions on the other hand, in particular close to the kinematic threshold, typically involve few partial waves. Furthermore, spin observables and decay parameters can be accessed in a straight-forward way in two-body reactions. This enables production dynamics studies as well as charge conjugation parity (CP) symmetry tests in the strange sector. The particle-antiparticle symmetric final state minimises systematic uncertainties. In principle, the aforementioned advantages apply also for baryon-antibaryon production in $e^+e^-$ colliders. However, the typically much smaller cross sections result in low production rates. The resulting data samples are therefore smaller and in order to obtain sufficiently many events, methods such as missing kinematics or single-tag analysis are common. This however limits the possibility to reduce the background and achieve good resolution. In $\bar{p}p$ experiments, one can obtain large data samples also in exclusive analysis, which increases the discovery potential.

The $\bar{p}p \to X$ process includes quark-antiquark annihilations, which result in gluons. Therefore, antiproton-proton annihilation provides a gluon-rich environment, where states with a gluonic component are likely to be produced if they exist. Gluon-rich environments exist also in radiative decays of charmonia and in central hadron-hadron collisions. However, in radiative decays, reconstruction of the properties of the resonant state of interest relies solely on detector information since the process is not a formation process. As a result, the resolution is limited by the detector. The same is true for central hadron-hadron collisions, where the final state consists of the scattered hadrons and the produced resonance. The spin and parity of the resulting multi-particle final state is complicated to reconstruct without assumptions about the underlying production mechanism. This in turn leads to model-dependent ambiguities. The process $\bar{p}p \to X$, where $X$ refers to a single resonance, is less complicated in this regard.

The momentum range (1.5$-$15~GeV/$c$), momentum precision ($\Delta p/p$$\approx$10$^{-4}$$-$10$^{-5}$) and intensity ($\sim$10$^{31}$~cm$^{-2}$s$^{-1}$)
of the antiproton beam in
PANDA is tailored for strong interaction studies. PANDA will give access to the
mass regime up to 5.5~GeV/$c^2$ whereby recently new and interesting forms of hadronic matter have been observed ($XYZ$ states), it can study the
hadron-antihadron formation close to their production threshold, such as for the reactions
$\bar{p}p\rightarrow \Lambda \bar{\Lambda}, \Sigma \bar{\Sigma}, \Xi \bar{\Xi}, \Omega\bar{\Omega}, D\bar{D}, D_s\bar{D}_s,
\Lambda_c\bar{\Lambda}_c, \Sigma_c\bar{\Sigma}_c, \Xi_c\bar{\Xi}_c, \Omega_c\bar{\Omega}_c$, and it has the resolution to measure the line-shape of states
very accurately. 


\section{The PANDA experiment at FAIR}

The PANDA experiment is one of the four pillars of FAIR~\cite{Durante19}. PANDA will be a fixed-target experiment where the antiproton beam will impinge on a
cluster jet ($\bar{p}p$ or $\bar{p}A$) or pellet target ($\bar{p}p$) or target foils ($\bar{p}A$). The High Energy Storage Ring (HESR)~\cite{hesrtdr} can provide antiprotons with momenta from 1.5~GeV/$c$ up to 15~GeV/$c$. The physics goals of PANDA outlined in this paper require a detector system with nearly full solid-angle coverage, high-resolution tracking, calorimetry and particle identification over a broad momentum range as well as vertex reconstruction. 

The success of the physics programme will depend not only on the detector performance but also on the quality and intensity of the antiproton beam. Antiprotons are produced from reactions of 30~GeV/$c$ protons on a nickel or copper target. The source of these protons will be a dedicated high-power proton Linac followed by the existing SIS18 synchrotron and the new SIS100 synchrotron. Produced antiprotons are focused by a pulsed magnetic horn and selected in a magnetic channel at a momentum of around 3.7~GeV/$c$. After phase-space cooling in the Collector Ring (CR), packets of about $10^8$ antiprotons are transferred to the HESR for accumulation and subsequent acceleration or deceleration necessary for measurements in PANDA. In this mode of operation, the HESR is able to accumulate up to $10^{10}$ antiprotons from 100 injections within a time span of 1000~s. In a later stage of FAIR, the accumulation will take place in a dedicated ring, \textit{i.e.} the Recuperated Experimental Storage Ring (RESR), allowing for up to $10^{11}$ antiprotons to be injected and stored in the HESR. An important feature of the HESR is the versatile stochastic cooling system operating during accumulation and target operation. It is designed to deliver a relative beam-momentum spread ($\Delta{p}/p$) of better than $5\cdot 10^{-5}$. Furthermore, it includes a barrier bucket cavity that compensates for the mean energy loss in the thick target and that fine-tunes the absolute beam energy. This enables precise energy scans around hadronic resonances and kinematic thresholds. The centre-of-mass resolution will be about 50~keV, which to date is unreachable by other accelerators using different probes.




\subsection{Staging of the experiment}

The PANDA experiment will follow a staged approach in the construction of the detector and in the usage of the antiproton beam. 
It comprises four phases, briefly outlined below. 

The first phase, {\em Phase Zero}, started in 2018 and it refers to physics activities where PANDA detectors and analysis methods are used at existing and running facilities. One example is the usage of PANDA tracking stations in the upgraded HADES at GSI~\cite{Agakishievetal:2009}, another is the deployment of parts of the PANDA calorimeter for experiments with A1 at MAMI~\cite{Blomqvist:1998}. 

\begin{figure*}[t]
\includegraphics[width=0.9\textwidth]{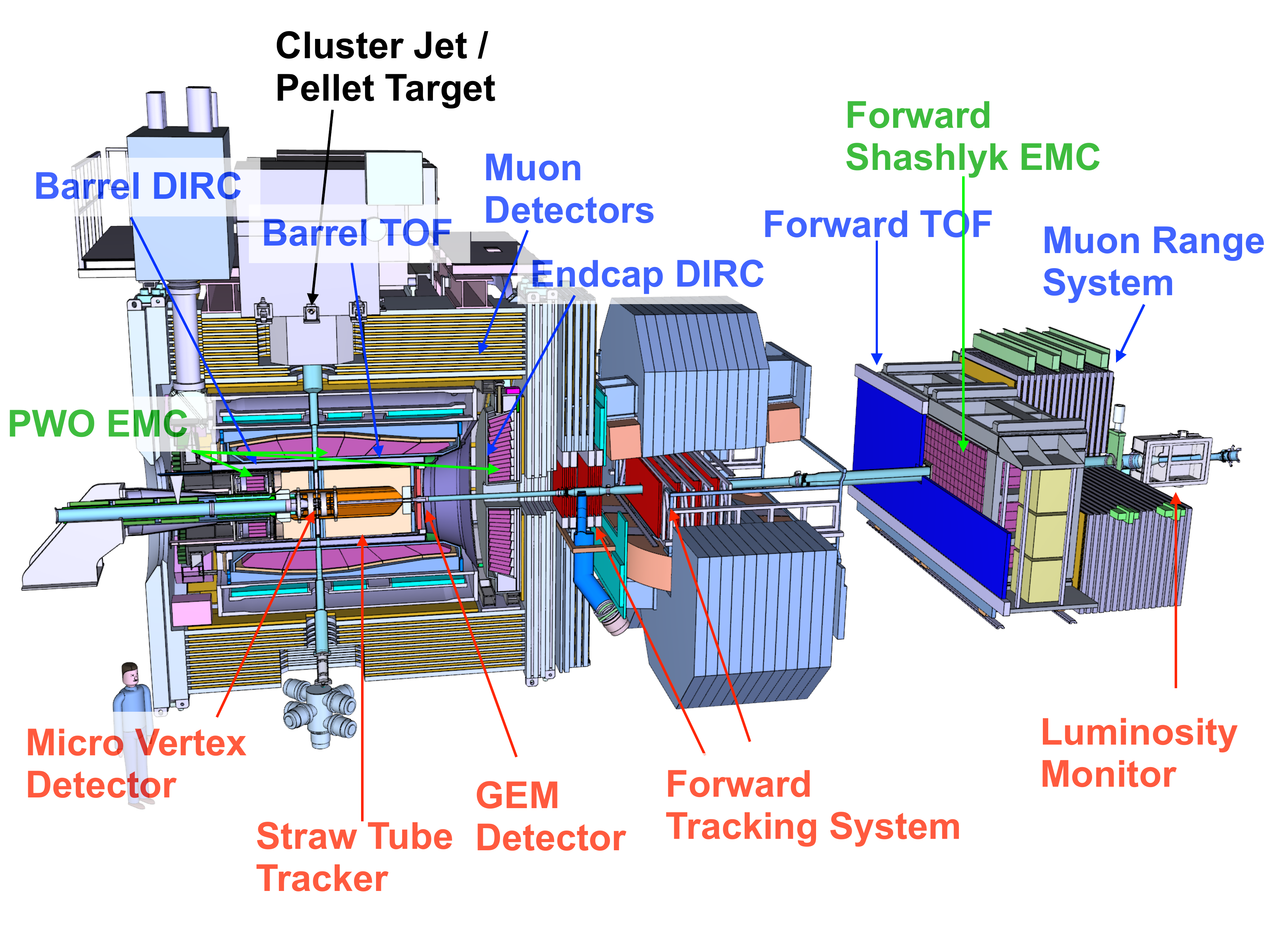}
\vspace*{-0.5cm}
\caption{Schematic overview of the start setup of PANDA. The various tracking detectors are indicated in red, the components for particle identification in blue, the electromagnetic calorimeters in green, and the target system in black. 
}
\label{setup}
\end{figure*}

The installation of the first major detector components of PANDA, including the two spectrometer magnets, will follow 
Phase Zero. 
This installation phase will be completed with a commissioning of the detectors using a proton beam at the HESR. 
The start of \textit{Phase One} will be marked with the usage of antiprotons together with the commissioned detectors. The corresponding physics programme is outlined in this paper. During Phase One, the HESR will be capable of accumulating at most 10$^{10}$ antiprotons in 1000~s. The luminosity is expected to rise gradually from about $10^{30}$ cm$^{-2}$s$^{-1}$ to the maximum of $2\times 10^{31}$~cm$^{-2}$s$^{-1}$ (at 15~GeV/$c$) during Phase One. The cluster jet target available from the beginning allows the use of hydrogen, deuterium or noble gases as target materials.
The available PANDA detector of Phase One will be referred to as the {\em start setup} and includes most of the major components as shown in Fig.~\ref{setup}. A description of the various available detector components will be given in section~\ref{start_setup}. The total integrated luminosity for Phase One is expected to be about 0.5~fb$^{-1}$.

The detector will be completed according to the final design in {\em Phase Two}. The main components beyond the start setup are the detector for charged particle identification in the forward region ({\it e.g.} forward RICH detector) and the completion of the Gas Electron Multiplier (GEM) tracker and forward trackers. Moreover, a pellet target system will become available. The corresponding setup will be referred to as the {\em full setup}. In {\em Phase Three}, the RESR will be available at FAIR which provides an increase in luminosity at HESR by a factor of approximately 20. 

\subsection{The PANDA start setup}
\label{start_setup}

To achieve the full physics potential of PANDA, the complete set of detector systems is needed. In Phase One, not all of these will be available and the focus is therefore on reactions with large expected cross sections and good signal-to-background ratios as well as relatively small multiplicities of final-state particles. 

In this section, we primarily describe the hardware systems to be installed as part of the {\em start setup}. 
The PANDA detector consists of two main parts: 
\begin{itemize}
\item The \textit{Target Spectrometer} (TS) for the detection of particles at large scattering angles ($>$~10$^{\circ}$ in the horizontal direction
  and $>$~5$^{\circ}$ in the vertical direction). The momentum measurement of charged particles is based on a superconducting solenoid magnet with a field strength of 2~T.
\item The \textit{Forward Spectrometer} (FS) for particles emitted in the forward direction ($\pm$5$^\circ$ vertically and $\pm$10$^\circ$ horizontally).
  The momentum measurement is based on a dipole magnet with a bending power of up to 2~Tm. 
\end{itemize}
The magnet system is described in Ref.~\cite{Magnet-TDR}. Both spectrometers are integrated with devices to perform tasks such as high resolution tracking, particle identification (PID), calorimetry and muon detection. 

The internal target operation of PANDA will employ a cluster jet target that can be operated with hydrogen as well as heavier gases. With hydrogen, an average luminosity of $10^{31}$cm$^{-2}$s$^{-1}$ can be reached in the experiment~\cite{Target-TDR}.

\subsubsection{The Target Spectrometer}
The beam-target interaction point will be enclosed by the Micro Vertex Detector (MVD) that will measure the interaction vertex position. It will consist of hybrid silicon pixels and silicon strip sensors. The vertex resolution is designed to be about 35~$\mu$m in the transverse direction and 100~$\mu$m in the longitudinal direction. Moreover, the MVD significantly contributes to the reconstruction of the transverse momentum of charged tracks~\cite{MVD-TDR}. The Straw Tube Tracker (STT) will surround the MVD with the primary purpose of measuring the momenta of particles from the curvature of their trajectories in the solenoid field. The low-mass (1.2\% of a radiation length) STT detector will consist of gas-filled straw-tubes arranged in cylindrical layers parallel to the beam direction. From these straws, a resolution better than 150~$\mu$m in the transverse $x$ and $y$ coordinates can be achieved. Some straw tube layers will be skewed with respect to the beam direction which enables an estimation of the $z$ coordinate along to the beam. The $z$ resolution will be approximately 3~mm. The STT will also contribute to the charged particle identification by measuring the energy loss $dE/dx$. Details of the STT can be found in Ref.~\cite{STT-TDR}. 
The PANDA Barrel DIRC~\cite{DIRC-TDR}, surrounding the STT, will cover the polar angle region between 22$^\circ$ and 140$^\circ$. 
The DIRC will be surrounded by a barrel-shaped Time of Flight (TOF) detector consisting of scintillating tiles read out by silicon photomultipliers.
The expected time resolution~\cite{TOF-TDR}, better than 100~ps, will allow precise particle identification of low momentum particles.
In addition it is used to assign a precise time information to reconstructed tracks to combine several tracks to one event of the same
antiproton-target interaction. The high time resolution and the long mean time between events during the phase-one operation of PANDA
will ensure that the mixing of events can be suppressed to a negligible amount.
The electromagnetic calorimeter (EMC), that will measure the energies of charged and neutral particles, will consist of three main parts: The barrel, the forward endcap and the backward endcap. The expected high count rates and the geometrically compact design of the target spectrometer require a fast scintillator material with a short radiation length and small Moli\`ere radius. Lead-tungstate (PbWO$_4$) fulfils the demands for photons, electrons and hadrons in the energy range of PANDA. The signals from the lead-tungstate crystals are read out by large-area avalanche photodiodes, except in the central part of the forward endcap where vacuum photo-tetrodes are needed for the expected higher rates. The EMC also plays an important role in the particle identification. In particular for electron/positron identification, it can suppress background from charged pions with a factor of about 1000 for momenta above 0.5~GeV/$c$.  A detailed description of the detector system can be found in Ref.~\cite{EMC-TDR}.  
The laminated yoke of the solenoid magnet, outside the barrel EMC, is interleaved with sensitive layers to act as a range system for the detection and identification of muons. Rectangular aluminium Mini Drift Tubes (MDT) are foreseen as sensors between the absorber layers. Details of this system are described in Ref.~\cite{Muon-TDR}. 
Downstream of the target, within the TS, a system of GEM foils will be located. The GEM planes will offer tracking of particles emitted with polar angles below 22$^\circ$, a region that the STT in the target spectrometer will not cover. In the start setup, two GEM stations will be installed. Part of the particles that pass the GEM tracking detector will be further registered by the Forward Spectrometer (FS) rather than the TS.

\subsubsection{The Forward Spectrometer}

The FS detector systems are conceptually similar to those of the TS, but will have a planar geometry instead of a cylindrical one. The detector planes will be arranged perpendicular to the beam pipe and thereby measure the deflection of particle trajectories in the field of the dipole magnet. 
Downstream of the GEMs, two pairs of straw tube tracking stations are foreseen for the start setup \cite{FTS-TDR}. One will be placed in front of the dipole magnet and the other inside its field. Particle identification will be provided by the Forward TOF wall consisting of scintillating slabs. The signals from the latter will be read out by photomultiplier tubes offering a time resolution better than 100~ps \cite{FTOF-TDR}. Forward-going photons and electrons will be detected and identified by a Shashlyk-type calorimeter with high resolution and efficiency. The detection is based on lead-scintillator sandwiches read out with wave-length shifting fibres passing through the block and coupled to photomultiplier tubes.  The system is described in detail in Ref.~\cite{FSC-TDR}. At the end of the FS, a muon range system is placed using sensors interleaved with absorber layers similar to the TS.

\subsubsection{Luminosity determination}

The luminosity at PANDA will be determined by using elastic antiproton-proton scattering as the reference channel. 
Since the Coulomb part of the elastic scattering can be calculated precisely and dominates at small momentum transfers, the polar angle of 3-8~mrad is chosen for the measurement. The track of each scattered antiproton and therefore the angular distribution of the tracks will be measured by the luminosity detector made of four layers of thin monolithic silicon pixel sensors (HV-MAPS)~\cite{LMD-TDR}.
An absolute precision of 5\% for the time integrated luminosity is expected and a relative precision of 1\% during the energy scans.

\subsubsection{Data acquisition}

The PANDA data acquisition concept is being developed to match the complexity of a next-generation hadron physics experiment. 
It will make use of high-level software algorithms for the on-line selection of events within the continuous data stream. This so-called software-based trigger system replaces the more traditional hardware-driven trigger systems that have been a common standard in the past.
In order to handle the expected Phase One event rate of  $2$~MHz, every subdetector system is a self-triggering entity.  Signals are detected autonomously by the sub-systems and are pre-processed in order to transmit only the physically relevant information. The online event selection occurs in computing nodes, which first perform event-building followed by filtering of physical signatures of interest for the corresponding beam-target settings. This concept provides a high degree of flexibility in the choice of trigger algorithms and hence a more sophisticated event selection based on complex trigger conditions, compared to the standard approach of hardware-based triggers.

\subsection{The simulation and analysis framework}


The feasibility studies presented in this paper have been carried out using a common simulation and analysis framework named \textit{PandaROOT}~\cite{PANDAROOT}. This framework provides a complete simulation chain starting from the Monte Carlo event generation, followed by particle propagation and detector response, signal digitisation, reconstruction and calibration, and finally the physics analysis.

PandaROOT is derived from the FairROOT framework~\cite{FAIRROOT} which is based on ROOT~\cite{ROOT}. FairROOT offers a large set of base classes which enables a straight-forward customisation for each individual detector setup. It offers an input-output manager, a run manager, database handling, an event display and the Virtual Monte Carlo (VMC) interface which allows to select different simulation engines. In addition, it uses the task system of ROOT to combine and exchange different algorithms into a simulation chain.

The first part in the simulation chain is the event generation. Here, the initial interaction of the antiproton beam with the target material is simulated using a Monte Carlo approach. Different generators exist for different purposes. Dedicated reactions and their subsequent decays are generated by the standard signal generator {\em EvtGen} \cite{EvtGen}. For the generic background, the {\em Dual Parton Model} (DPM) \cite{Capella1994} and the {\em Fritiof} (FTF) model \cite{FTF} can be chosen. Both include all possible final states and are tuned to an exhaustive compilation of experimental data. For detector- and software performance studies, the {\em BoxGenerator} creates single types of particles within user-defined momentum and angular ranges.

The generated particles are propagated through a detailed detector model, simulating the reactions with the detector material and possible decays in-flight. For this purpose, Geant3 and Geant4 are available to the user. The level of detail in the virtual detector description varies between the different subdetectors but all active components, as well as most of the passive material, are included. Separate descriptions are prepared for the start setup and the full setup. From this stage, the energy deposit, the position and the time of a given interaction in a sensitive detector element is delivered as output, all with infinite resolution. Real data will however consist of electronic signals with finite spatial- and time resolution. Therefore, the digitisation converts the information from the particle propagation stage into signals that mimic those of a real experiment. This includes noise and effects from discriminators and electronics. For some detector systems, the final electronics is not yet defined. In those cases, the digitisation procedure is based on realistic assumptions. 

In the reconstruction, the signals from the digitisation stage are combined into tracks. The procedure is divided into two steps: a local and a global part. In the local part, detector signals in a given tracking subdetector are combined into tracklets. Furthermore, the signal information is converted back to physical quantities such as position, energy deposit and time. In the global reconstruction, the tracklets from different tracking detectors are combined into tracks. Different algorithms are applied in the barrel part and the forward part. The track finding is followed by track fitting using a Kalman filter, where effects from different particle species and materials are taken into account. PANDA simulations thereby achieve a momentum resolution of about 1\%.

At the particle identification stage, the information from the dedicated PID detectors and the EMCs are associated with a charged track based on the distance between the predicted flight path and the hit position in the detector. Hits in the EMC without a corresponding charged track are regarded as neutral particles. The probabilities for various particle types of the different subdetectors are then combined into an overall probability of a given particle species.

The selection of events for partial or complete reaction channels, referred to as {\em Physics Analysis}, is performed based on the combined tracking, PID and calorimetry data using the Rho package, an integrated part of PandaROOT. With Rho, various constrained fits such as vertex fits, mass fits and tree fits are available.




\section{Nucleon structure}
\label{sec:nucleonstructure}
Hadron structure observables provide a way to test QCD and phenomenological approaches to the strong interaction in the confinement domain. Electromagnetic probes are particularly convenient and have been used extensively over the past 60 years. The structure is parameterised in terms of observables like \textit{form factors} or \textit{structure functions}.  

\begin{figure*}
\centering
\includegraphics[width=1.0\textwidth]{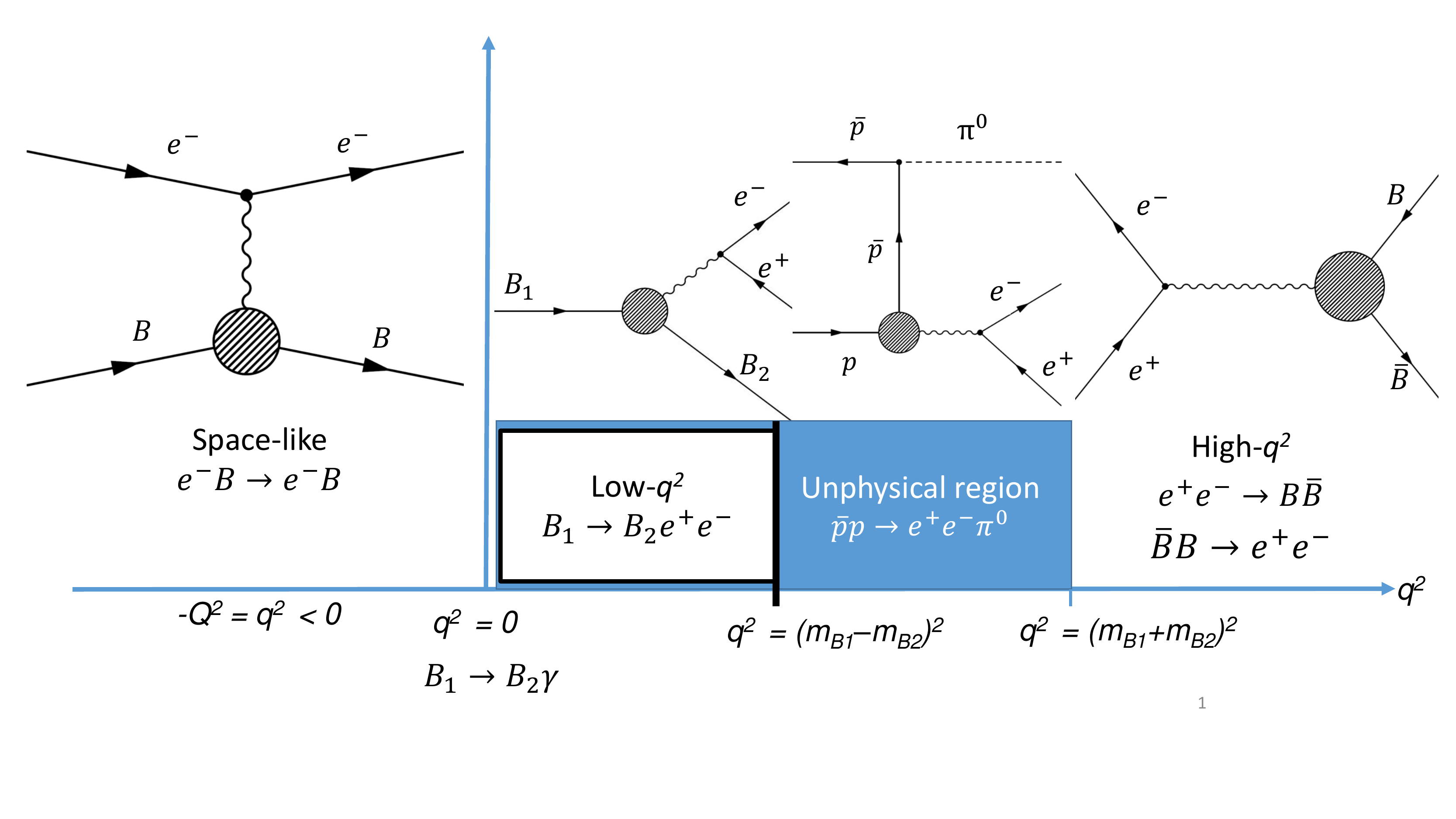}
\vspace*{-1.5cm}
\caption{Processes for extracting EMFF in the space-like (left) and time-like (right) regions. The low-$q^2$ ($q^2 < (M_{B1}-M_{B2})^2$) part of the time-like region is studied by Dalitz decays, the unphysical region ($4m_e^2 < q^2 < (M_{B1}+M_{B2})^2$) by $\bar p p \to \ell^+ \ell^- \pi^0$ and the high-$|q^2|$ region ($q^2 > (M_{B1}+M_{B2})^2$) by $B\bar{B} \leftrightarrow e^+e^-$. 
}
\label{fig:q2EMFF}
\end{figure*}

Electromagnetic form factors (EMFFs) quantify the hadron structure as a function of the four-momentum transfer squared $q^2$. At low energies, they probe distances of about the size of a hadron. EMFFs are defined on the whole $q^2$ complex plane and for $q^2 < 0$, they are referred to as \textit{space-like} and for $q^2 > 0$ as \textit{time-like}. Space-like EMFFs are real functions of $q^2$ and can be studied in elastic electron-hadron scattering. Assuming one-photon exchange (OPE) being the dominant process, protons and other spin-1/2 particles are described  by two EMFFs: the electric $G_E(q^2)$ and the magnetic $G_M(q^2)$ form factor. In the so-called \textit{Breit frame}, these are the Fourier transforms of the charge and magnetisation density, respectively. 
Time-like EMFFs are complex and can be studied using different processes in different $q^2$ regions. In the following, we consider baryons, denoted $B$, $B_1$ and $B_2$. For unstable baryons, the low-$q^2$ ($q^2 < (M_{B1}-M_{B2})^2$) part of the time-like region is probed by Dalitz decays, \textit{i.e.} $B_1 \to B_2 \ell^+ \ell^-$. For the proton, the so-called \textit{unphysical} region ($4m_l^2 < q^2 < (M_{B1}+M_{B2})^2  = 4M_p^2$ with $m_l$ the mass of the lepton $l$) can be probed by the reaction $\bar p p \to \ell^+ \ell^- \pi^0$. For all types of baryons, the high-$q^2$ region ($q^2 > (M_{B1}+M_{B2})^2$) can be accessed by $B\bar{B} \leftrightarrow e^+e^-$. If $B_1 = B_2 = B$, then the form factors are \textit{direct}, whereas if $B_1 \neq B_2$, \textit{transition} form factors are obtained. Being analytic functions of $q^2$, space-like and time-like form factors are related by dispersion theory. The processes for studying EMFFs at different $q^2$ are summarised in Fig.~\ref{fig:q2EMFF}.

At high energies, corresponding to distances much smaller than the size of a hadron, individual building blocks are resolved
rather than the hadron as a whole. Here, the \textit{factorisation theorem} applies, stating that the interaction can be
factorised into a hard, reaction-specific but perturbative and hence calculable part and a soft, reaction-universal and measurable part.
In the space-like region, probed by deep inelastic lepton-hadron scattering, the structure is described by parton distribution
functions (PDFs) \cite{Collins1989}, generalised parton distributions (GPDs) \cite{Ji:1996nm,Radyushkin:1996nd,Mueller:1998fv,Collins:1996fb,Berger:2001xd,Radyushkin:1998rt,Diehl:1998kh}, 
transverse momentum dependent parton distribution functions (TMDs) \cite{Barone2010}, and transition distribution amplitudes (TDAs) \cite{Frankfurt:1999fp,Pire:2004ie}.
These non-perturbative objects are complementary tools to explore the structure of the nucleon at the partonic level. They extend the
information given by the EMFFs and provide more detailed descriptions of the spatial and momentum distributions of the constituent partons
and the spin structure. In the time-like region, they can be accessed experimentally in hard hadron-antihadron annihilations.
Detailed studies to access $\pi N$ TDAs at PANDA in the reactions $\bar p p \to \gamma \pi^0 \to e^+ e^- \pi^0$ and
$\bar p p \to J/\Psi \pi^0 \to e^+ e^- \pi^0$
have been presented in Refs.~\cite{Singh:2014pfv,Singh:2016qjg}.
For these measurements, as well as for the TMD studies, the designed high luminosity of PANDA is needed to accumulate reasonable statistics.
The counterparts of the GPDs in the annihilation processes are the generalized distribution amplitudes (GDAs).
They can be measured in the hard exclusive processes $\bar p p \to \gamma \gamma$ \cite{Freund:2002cq}
and $\bar p p \to \gamma M$ \cite{Kroll:2005ni,Kroll:2013kra}, where $M$ could be a pseudo-scalar or vector meson (e.g. $\pi^0,~\eta,~\rho^0,~\phi$).
Cross section measurements as a function of $s$ and $t$ that allow us to test the theoretical models are expected to start in Phase Two of PANDA.
In the following, we focus on EMFFs, in line with the emphasis of Phase One.

\subsection{State of the art}
 
Elastic electron-proton scattering has been studied since the 1950s \cite{Hofstadter:1956qs}. During the first decades, unpolarised electron-nucleon scattering was analysed using the Rosenbluth separation method \cite{rosenbluth}. Modern facilities, offering high-intensity lepton beams and high-resolution detectors, gave rise to a renewed interest in the field \cite{Perdrisat:2006hj,Puckett:2010ac}.
In particular, the polarisation transfer method \cite{Akhiezer:1968ek} applied by the JLab-GEp collaboration
(see \cite{Puckett:2010ac} and references therein) revealed the surprising result that the ratio $G_E/G_M$ decreases almost linearly with $Q^2 = -q^2$.
This result is in contrast to previous measurements based on a Rosenbluth method using unpolarised elastic $ep$ scattering which do not reveal such a dependency.
The correction of the unpolarised elastic $ep$ cross section by the two-photon exchange (TPE) contribution has been suggested to solve this
discrepancy~\cite{guichon:2003}. The TPE correction does not impact the polarization transfer extraction of $G_E/G_M$ in a significant way.
The large amount of high-quality data inspired extensive activity also on
the theory side, from which we have learned about the importance of vector dominance at low $q^2$ \cite{vectordom,hammer2020}.

Until recently, measurements in the time-like region have not achieved precisions comparable to the corresponding space-like data, partly because most $e^+e^-$ colliders have been optimised in different $q^2$ regions \cite{Denig2013,Pacetti2015}. In $\bar p p$ annihilation experiments, the clean identification of $e^+e^-$ pairs has been a challenge. Among the few experiments that so far have provided a separation between $G_E$ and $G_M$ of the proton, the results at overlapping energies disagree. The ratio $R=|G_E|/|G_M|$, accessible from the final state angular distribution, has been measured below $q^2=9$ (GeV/$c$)$^2$ by PS170 at LEAR \cite{Bardin:1994am}, BABAR \cite{Lees:2013uta} and more recently by BESIII \cite{Ablikim:2015vga, Ablikim:2019njl, Ablikim:2021kjh} and CMD-3  \cite{Akhmetshin:2015ifg}. The PS170 and BABAR data differ up to 3$\sigma$, while the BESIII and CMD-3 measurements have large total uncertainties. In the limit $|q^2| \to \infty$, the space-like and the time-like form factors should approach the same value as a consequence of the Phragm{\'e}n-Lindel\"of theorem \cite{phragmen}. Experimentally, the onset of this scale has not been established (see Ref. \cite{Pacetti2015} for a recent review). In measurements just below $|q^2| = 20.25$ (GeV/$c$)$^2$, the time-like magnetic form factor is about two times larger than the corresponding space-like one. A recent analysis of BaBar data above $|q^2| = 20.25$ (GeV/$c$)$^2$, indicates a decreasing difference, but the uncertainties are large \cite{Lees:2013uta}. 

In 2019, the BESIII collaboration measured the Born cross section of the process $e^+ e^- \to \bar{p} p$ and the proton EMFFs at 22 centre-of-mass energy points from $q^2 = 4$ (GeV/$c$)$^2$  to $q^2 = 9.5$ (GeV/$c$)$^2$ with an improved accuracy \cite{Ablikim:2019eau}, comparable to data in the space-like region. Uncertainties on the form factor ratio $|G_E|/|G_M|$ better than $10\%$ have been achieved  at different $q^2$ values below $~5$ (GeV/$c$)$^2$.  The BESIII data on the proton effective form factor confirm the structures seen by the BABAR Collaboration. These structures are currently the subject of several theoretical studies~\cite{Haidenbauer:2014kja,Lorentz2015,Bianconi2016}.

The PANDA experiment aims to improve the current situation of the time-like EMFFs by providing data in a large kinematic region between $5.08$ (GeV/$c$)$^2$ and $\sim 30$ (GeV/$c$)$^2$. Precisions in this region of at least a factor 3 better than the current data, as well as measurements in the unphysical region below $(2M_p)^2$, whereby $M_p$ is the mass of the proton, are called for to constrain the theoretical models and to resolve the aforementioned issues.

\subsection{Potential of Phase One}

The PANDA experiment in Phase One offers the opportunity to measure the proton form factor in the process
$\bar p p \to \ell^+ \ell^-,~(\ell=e,~\mu)$ over a wide energy range, including the high $|q^2|$
region~\cite{Singh:2016dtf, Zimmermann2020}. The $\bar p p \to \mu^+ \mu^-$ reactions can be studied for the first time.
The interest for $\bar{p}p$ annihilation into heavy leptons ($\mu$ and $\tau$) has been discussed in several theory
studies~\cite{Zichichi:1962ni,Gakh:2005wa,Dbeyssi:2011tv}.
In contrast to the $\bar p p \to e^+ e^-$ process,
the $\bar p p \to \mu^+ \mu^-$ reaction has the advantage that
corrections due to final state radiation are expected to be smaller. Measuring both channels should therefore
allow the formalism for radiative corrections to be tested.

Furthermore, the unphysical region of the proton EMFFs can be accessed through the measurement of  the $\bar p p \to \ell^+ \ell^- \pi^0$ process \cite{dubnickova96,Adamuscin:2006bk,Guttmann:2012sq}.  
These measurements by PANDA are unique and will provide the possibility to test models for this process that contain EMFFs~\cite{Boucher:2011}.

\subsubsection{EMFFs in $\bar p p \to e^+ e^-$}
A previous simulation study of the process $\bar p p \to e^+ e^-$ within the PandaROOT framework demonstrates the excellent prospect of nucleon structure studies with the PANDA design luminosity  \cite{Singh:2016dtf}. The simulations were performed applying an integrated luminosity of 2~fb$^{-1}$ for each energy-scan point and the full PANDA setup corresponding to the conditions that are foreseen for Phase Three of PANDA. A new, dedicated simulation study with the Phase One conditions has recently been performed at $q^2=5.08$ and  $8.21$ (GeV/$c$)$^2$ ($p_{\rm lab}=1.5$ and 3.3 GeV/$c$, respectively). The difficulty of the measurement is related to the hadronic background, mostly annihilation with the subsequent production of two charged pions. This reaction has a cross section about five to six orders of magnitude larger than that of the production of a lepton pair. In the energy scale of the PANDA experiment, the mass of the electron is sufficiently close to the pion mass for this to be an issue. Therefore, the signal and the main background reactions have very similar kinematics. The signal events are generated according to the differential cross section parameterised in terms of proton EMFFs from Ref.~\cite{FFtomasi} with setting $R=|G_E|/|G_M|=1$. The same event selection criteria as in Ref.~\cite{Singh:2016dtf} were applied. The output of the PID and tracking subdetectors as EMC, STT, MVD, and barrel DIRC have been used to separate the signal from the background. These resulted in signal efficiencies of 40\% at $p_{\rm lab}$ = 1.5~GeV/$c$ and 44\% at $p_{\rm lab}$ = 3.3~GeV/$c$. The suppression factor of the main background process $\bar p p \to \pi^+ \pi^-$ was found to be of the order $\sim10^{8}$. The proton form factors $|G_E|$, $|G_M|$, and their ratio $R=|G_E|/|G_M|$ are extracted from the electron angular distribution, after reconstruction and efficiency correction. The proton effective form factor $|G^e_{\rm eff}|$ is extracted from the determined cross section of the signal ($\sigma$) integrated over the electron polar angle. The resulting precision for different $q^2$ is summarised in Table~\ref{tab:PEMFF} and shown in Fig.~\ref{figEMFF}, together with existing experimental data. Systematic uncertainties arise due background contamination and uncertainties in the luminosity measurement. These effects can be quantified by MC simulations. From these we conclude that the proton EMFFs can be measured with an overall good precision and accuracy. At low $q^2$, the signal event yield is relatively large. However, at higher $q^2$, the cross section of the process reduces significantly which leads to a smaller event yield and thus larger statistical uncertainties for a given integrated luminosity. Previous studies show that the efficiency at larger $q^2$ is sufficient for precise cross section measurements \cite{Singh:2016dtf}.

\begin{table*}
\centering
\caption{Results from simulation studies of $\bar p p \to e^+ e^-$ and $\bar p p \to \mu^+ \mu^-$. }
\label{tab:PEMFF}       
\begin{tabular}{lllllll}
\hline
$q^2$ / $({\rm GeV}/c)^2$ & Reaction & $L$ / fb$^{-1}$ & $\sigma_{\sigma}$ (\%) & $\sigma_{R}$ (\%) & $\sigma_{G_E}$ (\%) & $\sigma_{G_M}$  (\%)\\\hline
5.08 & $\overline{p}p \rightarrow e^+ e^- $ & 0.1 & 5.2 & 4.2 & 3.3 & 3.2  \\\hline
8.21 & $\overline{p}p \rightarrow e^+ e^- $ & 0.1 & 5.2 & 26 & 21 & 5.9  \\\hline
5.08 & $\overline{p}p \rightarrow \mu^+ \mu^- $ & 0.1 & 5.0 & 21 & 14 & 6.9  \\\hline
\end{tabular}
\end{table*}

\subsubsection{EMFFs in $\bar p p \to \mu^+ \mu^-$}
\begin{figure*}
\centering
\subfloat[\label{ff1}]{\resizebox{0.5\textwidth}{!}{\includegraphics[width=.5\textwidth]{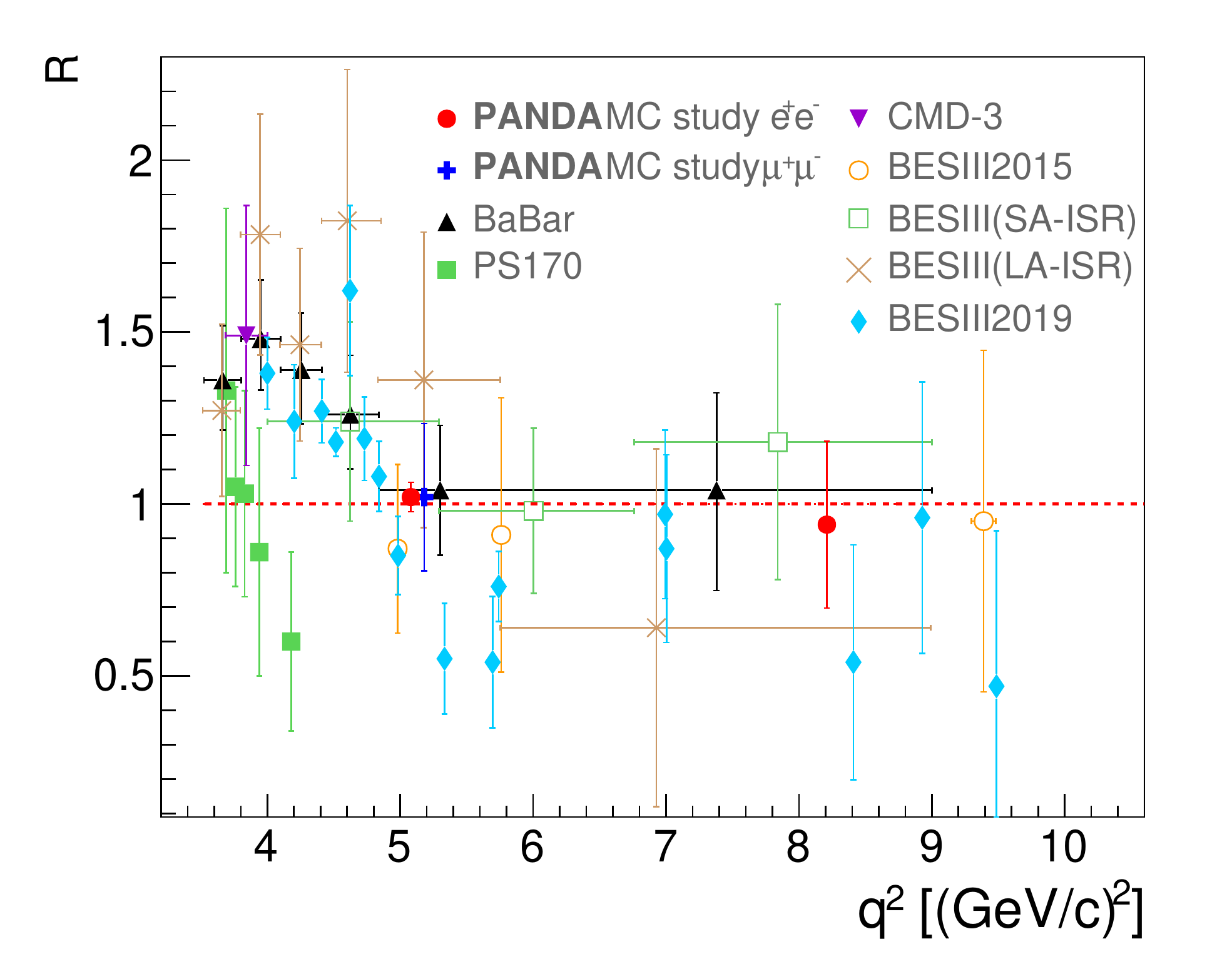}}}
\subfloat[\label{ff2}]{\resizebox{0.5\textwidth}{!}{\includegraphics[width=.5\textwidth]{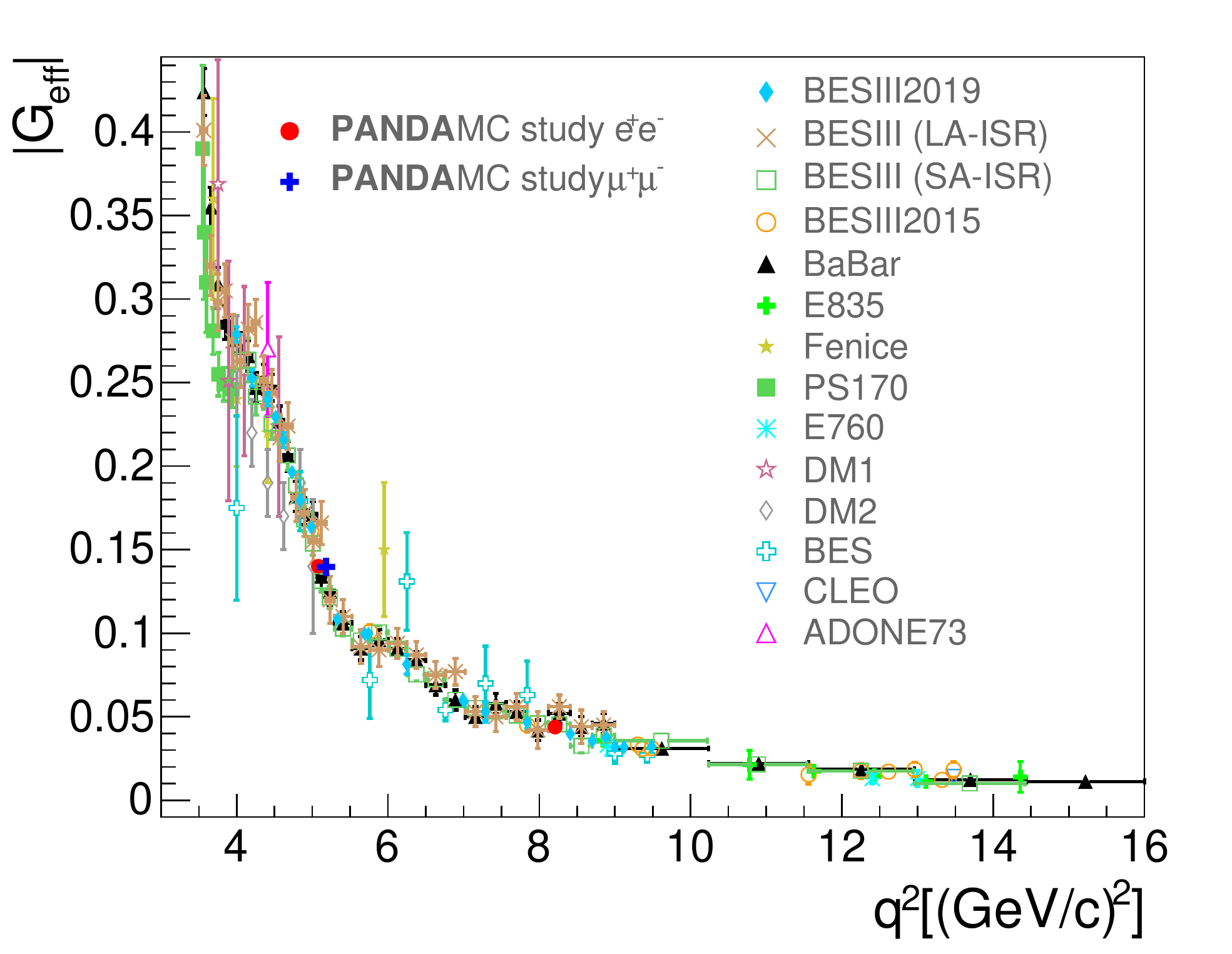}}} \\
\subfloat[\label{ff3}]{\resizebox{0.5\textwidth}{!}{\includegraphics[width=.5\textwidth]{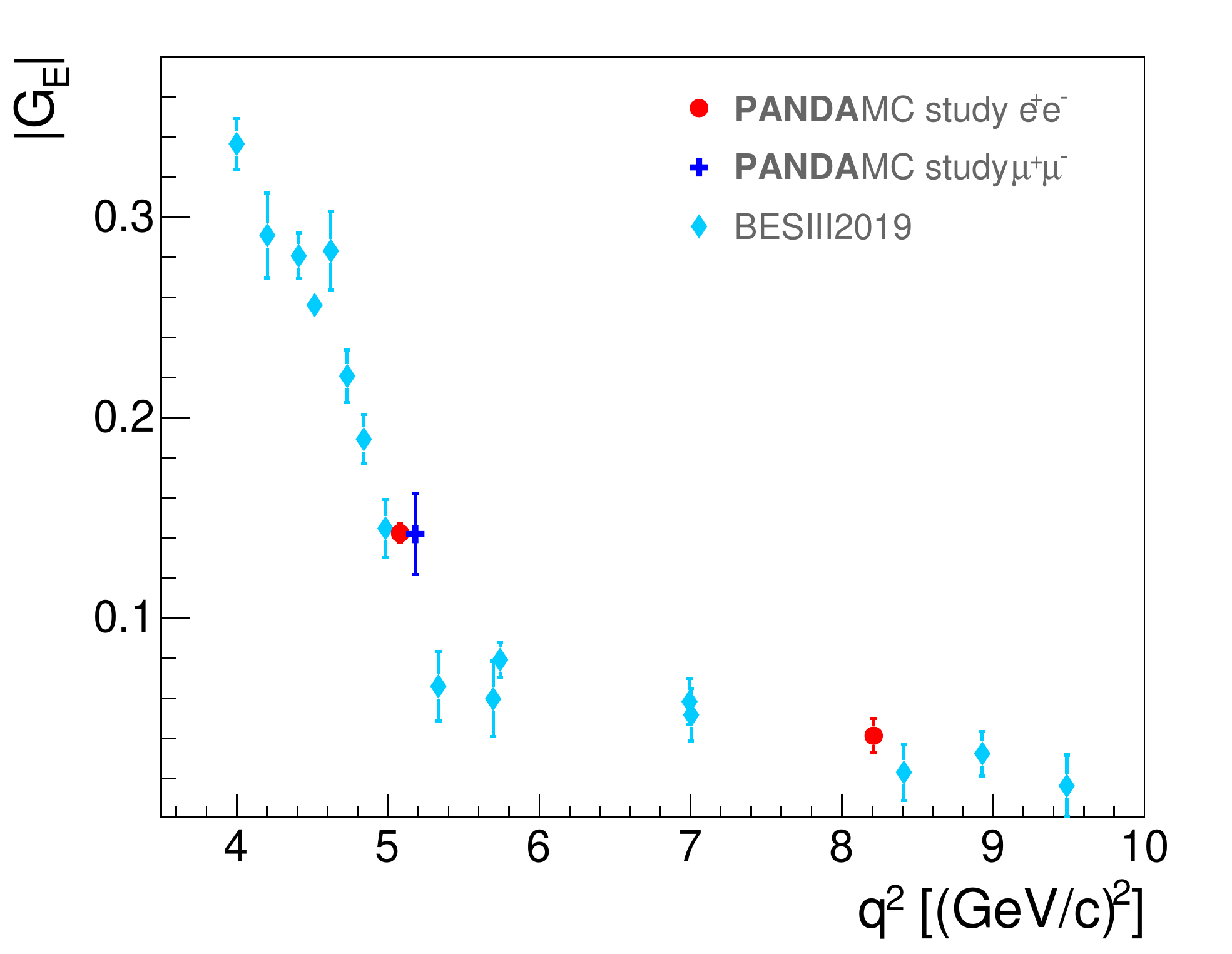}}}
\subfloat[\label{ff4}]{\resizebox{0.5\textwidth}{!}{\includegraphics[width=.5\textwidth]{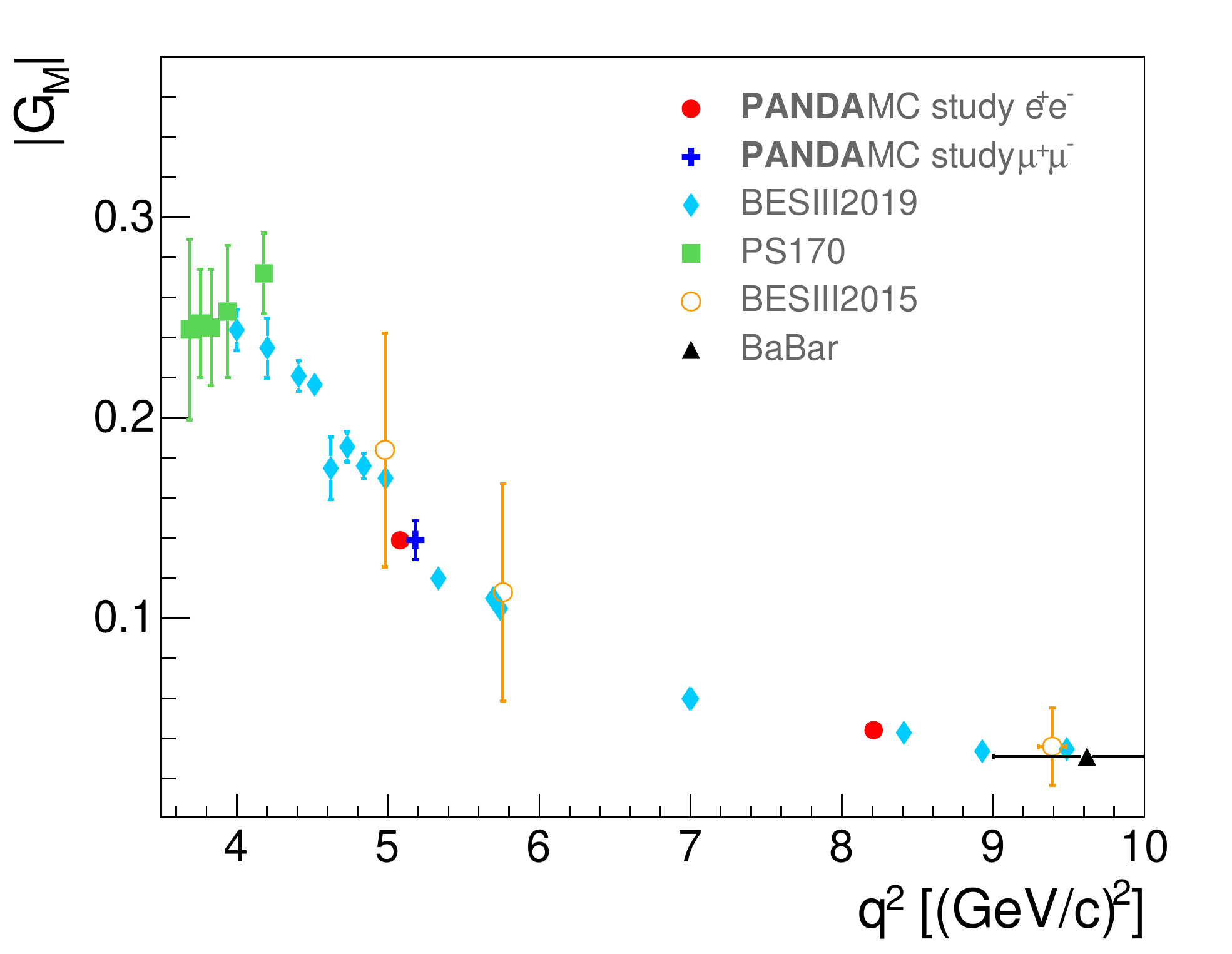}}}
\caption{Expected total precisions, indicated by the red and blue error bars, on the determination of (a) the proton form factor ratio, (b) the proton effective form factor, (c) the proton electric form factor, and (d) the proton magnetic form factor, from the present simulations for PANDA Phase One as a function of $q^2$. Also shown are data from  PS170~\cite{Bardin:1994am}, BaBar~\cite{Lees:2013b,Lees:2013uta},
  BESIII~\cite{Ablikim:2015vga, Ablikim:2019eau, Ablikim:2019njl, Ablikim:2021kjh},
  CMD-3~\cite{Akhmetshin:2015ifg}, E835~\cite{e835},  Fenice~\cite{fenice},  E760~\cite{e760},  DM1~\cite{dm1},  DM2~\cite{dm2},  CLEO~\cite{cleo},   and ADONE73~\cite{adone73}. The results indicated by
  PANDA MC study are based on an integrated luminosity of 0.1~fb$^{-1}$ for each point in $q^2$.}
\label{figEMFF}
\end{figure*}

An independent Monte Carlo simulation study of the $\bar p p \to \mu^+ \mu^-$ reaction has been carried out at $q^2 = 5.08$ (GeV/$c$)$^2$.
The di-muon channel provides a clean environment, where radiative corrections from final state photon emissions are reduced thanks to the
larger mass of the muon. However in case of muons, the suppression of the hadronic background  $\bar p p \to \pi^+ \pi^-$ is more challenging.
Muon identification is mainly based on the information from the Muon System, since other subdetectors show less separation power which complicates
the background separation considerably.  Monte Carlo samples of $10^8$ events were generated for the background process $\bar p p \to \pi^+ \pi^-$.
They were  used for the determination of the background suppression factor and for the calculation of the pion contamination,
which will remain in the signal events after the application of all selection criteria. The separation of the signal from the background
has been optimised through the use of multivariate classification methods (Boosted Decision Trees). The event selection is described in
Ref.~\cite{Zimmermann2020}. A background rejection factor of $1.2 \times 10^{-5}$ was achieved, resulting in a signal-to-background ratio of 1:8.
The total signal efficiency is $31.5\%$.  Due to the insufficient background rejection, the pion contamination needs to be subtracted
from the signal and the corresponding angular distributions by Monte Carlo modelling and subsequent subtraction.
This has been taken into account in our feasibility studies. The angular distributions from the pion contamination
are reconstructed with both the expected magnitude and shape. The sensitivity of the EMFFs to the shape was investigated and from that,
the systematic uncertainty was estimated. The ratio $R$, and consequently $|G_E|$ and $|G_M|$, were extracted from the angular distribution
of the muons after background subtraction and  efficiency correction. The results are summarised in Fig.~\ref{figEMFF} and Table~\ref{tab:PEMFF}.
The uncertainty of the signal cross section is dominated by the luminosity uncertainty. The simultaneous but independent measurements
of the effective EMFFs $G_{\rm eff}^e$ and $G_{\rm eff}^{\mu}$ from the $e^+e^-$ final state and the $\mu^+\mu^-$ final state, respectively,
enable a test of radiative corrections that are applied in the $e^+e^-$ channel. The expected uncertainty in the ratio
$G_{\rm eff}^e/G_{\rm eff}^{\mu}$ is estimated to be 3.2\% already during Phase One. It should be noted that although the
uncertainties from radiative corrections are not yet taken into account, these are expected to contribute with only
a small fraction to the total uncertainty.

\subsubsection{EMFFs in $\bar p p \to e^+ e^- \pi^0$}
Some information about the unphysical region can be obtained from the $\bar p p \to e^+ e^- \pi^0$ process, when studied in different intervals of the pion angular distribution. In the time-like region, the EMFFs are complex, hence they have a relative phase. This phase is generally inaccessible for protons in an experiment with an unpolarised beam or target. However, the cross section of $\bar p p \to e^+ e^- \pi^0$ channel can provide some information, as outlined in Refs.~\cite{dubnickova96,Guttmann:2012sq}.

The validity of the theoretical models used to describe the cross section of the process $\bar{p} p \to e^+ e^- \pi^0$ needs to be tested experimentally. Since PANDA has almost $4\pi$ coverage, the measurement of the final state angular distributions in the processes $\bar{p} p \to e^+ e^- \pi^0$ and $\bar p p \to \gamma \pi^0$ will provide a sensitive check of these models. The EMFFs extracted at threshold via $\bar p p \to e^+ e^- \pi^0$ and $\bar{p} p \to e^+ e^-$ or  $ e^+  e^- \to \bar{p} p$  can be compared and used as an additional test. We note that the process $\gamma p \to p  e^+ e^- $ may give access to the EMFFs of the proton in the unphysical region as well.
The feasibility to extract EMFFs from such challenging measurements has so-far not been demonstrated as discussed in
theoretical studies in Refs.~\cite{Schafer:1994vr,Dieperink:1997jh}.

For an ideal detector ($100\%$ acceptance and efficiency) and an integrated luminosity of 0.1~fb$^{-1}$, the expected count rate for this reaction for $q^2 < 2$ (GeV/$c$)$^2$ has been found to be up to $10^5$ events in different intervals of the pion angular distribution \cite{Boucher:2011,Singh:2014pfv}. This number is about a factor two larger than the corresponding value for $\bar p p  \to e^+ e^-$ at $q^2=5.08$ (GeV/$c$)$^2$. The large expected count rate of $\bar p p \to e^+ e^- \pi^0$ and the clean separation between this channel and background \cite{Boucher:2011}, indicate good prospects for EMFF measurements in the unphysical region already in PANDA Phase One. Full simulation studies to investigate the possibility to extract the proton EMFFs in this region at PANDA are currently being carried out.

\subsection{Impact and long-term perspective}

The simulation studies presented in the previous sections show that PANDA will improve the precision of the proton EMFF measurements
for $q^2 > 5.08$ (GeV/$c$)$^2$. The measurement of the effective form factor can be extended to higher $|q^2|$ values in Phase Three of the experiment
when the luminosity reaches its design value. This enables systematic comparisons of space-like and time-like EMFFs at large $|q^2|$ and hence,
the onset of the convergence scale of the space-like and time-like form factors may be deduced. Furthermore, the foreseen
PANDA studies of the $\bar p p \to \mu^+ \mu^-$ are unique. Since the effects from final state radiation are negligible for muons,
this channel provides an important cross check of the $\bar p p \to e^+ e^-$ results.
Finally, in PANDA, the unphysical region of the proton EMFF will be accessed for the first time through the $\bar p p \to e^+ e^- \pi^0$ process. 

In general, the relative phase between the electric and the magnetic form factors is inaccessible in unpolarised cross section measurements. To measure the phase, either a polarised antiproton beam and/or a polarised proton target is required. The feasibility of implementing a transversely polarised proton target in the PANDA detector is under investigation. If feasible, the PANDA experiment will offer a first direct measurement of the relative phase between $G_E$ and $G_M$.  

In the next stages of PANDA, we will be able to extend our studies to the hard exclusive processes and perform complementary measurements of the
nucleon structure objects.

\section{Physics with strangeness}
 The key question in hyperon physics is ``What happens if you replace one (or several) light quark(s) in the nucleon with one (or several) heavier one(s)?''.  Strangeness serves as a diagnostic tool for various phenomena in subatomic physics:
\begin{enumerate}
\item  Hyperons provide a new angle to the structure and excitations of the nucleon, since the strange quark is sufficiently light to relate the knowledge about hyperons to nucleons and vice versa.
\item Hyperon decays, where the spin is experimentally accessible, provide an ideal testing ground for CP violation and thereby searches for physics beyond the SM at the precision frontier. Furthermore, it can give clues about Baryogenesis \cite{Sakharov:1967dj}. 
\item In hypernuclei, strangeness provides an additional degree of freedom which plays a key role in understanding \textit{e.g.} neutron stars \cite{Vidana2018}.
\item Enhancement of strangeness in relativistic heavy-ion collisions was one of the first proposed signals of Quark-Gluon Plasma \cite{rafelski}.
\end{enumerate}

\noindent Number 1 will be explored with PANDA Phase One within the subtopics \textit{hyperon production} and \textit{hyperon spectroscopy}. Number 2, \textit{i.e. hyperon decays} will be studied extensively in Phases Two and Three. However, a good understanding of the production mechanism has proven crucial to decay measurements \cite{bes3nature} and the planned hyperon production studies within Phase One are therefore an important milestone in the search for CP violation in baryon decays. Number 3 will be investigated during Phases Two and Three within our programme for hadrons in nuclei.
Number 4 is currently studied at ALICE \cite{alicehyp} and is not within the scope of PANDA. However, precision studies of strangeness production in elementary $\bar{p}p$ reactions contribute to a more general understanding of strangeness production, which can be useful also in more complex reactions at higher energies. The same is true for the planned studies of hyperon-antihyperon pair production in $\bar{p}N$ reactions. These will provide information on absorption and rescattering of hyperons as well as antihyperons under well-defined conditions in cold nuclei. In this chapter, we discuss the subtopics hyperon production and hyperon spectroscopy in the context of what can be achieved at Phase One. In section \ref{sec:antihyp} we also discuss anti-strange hadrons in nuclei.

\begin{figure*}[ht]
\begin{minipage}{.4\textwidth}
\centering
\hspace{-10mm}
\includegraphics[height=0.18\textheight]{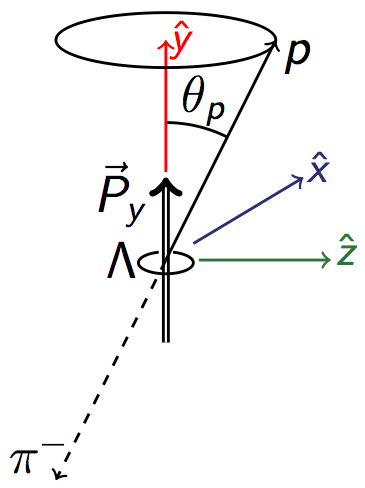}\\
\textbf{(a)}
\end{minipage}
\hspace{-20mm}
\begin{minipage}{.6\textwidth}
\centering
\includegraphics[height=0.18\textheight]{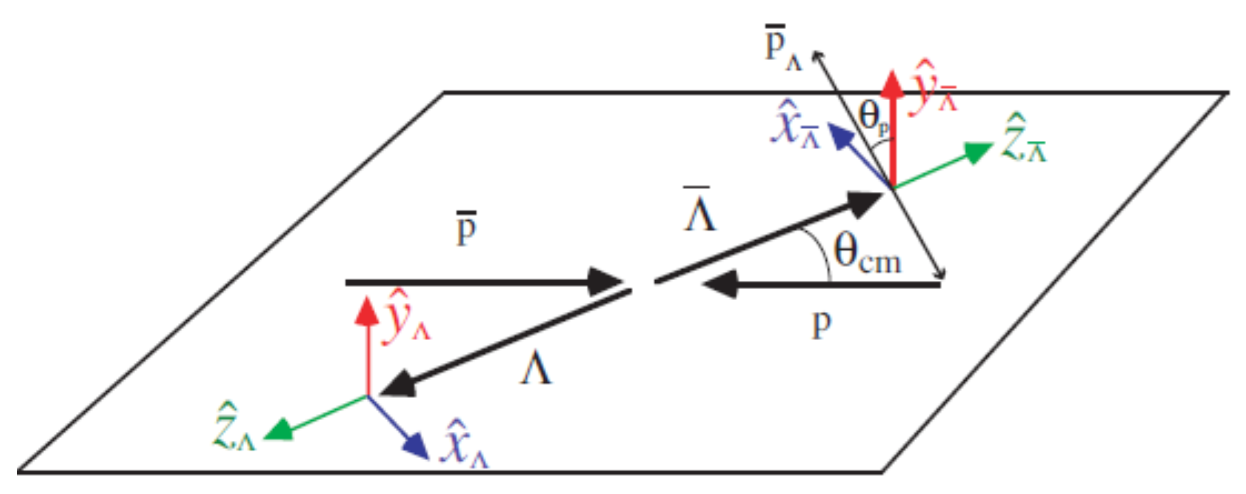}\\
\textbf{(b)}
\end{minipage}
\caption{\label{prod}\textbf{(a)} The $\Lambda$ decay frame. The opening angle between the polarisation axis and the outgoing proton $\theta_p$ is shown. \textbf{(b)} Production plane of the $\overline{p}p \to \overline{\Lambda}\Lambda$ reaction. The $y$-axis of the $\Lambda$ decay frame is perpendicular to the production plane. The $z$-axis is in the direction of the outgoing $\Lambda$ with respect to origin in the centre-of-mass frame.}
\label{fig:prodplane}
\end{figure*}

\subsection{Hyperon production}
\label{sec:hyperprod}
The scale probed in a hadronic reaction is influenced by the energy and, therefore, by the mass of the produced quarks. The strange quark mass is \ $m_s \approx 100$ MeV/$c^2$ which corresponds to the scale where quarks and gluons form hadrons. Therefore, the relevant degrees of freedom are unclear --- quarks and gluons, or hadrons? It is challenging to solve QCD in this energy regime. Guidance by experimental data is needed to improve the theory such that quantitative predictions can be accurately tested. As an intermediate step phenomenological models are developed which are constrained by experimental data. Exclusive hyperon-antihyperon production provides the cleanest environment to constrain these models. Phenomenological models based on quark-gluon degrees of freedom \cite{quarkgluon}, meson exchange \cite{kaonexchange} and a combination of the two \cite{quarkgluonhadron} have been developed for single-strange hyperons. The quark-gluon approach and the meson exchange approach have also been extended to the multi-strange sector \cite{XiQG,XiMEX,kaidalov}. Here, the interaction requires either annihilation of two quark-antiquark pairs, or in the meson picture, exchange of two kaons. This means that the interactions occur at shorter distances which make double-strange production more suitable for establishing the relevant degrees of freedom. The clearest difference between the quark-gluon picture and the kaon exchange picture is typically found in the predictions of spin observables \textit{e.g.} polarisation and spin correlations. 

Understanding the mechanism of hyperon production is also important in order to correctly interpret experimental data on other aspects of hyperons. One example is that of the recent theoretical and experimental studies of the hyperon structure in $e^+e^- \to \Lambda \bar{\Lambda}$. In Ref.~\cite{HaidenbauerHyp}, the time-like form factors $G_E$ and $G_M$ were predicted, including their relative phase $\Delta\Phi = \Phi(G_E)-\Phi(G_M)$ that manifests itself in a polarised final state. Different potential models were applied, using $\bar{p}p \to \bar{\Lambda}\Lambda$ data from PS185 \cite{ps185hyp} as input. In the model predictions for the channel $e^+e^- \to \Lambda \bar{\Lambda}$, the total cross section and the form factor ratio $R=|G_E/G_M|$ differ very little for different potentials. However, the relative phase $\Delta\Phi$ and hence the $\Lambda$ polarisation showed large sensitivity. New data from BESIII \cite{bes3hyp} provide an independent test of the $\Lambda\bar{\Lambda}$ potentials. Another example is hyperons and antihyperons in atomic nuclei, since understanding the elementary $\bar{p}p \to \bar{Y}Y$ reactions is crucial in order to correctly interpret data from $\bar{p}A$ collisions.

 Spin observables are straight-forward to measure for ground-state hyperons thanks to their weak, self-analysing decays. This means that the decay products are preferentially emitted along the spin direction of the parent hadron. Consider a spin $\frac{1}{2}$ hyperon $Y$ decaying into a spin $\frac{1}{2}$ baryon $B$ and a pseudoscalar meson $M$. The angular distribution of the daughter baryon $B$ is related to the hyperon polarisation by 
 \begin{equation}
 I(\cos\theta_B) = \frac{1}{4\pi}(1+\alpha_{Y} P_y \cos\theta_B)  
 \label{eq:polhyp}
 \end{equation}
 as illustrated in Fig.~\ref{fig:prodplane}a, where $\alpha_{Y}$~\cite{PDG} is the asymmetry parameter of the hyperon decay related to the interference between the parity conserving and the parity violating decay amplitudes. The polarisation $P_y$ is related to the production dynamics, hence it depends on the centre-of-mass (CMS) energy / beam momentum and on the hyperon scattering angle. In strong production processes, such as $\bar{p}p \to \bar{Y}Y$, with unpolarised beam and target, the polarisation can be non-zero normal to the production plane, spanned by the incoming antiproton beam and the outgoing antihyperon as shown in Fig.~\ref{fig:prodplane}b. Spin correlations between the produced hyperon and antihyperon are also accessible \cite{PaschkeQuinn} and from these, the \textit{singlet fraction} can be calculated, \textit{i.e.} the fraction of the hyperon-antihyperon pairs that are produced in a spin singlet state. Additional information can be obtained from hyperons that decay into other hyperons, \textit{e.g.} the $\Xi$. In the sequential decay $\Xi^- \rightarrow \Lambda \pi^-, \Lambda \rightarrow p \pi^-$, the additional asymmetry parameters $\beta$ and $\gamma$ of the $\Xi^-$ hyperon are accessible \textit{via} the joint angular distribution of the $\Lambda$ hyperons and the protons \cite{koch,perotti}.
For spin $\frac{3}{2}$ hyperons, \textit{e.g.} the $\Omega^-$, the spin structure is more complicated. Only considering the polarisation parameters of individual spin $\frac{3}{2}$ hyperons, we find that spin $\frac{3}{2}$ hyperons produced in strong processes like $\overline{p}p \rightarrow \overline{\Omega}^+\Omega^-$ have seven non-zero polarisation parameters. Three of these can be extracted from the $\Lambda$ angular distribution in the $\Omega^- \rightarrow \Lambda K^-$ decay \cite{erikthesis}. The remaining four parameters can be obtained by studying the joint angular distribution $I(\theta_{\Lambda},\phi_{\Lambda},\theta_p,\phi_p)$  of the $\Lambda$ hyperons from the $\Omega^-$ decay and the protons from the subsequent $\Lambda$ decay \cite{perotti}.

\subsubsection{State of the art}
\label{exphyperprod}
The PS185 collaboration has provided a large set of high-quality data on single-strange hyperons \cite{ps185hyp,LEAR} produced in antiproton-proton annihilation. One interesting finding is that the $\bar{\Lambda}\Lambda$ pair is produced almost exclusively in a spin triplet state. This can be explained by the $\Lambda$ quark structure: the light $u$ and $d$ quarks are in a relative spin-0 state, which means that the spin of the $\Lambda$ is carried by the $s$ quark. Various theoretical investigations reproduce this finding \cite{quarkgluon,kaonexchange,quarkgluonhadron}, but no model has yet been formulated to describe the complete spin structure of the reaction. The extension of models into the double-strange sector \cite{XiQG,XiMEX} and even the triple-strange $\Omega$ \cite{kaidalov}, have not been tested due to the lack of data. For $\Xi^-$ and $\Xi^0$ from $\bar{p}p$ annihilations, only a few bubble-chamber events exist \cite{Musgrave1965}, whereas no data at all are available related to triple-strange hyperon production since no studies have been carried out so far. As a result, further progress of this field is still pending. New data on the spin structure of $\overline{p}p \rightarrow \overline{Y}Y$ for ground-state multi-strange and single-charmed hyperons would therefore be immensely important for the development of a coherent picture of the role of spin in strangeness production.

\subsubsection{Potential of Phase One} 
\label{simhyperprod}

\begin{table*}
\centering
\caption{Results from simulation studies of the various hyperon production channels. The efficiencies are exclusive, \textit{i.e.} all final state particles are reconstructed. The lower limits marked with an asterisk ($*$) denote a  90\% confidence level. }
\label{tab:hyperons}       
\begin{tabular}{lllllll}
\hline
$p_{\overline{p}}$ & Reaction & $\sigma$ ($\mu$b) & Reconstruction & Decay & S/B & Rate ($s^{-1}$)\\(GeV/$c$) &&& efficiency (\%)&&& at $10^{31}$cm$^{-2}$s$^{-1}$ \\\hline
1.64 & $\overline{p}p \rightarrow \overline{\Lambda}\Lambda$ & 64.0~\cite{ps185hyp} & 15.7 & $\Lambda \rightarrow p \pi^-$ & 114 & 44   \\\hline

1.77 & $\overline{p}p \rightarrow \overline{\Sigma}^0\Lambda$ & 10.9~\cite{ps185hyp} & 5.3 & $\Sigma^0 \to \Lambda \gamma$ & > 11* & 2.4  \\\hline

6.0 & $\overline{p}p \rightarrow \overline{\Sigma}^0\Lambda$ & 20.0~\cite{sigma6} & 6.1 & $\Sigma^0 \to \Lambda \gamma$ & 21 & 5.0  \\\hline

4.6 & $\overline{p}p \rightarrow \overline{\Xi}^+\Xi^-$ & 1.0~~\cite{kaidalov} & 8.2 & $\Xi^- \rightarrow \Lambda \pi^-$ & 274 & 0.3  \\\hline

7.0 & $\overline{p}p \rightarrow \overline{\Xi}^+\Xi^-$ & 0.3~~\cite{kaidalov} & 7.9 & $\Xi^- \rightarrow \Lambda \pi^-$ & 165 & 0.1  \\\hline

4.6 & $\overline{p}p \rightarrow \overline{\Lambda} K^+ \Xi^- + {\rm c.c}$ & 1 & 5.4 & $\Xi^- \rightarrow \Lambda \pi^-$ & > 19* & 0.2  \\
& & & & $\Lambda \rightarrow p \pi^-$ & \\\hline


\end{tabular}
\end{table*}

Previous studies of mainly single-, but also a few double strange hyperon-antihyperon pairs produced in antiproton-proton annihilations show remarkably large cross sections within the PANDA energy range \cite{LEAR}. This means that large hyperon data samples can be collected within a reasonable time even with the reduced luminosity of the Phase One setup. Simulation studies of exclusive hyperon production, using a simplified Monte Carlo framework, were performed and presented in detail in Refs. \cite{erikthesis,pandaphysics,grapethesis,pandahyperons}. New, dedicated simulation studies of hyperon production have been performed for this review:
\begin{itemize}
    \item $\overline{p}p \rightarrow \overline{\Lambda}\Lambda, \bar{\Lambda} \to \bar{p}\pi^+, \Lambda \to p \pi^-$ at $p_{\rm beam} = 1.64$ GeV/$c$.
    \item $\overline{p}p \rightarrow \overline{\Sigma}^0\Lambda, \bar{\Sigma}^0 \to \bar{\Lambda} \gamma, \bar{\Lambda} \to \bar{p} \pi^+, \Lambda \to p \pi^-$ at $p_{\rm beam} = 1.77$ GeV/$c$ and $p_{\rm beam} = 6.0$ GeV/$c$.
    \item $\overline{p}p \rightarrow \overline{\Xi}^+\Xi^-, \bar{\Xi}^+ \to \bar{\Lambda}\pi^+, \bar{\Lambda} \to \bar{p}\pi^+, \Xi^- \to \Lambda \pi^-, \Lambda \to p \pi^-$ at $p_{\rm beam} = 4.6$ GeV/$c$ and $p_{\rm beam} = 7.0$ GeV/$c$.
\end{itemize}

\noindent The beam momenta for the single-strange hyperons were chosen in order to coincide with those of other benchmark studies. For the double-strange $\Xi^-$, the chosen beam momenta coincide with the hyperon spectroscopy campaign (4.6 GeV/$c$, see Section~\ref{sec:hypspec}) and the $\chi_{c1}(3872)$ line-shape campaign (7 GeV/$c$, see Section~\ref{sec:scan}). In these new simulation studies, a realistic PandaROOT implementation of the Phase One conditions was used, though with some simplifications due to current limitation in the simulation software: i) ideal pattern recognition, with some additional criteria on the number of hits per track in order to mimic a realistic implementation of the track reconstruction ii) ideal PID matching, to reduce the run-time. It was however shown in Ref.~\cite{erikthesis} that the event selection can be performed without PID thanks to the distinct topology of hyperon events: since the hyperons have relatively long life-time ($10^{-10}$~s) they travel a measurable distance before decaying. This provides a challenge in the tracking but also makes the background reduction very efficient. 

Around $10^6$ events were generated for $\bar{\Lambda}\Lambda$ and $\bar{\Xi}^+\Xi^-$ \cite{walter,pandahyperons}, whereas $10^4$ events for $\bar{\Sigma}^0\Lambda$ \cite{gabriela}. The larger event samples in the $\bar{\Lambda}\Lambda$ and $\bar{\Xi}^+\Xi^-$ cases enable studies of spin observables. In the case of $\bar{\Sigma}^0\Lambda$, only a general feasibility study of cross section and angular distribution measurements has been carried out so far. The $\bar{\Lambda}\Lambda$ and $\bar{\Sigma}^0\Lambda$ final states were modelled using parameterisations based on data from Refs. \cite{ps185hyp,sigma6}, where it was found that single-strange antihyperons are very strongly forward-going in the CMS of the reaction. The $\bar{\Xi}^+\Xi^-$ final state has never been studied and was therefore generated both with an isotropic angular distribution and with a forward-peaking distribution. The results were found to differ only marginally.

The particles were propagated through the PandaROOT detector implementation and the signals were digitised, reconstructed and analysed. The signal events were selected by requiring all stable ($p$, $\bar{p}$ and $\gamma$) or pseudo-stable ($\pi^+$ and $\pi^-$) particles to be found: 
\begin{itemize}
    \item $\bar{\Lambda}\Lambda$: $p$, $\pi^-$, $\bar{p}$ and $\pi^+$.
    \item $\bar{\Sigma}^0\Lambda$: $p$, $\pi^-$, $\bar{p}$, $\pi^+$ and $\gamma$.
    \item $\bar{\Xi}^+\Xi^-$: $p$, 2$\pi^-$, $\bar{p}$ and 2$\pi^+$.
\end{itemize}
To reduce the number of background photon signals, additional energy cuts were applied to identify the photon from the $\bar{\Sigma}^0$ decay \cite{gabriela}. 
The $\Lambda$ and $\bar{\Lambda}$, that appear in all channels, were identified by combining the reconstructed pions and protons/antiprotons and applying vertex fits and mass window criteria on the combinations. Furthermore, the decay vertex of the $\Lambda/\bar{\Lambda}$ was required to be displaced with a certain distance from the interaction point. To identify $\bar{\Sigma}^0$ or $\Xi^-/\bar{\Xi}^+$, the $\Lambda/\bar{\Lambda}$ candidates were combined with the photons or remaining pions. In the case of $\bar{\Lambda}\Lambda$ and $\bar{\Sigma}^0\Lambda$ final states, four-momentum conservation was applied using kinematic fits constraining the momenta and total energy of the final-state particles with the initial antiproton-proton momenta and energy to further reduce the background. Since the $\Xi^-$ decays sequentially, a more elaborate method including a decay tree fitter was applied \cite{walter,pandahyperons}. 

The resulting signal efficiencies are given in Table \ref{tab:hyperons}, that also includes the results from the $\Xi^*$ study described in Section \ref{sec:exhyp}. The expected rates of reconstructed events are calculated based on the Phase One luminosity of $10^{31}$cm$^{-2}$s$^{-1}$ and cross sections from Refs. \cite{ps185hyp,sigma6} ($\bar{\Lambda}\Lambda$ and $\bar{\Sigma}^0\Lambda$) and Ref.\cite{kaidalov} ($\bar{\Xi}^+\Xi^-$). The signal-to-background ratios (S/B) were obtained by simulating $10^7$ events at each energy, generated with the Dual Parton Model \cite{Capella1994}. 


In this work, we have also investigated the feasibility of reconstructing spin observables such as the polarisation and spin correlations using the methods outlined in Ref. \cite{erikthesis}. For the analysis, the $\overline{p}p \rightarrow \overline{\Lambda}\Lambda, \bar{\Lambda} \to \bar{p}\pi^+, \Lambda \to p \pi^-$ sample was used, containing 157000 signal events surviving the selection criteria. A sample of this size can be collected within a few hours with the Phase One luminosity. The simulated events were weighted according to an input polarisation function $P_y = \sin2\theta_{\Lambda}$ and the spin correlation distributions $C_{ij} = \sin\theta_{\Lambda}$ ($i,j = x, y, z$). Symmetry implies $P_{Y}$ = -$P_{\bar{Y}}$ which means that the extracted polarisation from $\Lambda$ and $\bar{\Lambda}$ can be combined for better statistical precision. 

\begin{figure}[h]
\begin{minipage}{.5\textwidth}
\centering
\includegraphics[height=0.22\textheight]{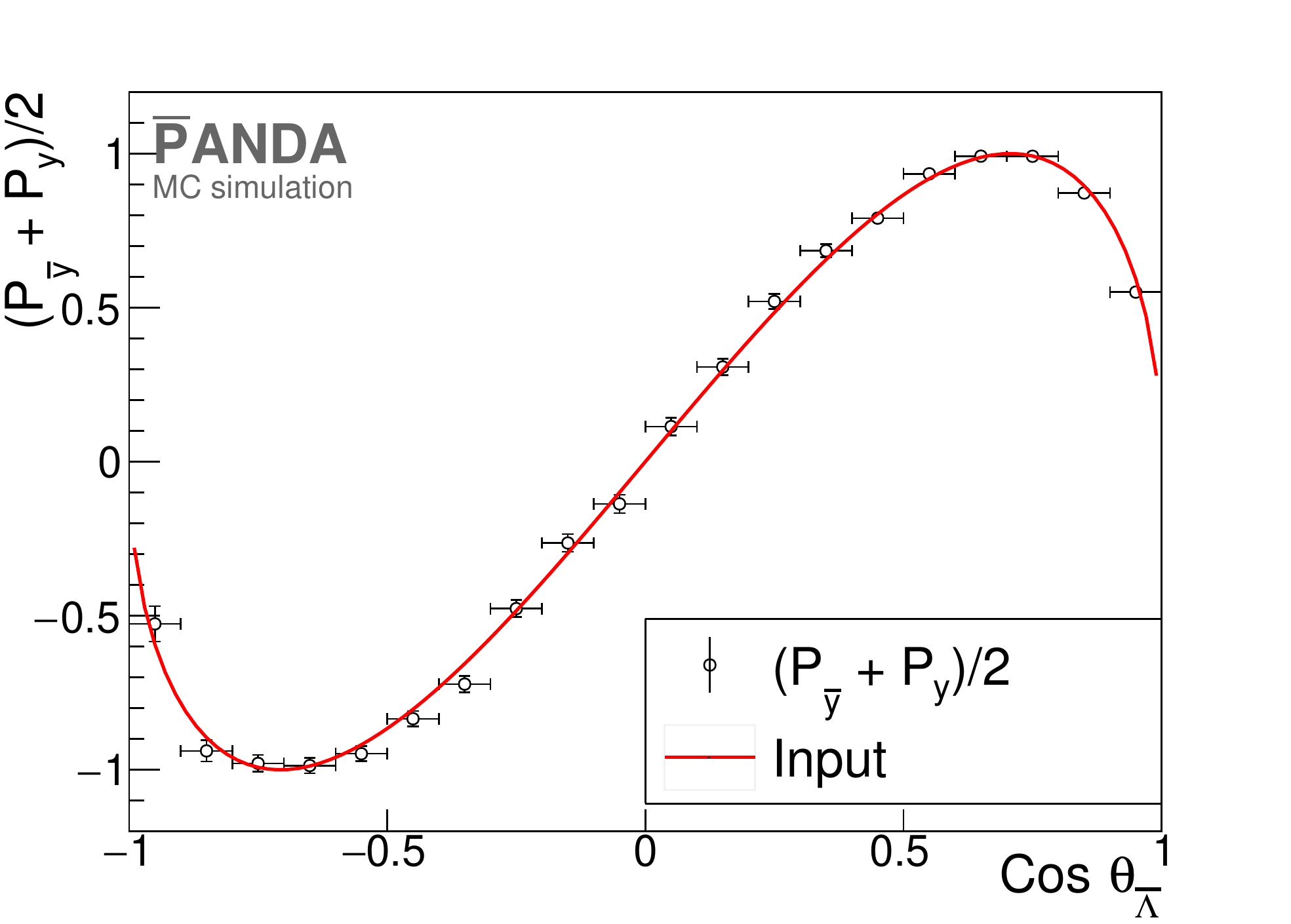}\\
\textbf{(a)}
\end{minipage}
\begin{minipage}{.5\textwidth}
\centering
\includegraphics[height=0.22\textheight]{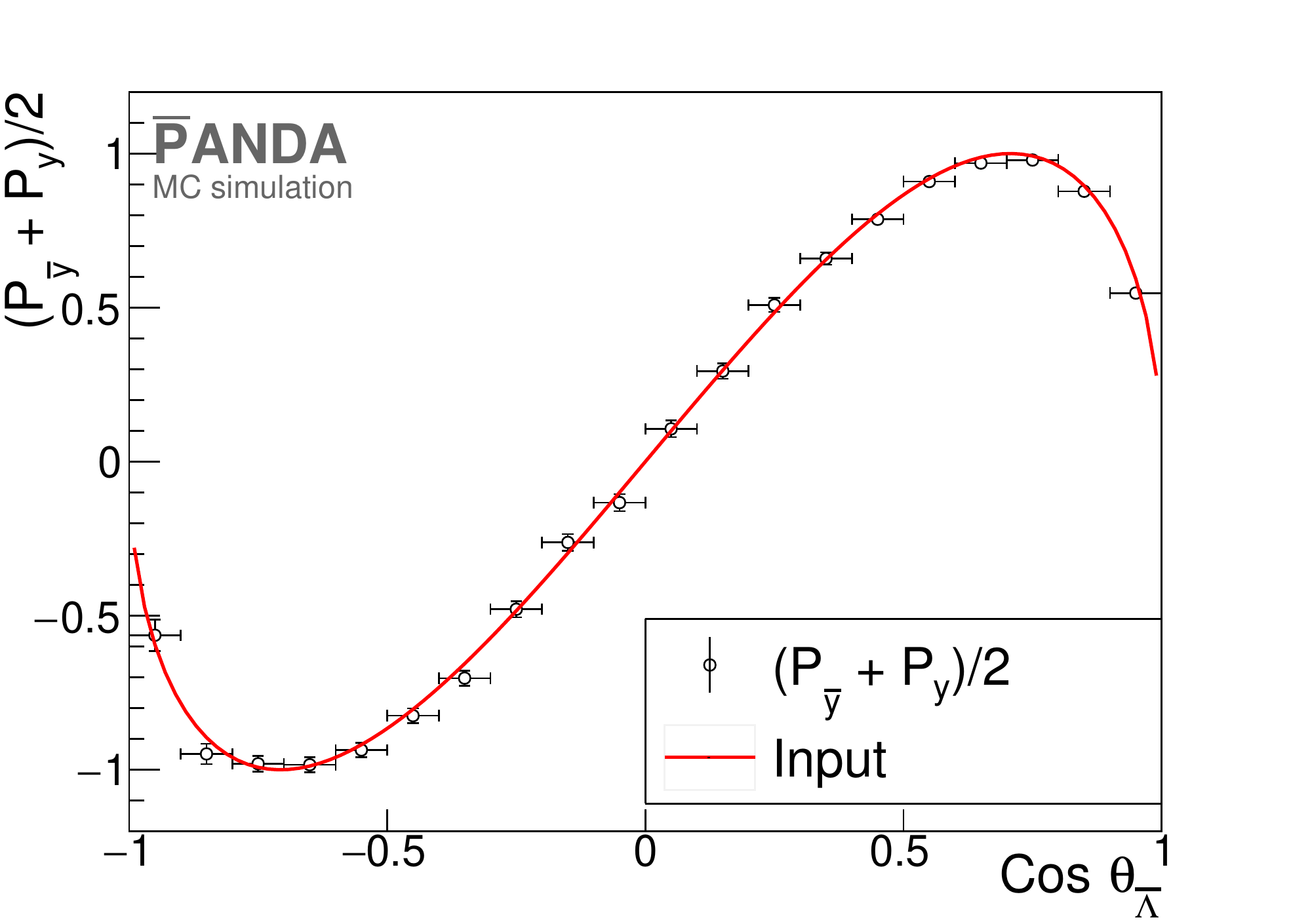}\\
\textbf{(b)}
\end{minipage}
\caption{\textbf{(a)} Average polarisation of the $\Lambda$/$\bar{\Lambda}$. \textbf{(b)} Average of the polarisations reconstructed without any acceptance correction. The vertical error bars are statistical uncertainties only correspong to a few days of data taking. The horizontal bars are the bin widths. The red solid line marks the input polarisation as a function of $\cos\theta_{\Lambda}$.}
\label{fig:resultpol}
\end{figure}

The reconstruction efficiency was accounted for using two different, independent methods: i) regular, multi-dimensional acceptance correction as in Ref. \cite{grapethesis} and ii) using the acceptance-independent method outlined in Ref. \cite{erikthesis}. The results of the MC simulations were divided into bins with respect to the $\bar{\Lambda}$ scattering angle. In each bin, the polarisation $P_Y$ and spin correlations $C_{ij}$ were reconstructed. The resulting polarisation distribution is shown in panel a) of Fig.~\ref{fig:resultpol} with acceptance corrections and in panel b) with the acceptance-independent method. The polarisation distributions extracted with the two independent methods agree with each other as well as with the input functions.

\begin{figure}[h]
\begin{minipage}{.5\textwidth}
\centering
\includegraphics[height=0.22\textheight]{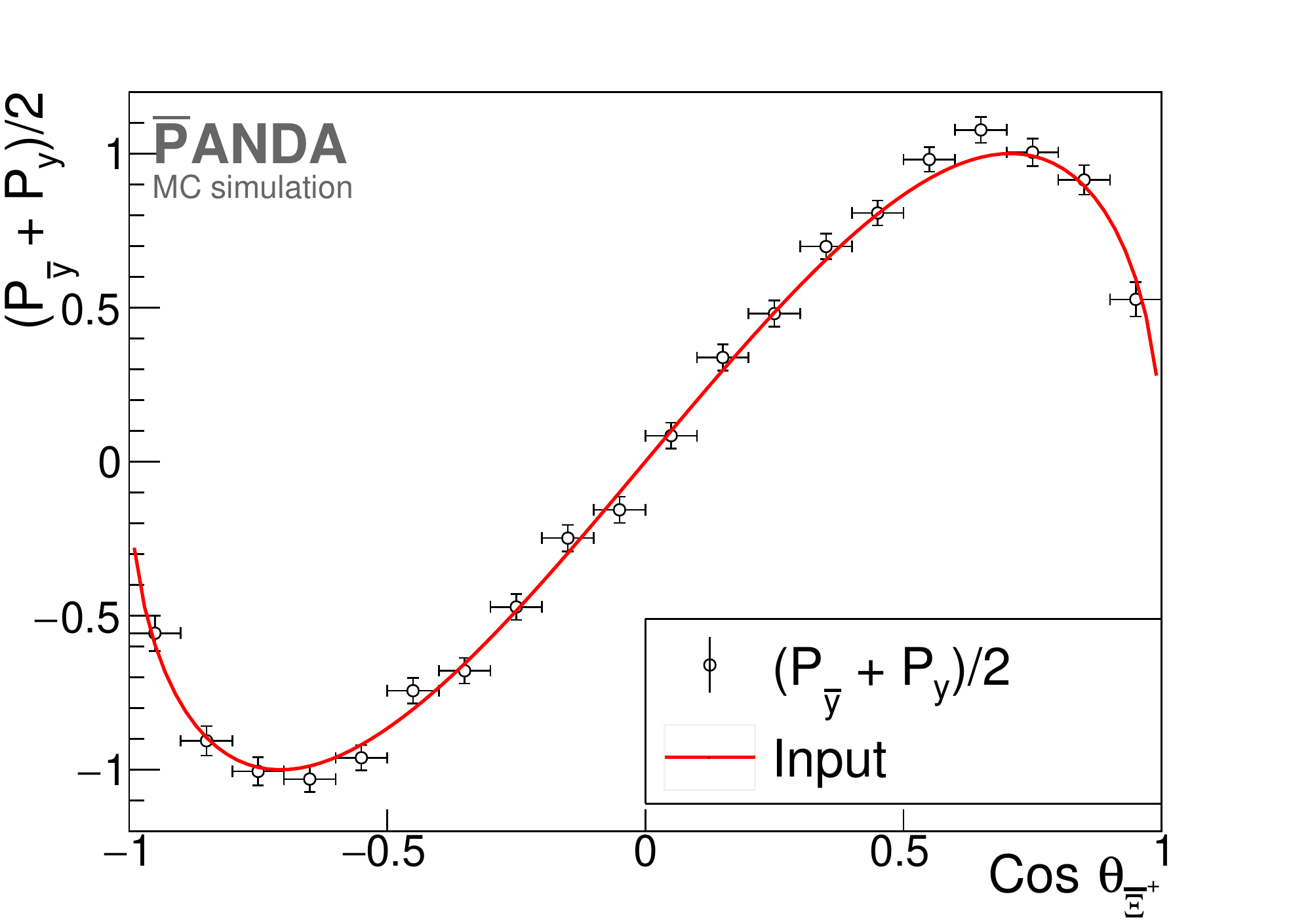}\\
\textbf{(a)}
\end{minipage}
\begin{minipage}{.5\textwidth}
\centering
\includegraphics[height=0.22\textheight]{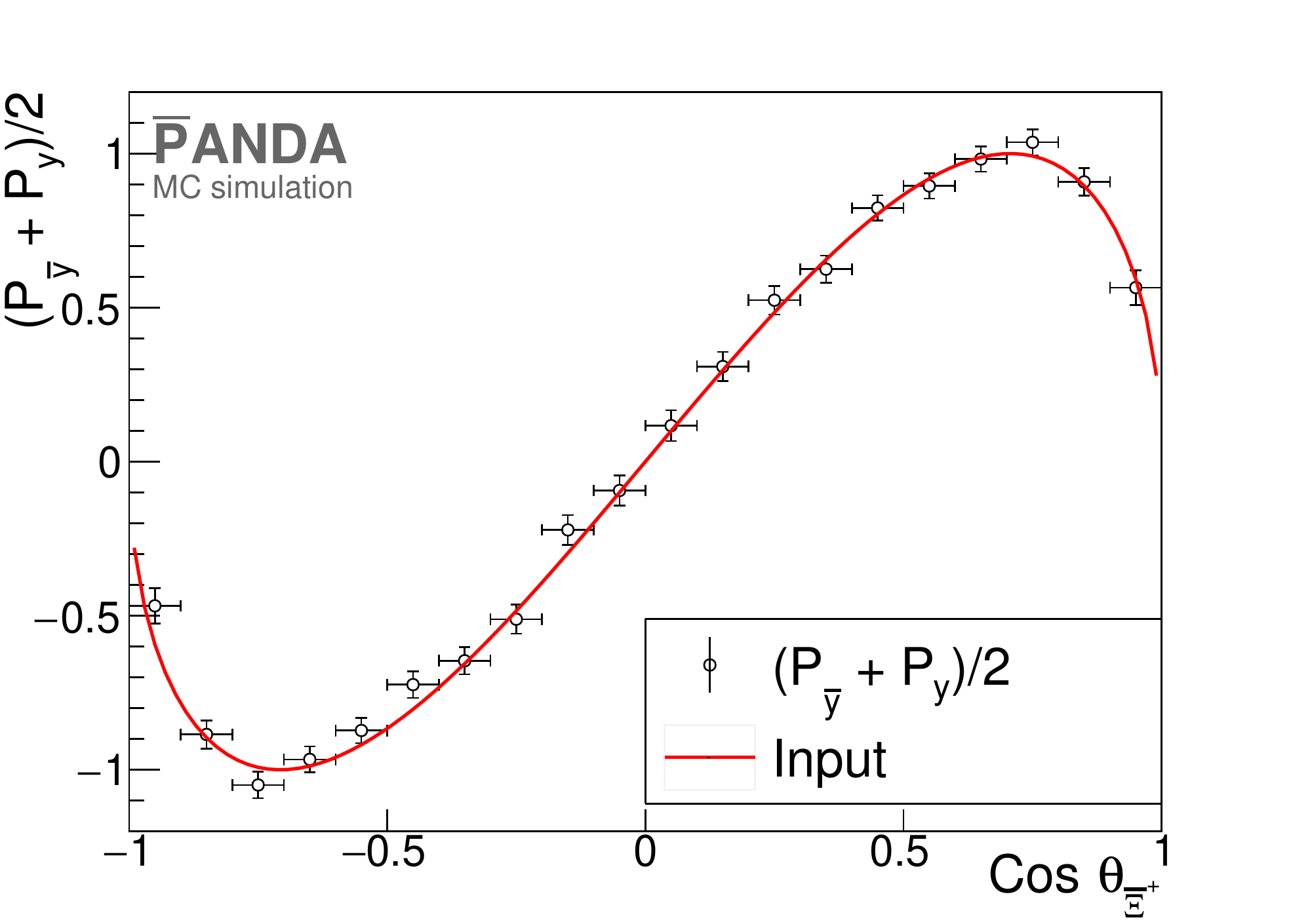}\\
\textbf{(b)}
\end{minipage}
\caption{\textbf{(a)} Average polarisation of the $\Xi^-$/$\bar{\Xi}^+$ at 4.6~GeV/$c$. \textbf{(b)} Average of the polarisation of $\Xi^-$/$\bar{\Xi}^+$ at 7.0 GeV/$c$. The vertical error bars are statistical uncertainties only corresponding to a few days of data taking. The horizontal bars are the bin widths. The red solid line marks the input polarisation as a function of $\cos\theta_{\Xi}$.}
\label{fig:resultpolXi}
\end{figure}

\begin{figure*}[ht]
	    \centering
	    \begin{minipage}{.5\textwidth}
	        \centering
	        \includegraphics[width=1.\linewidth]{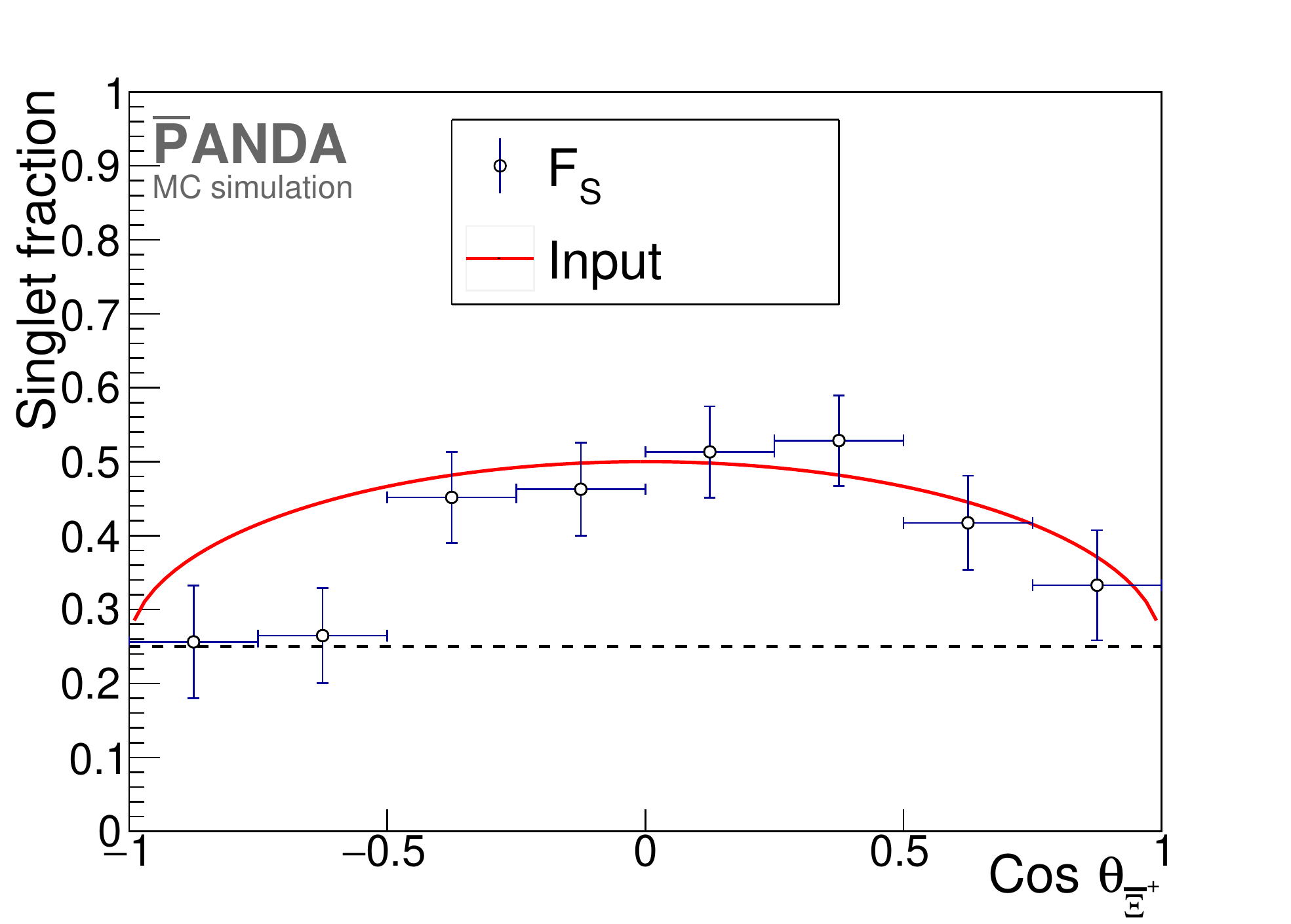}\\
			\textbf{(a)}
	    \end{minipage}%
	    \begin{minipage}{0.5\textwidth}
	        \centering
	        \includegraphics[width=1.\linewidth]{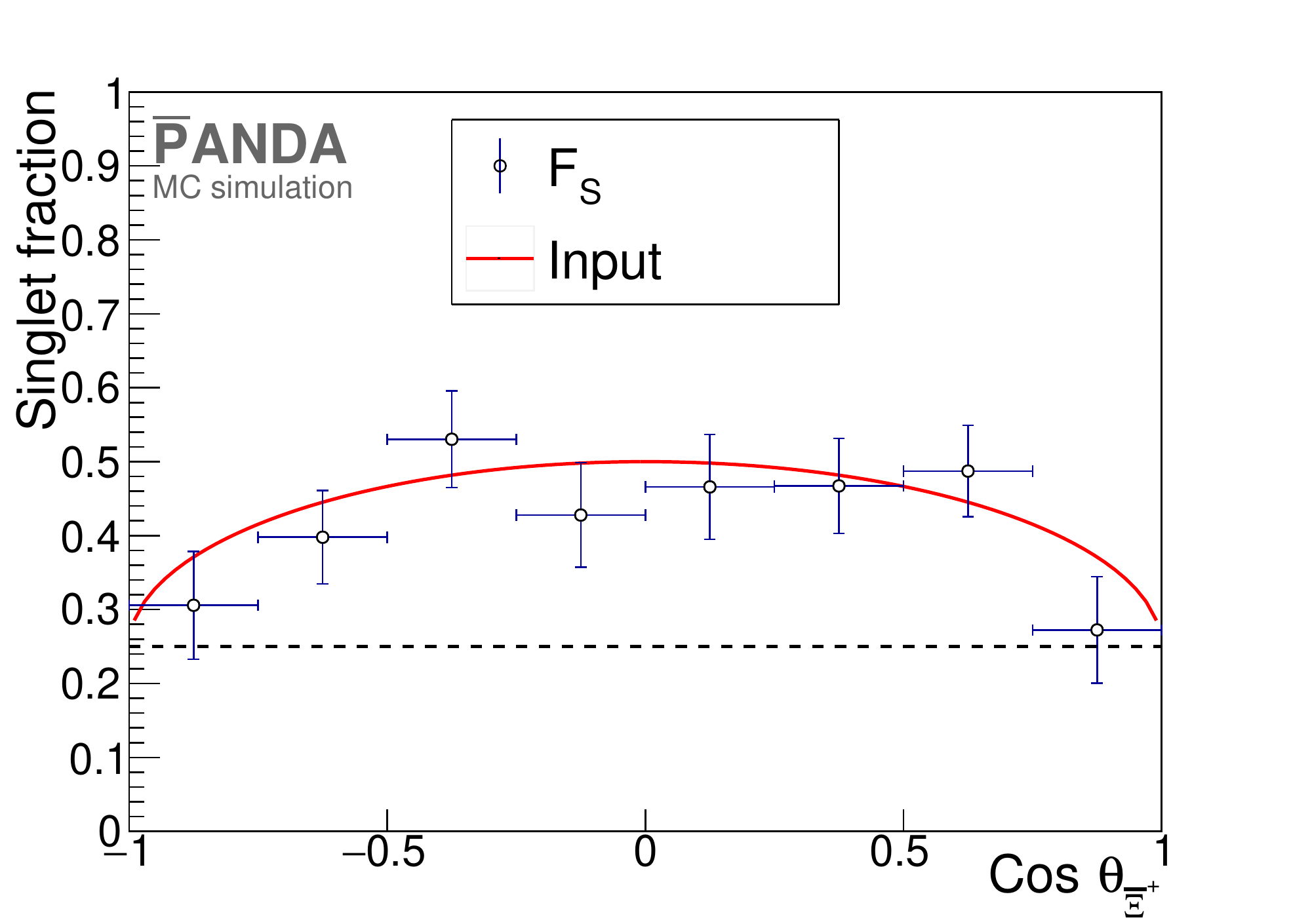}\\
			\textbf{(b)}
	    \end{minipage}
	    \caption{Reconstructed Singlet Fraction $F_S$ at \textbf{(a)} $p_{\mathrm{beam}}=4.6$~GeV/$c$ and \textbf{(b)} $p_{\mathrm{beam}}=7.0$~GeV/$c$.
              The vertical error bars correspond to statistical uncertainties and reflect the precision after a few days of data taking. The red curves are the input Singlet Fraction. The dashed line indicates values corresponding to a statistical mixture of singlet and triplet final states.}
		\label{fig:sf}
	\end{figure*}

In the same way, spin observables of the $\Xi^-$ hyperons were studied at both 4.6 GeV/$c$ and 7.0 GeV/$c$. The number of signal events were $7.2\cdot10^4$ and $6.7\cdot10^4$, respectively, samples that can be collected within a few days during Phase One. The resulting polarisation distributions as a function of $\cos\theta_{\Xi}$ obtained at each energy are shown in Fig.~\ref{fig:resultpolXi}. The singlet fractions were calculated from the spin correlations and are shown in Fig.~\ref{fig:sf}. A singlet fraction of 0 means that all $\Xi^-\bar{\Xi}^+$ states are produced in a spin triplet state, a fraction of 1 means they are all in a singlet state, and a fraction of 0.25 means the spins are completely uncorrelated. In Ref.~\cite{XiMEX}, the singlet fraction is predicted to be 0 for forward-going $\bar{\Xi}^+$ and closer to 1 in the backward region. This is in contrast to the single-strange case where the singlet fraction is almost independent of the scattering angle \cite{LEAR}. The results of the simulations shown in Fig.~\ref{fig:sf} indicate that the uncertainties in the singlet fraction will be modest at all scattering angles, which enables a precise test of the prediction from Ref. \cite{XiMEX}. 



Most systematic effects that are important in cross section measurements, \textit{e.g.} trigger efficiencies and luminosity, are expected to be isotropically distributed in a near $4\pi$ experiment like PANDA. This means that their impact on angular distributions, and parameters extracted from these, are expected to be small. Hyperon polarisation studies with BESIII (\textit{e.g.} \cite{bes3hyp}) instead indicate that imperfections in the Monte Carlo description of the data, due to for example gain drift in HV supplies, may be more important. Most of these effects can however only be studied once PANDA is operational and by careful Monte Carlo modelling, they can be minimised. In the simulation studies presented here, three basic consistency tests have been performed in order to reveal eventual sensitivity to detection- and reconstruction artefacts: i) comparison between generated and reconstructed distributions that are efficiency corrected ii) comparison between extracted hyperon and antihyperon parameters iii) comparison between two different efficiency correction methods. All three tests show differences that are negligible with respect to the small statistical uncertainties.

\subsection{Hyperon spectroscopy}
\label{sec:hypspec}

Baryon spectroscopy has been decisive in the development of our understanding of the microscopic world, the best example being the plethora of new states discovered in the 1950's and 1960's. It was found that these states could be organised according to ``the Eightfold Way", \textit{i.e.} SU(3) flavour symmetry, that led to formulation of the quark model by Gell-Mann and Zweig~\cite{quarkmodel}. Though successful in classifying ground-state baryons and describing some of their ground-state properties, the quark model fails to explain some features of the baryon excitation spectra. This indicates that the underlying picture is more complicated. In contemporary baryon spectroscopy, the most intriguing questions are i) Which effective degrees of freedom are adequate to describe the hadronic reaction dynamics? Which baryonic excitations are efficiently and well described in a three-quark picture and which are generated by coupled-channel effects of hadronic interactions? ii)  To which extent does the excitation spectra of baryons consisting of {\it u, d, s} obey SU(3) flavour symmetry? iii) Are there exotic baryon states, \textit{e.g.} pentaquarks or dibaryons? 

Among the theoretical tools available to study the spectra and internal properties of baryons, lattice QCD approaches have received a lot of attention thanks to the tremendous progress over the past years. Prominent examples are the mass prediction of the double charm ground state $\Xi_{cc}$ baryon \cite{alexandrou,lat1, lat2,lat3,lat4}, now confirmed by LHCb~\cite{doublecharm}, and accurate lattice calculations of the mass splitting of the neutron and proton~\cite{borsanyi2015}.  

However, for the excited states in the light-baryon sector differences between recent calculations~\cite{leinweber2015, sun2020} remain and unambiguous conclusions cannot yet be drawn. Other approaches to baryon excitation spectra are based on the Dyson-Schwinger framework~\cite{Eichmann:2016yit}, on the coupled-channel chiral Lagrangian~\cite{Kolomeitsev:2003kt, Hyodo:2020czb, Mai:2020ltx, Meissner:2020khl}, and on the AdS/QCD approach~\cite{Brodsky:2015,Deur:2015,Brodsky:2016}. 

The next step is systematic studies of the strange sector, in particular states with double and triple strangeness. These bridge the gap between the highly relativistic light quarks and the less relativistic heavy ones.

\subsubsection{State of the art}

So far, worldwide experimental efforts in baryon spectroscopy have been focused on $N^*$ and $\Delta$ resonances.
Most of the known states have masses smaller than 2 GeV/$c^2$ and were discovered in $\pi N$ scattering experiments. In recent years, many laboratories (JLab, ELSA, MAMI, GRAAL, Spring-8 etc) have studied these resonances in photon-induced reactions~\cite{Klempt,Crede}. As a result, the data bank on nucleon and $\Delta$ spectra has become significantly bigger and a lot has been learned. However, there are several puzzles that remain to be resolved. 

One example is the so-called \textit{missing resonance} problem of Constituent Quark Models (CQMs): many states that are expected from these phenomenological-driven models have not been observed experimentally. This is in contrast to the Dyson-Schwinger approach whose predictions agree almost one-to-one with the experimentally measured light baryon spectra below 2~GeV/$c^2$ \cite{dslight,dsstrange}. This observation demonstrates the shortcomings of CQMs, thereby motivating the necessity to experimentally establish the spectra of excited baryons. For a successful campaign, an experimental approach is needed in which these states are searched for and their properties are studied using various complementary initial probes such as $\pi N$, $\gamma N$, and, with PANDA, $\bar{p} N$. 


Another example of an unresolved conundrum is the \textit{level ordering}: The lightest baryon, \textit{i.e} the nucleon, has $J^P = \frac{1}{2}^+$ and the next-to-lightest baryon is expected to be its parity partner, with $J^P = \frac{1}{2}^-$. However, this is in contrast to experimental findings where the Roper $N^*(1440)$ resonance, with $J^P = \frac{1}{2}^+$, is significantly lighter than the lightest $J^P = \frac{1}{2}^-$ state, \textit{i.e.} the $N^*(1535)$.

A new angle to the aforementioned puzzles can be provided by studying how they carry over to strange baryons. In the single-strange sector, the missing CQM resonance problem remains. Regarding the level-ordering, the situation is very different regarding light baryons: the parity partner of the lightest $\Lambda$ hyperon is the $\Lambda(1405)$ which is indeed the next-to-lightest isosinglet hyperon~\cite{claslambda1405}. However, the $\Lambda(1405)$ is very light, and, therefore, it has been suggested to be a molecular state, see \textit{e.g.} Ref.~\cite{dalitz:1959,adelaide}. The existing world data on double- and triple-strange baryons are very scarce and do not allow for the kind of systematic comparisons with theory predictions that led to progress in the light and single-strange sector. Only one excited $\Xi$ state and no excited $\Omega$ states are considered well established within the PDG classification scheme~\cite{PDG}. It is also worth pointing out that even for the ground state $\Xi$ and $\Omega$, the parity has not been determined experimentally. Furthermore, the spin determination of the $\Omega$ is not model-independent but inferred by assumptions on the $\Xi_c$ and $\Omega_c$ spin \cite{babaromega}. It would be very illuminating to study the features of the double- and triple-strange hyperon spectra since it enables a systematic comparison of systems containing different strangeness.


\begin{figure*}[htbp]
	\begin{center}
		\includegraphics[width=0.6\textwidth]{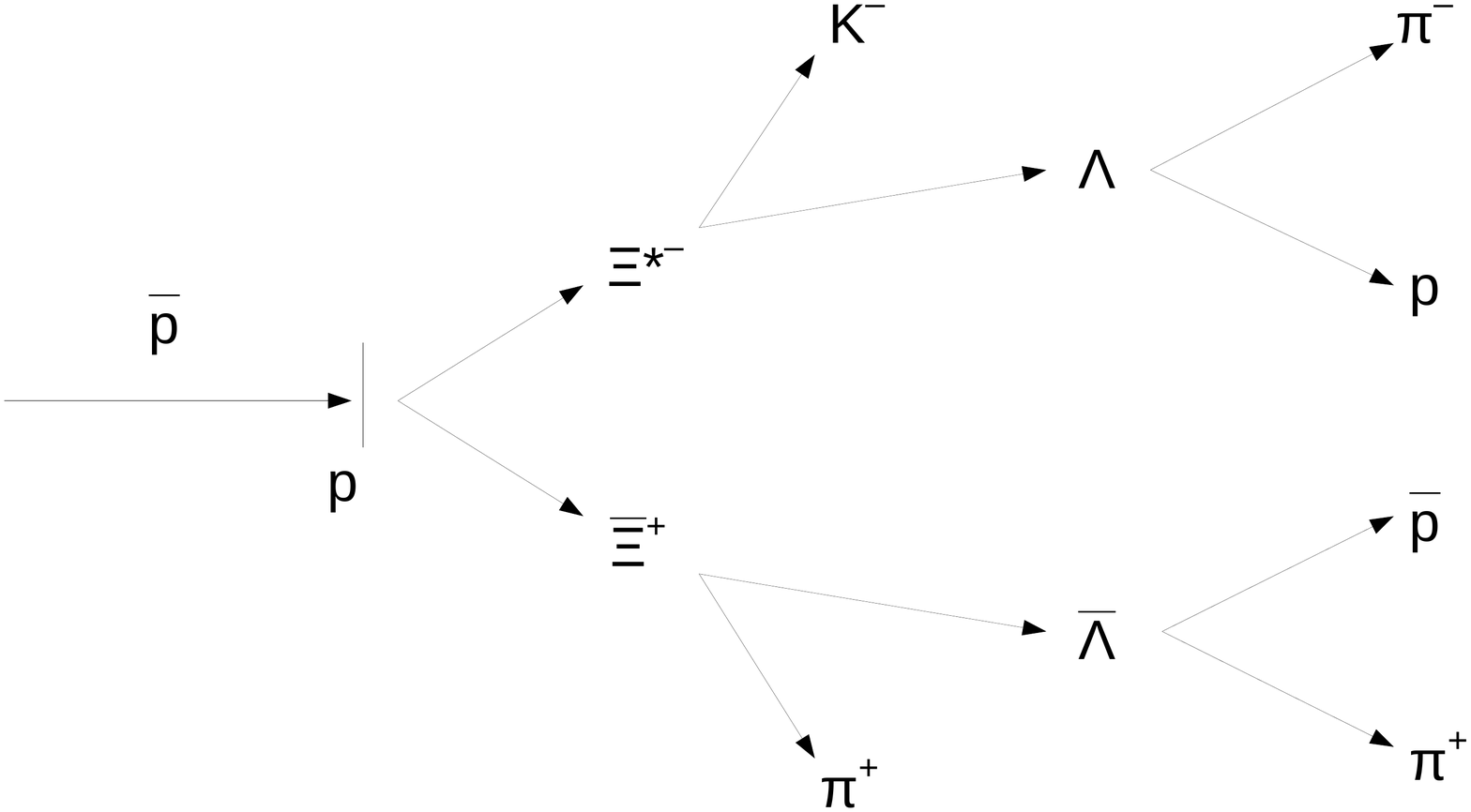}
	\end{center}
	\caption{A schematic view of the reaction topology used for the generation of Monte Carlo events.
	}
	\label{fig:decaytree}
\end{figure*}

\subsubsection{Potential for Phase One}
\label{sec:exhyp}
A dedicated simulation study has been performed of the $\bar{p} p \rightarrow \Lambda K^- \overline{\Xi}^+ + c.c.$ reaction at a beam momentum of 4.6~GeV/$c$.
In the following, the inclusion of the charge conjugate channel is implicit. In spectroscopy, parameters like mass, widths and Dalitz plots are essential.
Therefore, the focus of this study is to estimate how well such parameters can be measured with PANDA. 
The simulated data sample of 4.5$\cdot 10^6$ events includes the $\Xi(1690)^\pm$ and $\Xi(1820)^\pm$ resonances, decaying into $\Lambda K^-$ + c.c.
(each 40\% of the total generated events), as well as non-resonant $\Lambda K^- \overline{\Xi}^+$ + c.c. production (20\% of the generated sample). 
The simulated widths of the $\Xi(1690)^-$ and $\Xi(1820)^-$ resonances were 30 MeV/$c^2$ and 24 MeV/$c^2$, respectively, in line with the PDG \cite{PDG}.
The event generation was performed using EvtGen \cite{Lange2001} with the reaction topology as illustrated in Figure~\ref{fig:decaytree}.
The angular distribution of the produced $\Xi^*$ resonance are isotropically generated since no information from experimental data exist.
Further technical details related to the simulation study can be found in Ref.~\cite{puetz:2021}.



The analysis was performed in the same way as described in Section~\ref{simhyperprod}. The final state is required to contain $p$, $\bar{p}$, $\pi^{-}$, $\pi^{+}$, $K^{-}$ and $K^{+}$. The $\Lambda$ candidates were identified by combining $p$ and $\pi^-$ into a common vertex and applying a mass window criterion.  The $\Xi^-$ ($\Xi^*$) hyperons were identified by combining $\Lambda$ candidates with the remaining pions (kaons). Background was further suppressed by a decay tree fit in the same way as in Section~\ref{simhyperprod}. The exclusive reconstruction efficiency was found to be 5.4\%. We assume a $\bar{p}p \to \bar{\Lambda}K\Xi + c.c.$ cross section of 1~$\mu$b, where the production mainly occurs through a $\Xi^-\Xi^* + c.c.$ pair and where the excited cascade could be either $\Xi^*(1690)$ or $\Xi^*(1820)$. With this assumption, the reconstruction rate is 0.2 $s^{-1}$ or 18000 events per day. These cross sections have never been measured and assumed of the same order as the ground-state $\bar{\Xi}^+\Xi^-$ \cite{Flamino1984} that was measured by Ref.~\cite{Musgrave1965} to be around $1~\mu$b.

The background was studied using a DPM sample containing $10^8$ events and the signal events were weighted assuming a total cross section of 50~mb. No background events survived the selection criteria and we therefore conclude that on a 90\% confidence level, the signal-to-background is $S/B > 19$. The numbers are summarised in Table \ref{tab:hyperons}.

\begin{figure}
\begin{minipage}{.5\textwidth}
\centering
\includegraphics[height=0.22\textheight]{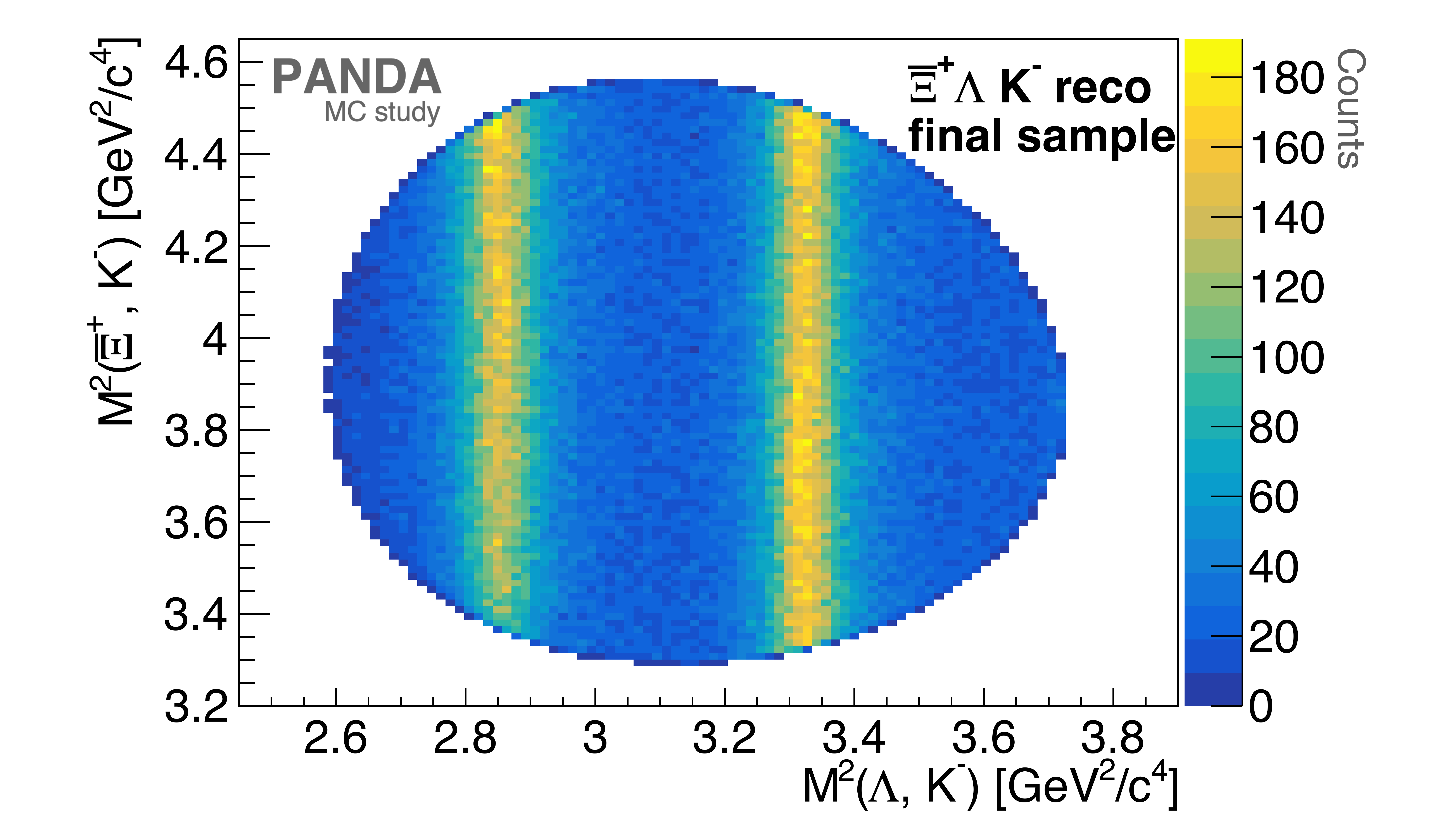}\\
\textbf{(a)}
\end{minipage}
\begin{minipage}{.5\textwidth}
\centering
\includegraphics[height=0.2\textheight]{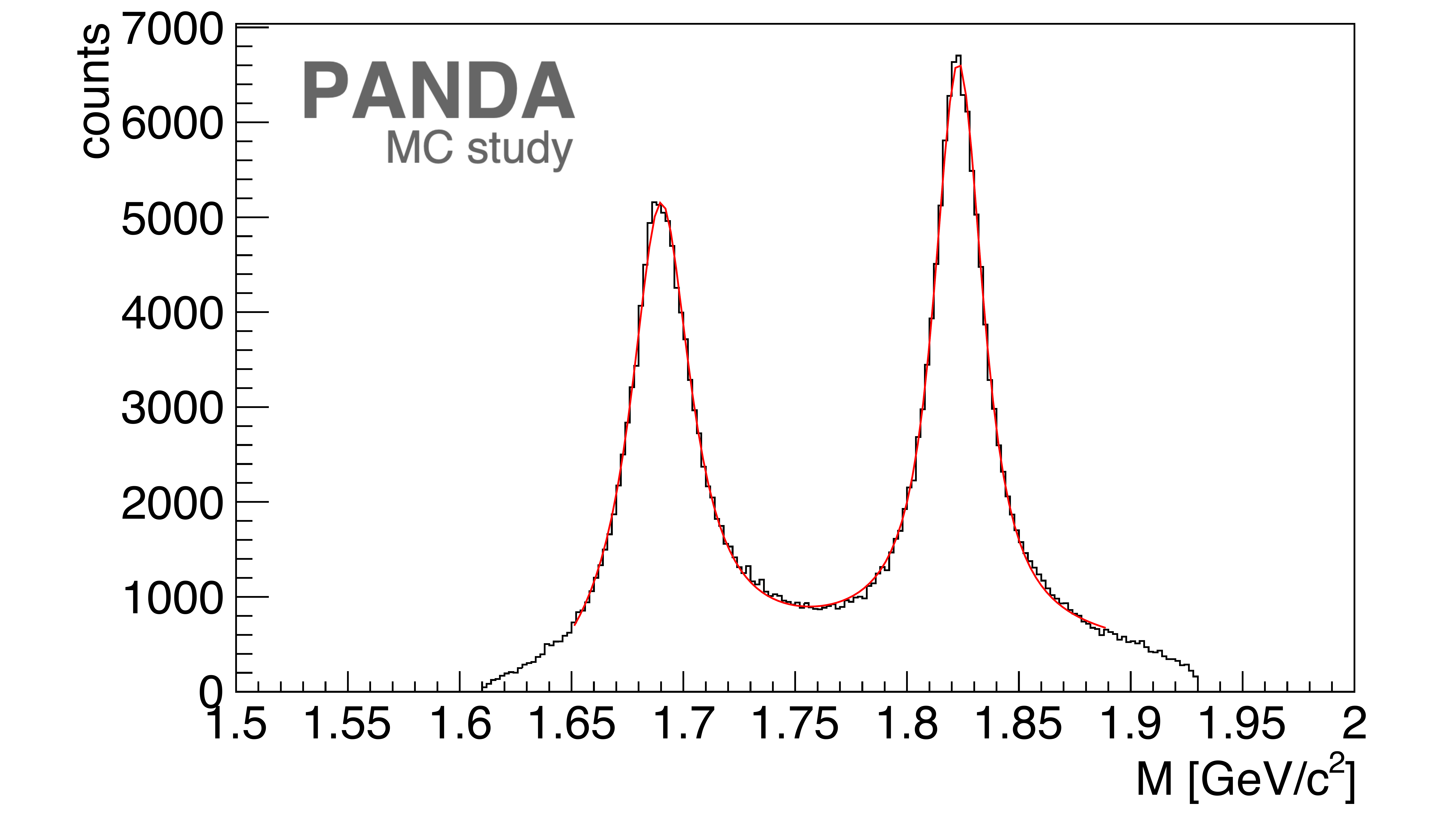}\\
\textbf{(b)}
\end{minipage}
\caption{\textbf{(a)} The reconstructed Dalitz plot of the $\Lambda K^- \overline{\Xi}^+$ final state.\textbf{(b)} The $\Lambda K^-$ invariant mass of the reconstructed MC data.}
\label{fig:dalitzreco}
\end{figure}

The reconstructed Dalitz plot and $\Lambda K^-$ invariant mass are shown in Figure \ref{fig:dalitzreco}. The reconstruction efficiency distribution is flat with respect to the Dalitz plot variables and the angles. This is a necessary condition in order to minimise systematic effects in the planned partial-wave analysis of this final state.

In order to evaluate the $\Xi$ and $\bar{\Xi}$ resonance parameters, the $\Lambda{}K^-$ and $\bar{\Lambda}K^+$ mass distributions have been fitted with two Voigt functions combined with a polynomial. By comparing the reconstructed $\Lambda K^-$ and $\bar{\Lambda}K^+$ widths to the generated ones, the mass resolution was estimated to $\sigma_{M}=4.0\,{\rm{}MeV}/c^2$ for the $\Xi(1690)^-$ and $\sigma_{M}=6.7\,{\rm{}MeV}/c^2$ for the $\Xi(1820)^-$.
\begin{table*}
	\caption{Fit values for $\Lambda\,K^{-}$ and $\overline{\Lambda}\,K^{+}$.}
	\label{tab:FitValuesFullTree}
	\centering
		\begin{tabular}{lrrrr}
		\hline
		& \multicolumn{2}{c}{$\Lambda\,K^{-}$} & \multicolumn{2}{c}{$\overline{\Lambda}\,K^{+}$} \\
		\hline
		& \multicolumn{1}{c}{$\Xi\left(1690\right)^{-}$} & \multicolumn{1}{c}{$\Xi\left(1820\right)^{-}$} & \multicolumn{1}{c}{$\overline{\Xi}\left(1690\right)^{+}$} & \multicolumn{1}{c}{$\overline{\Xi}\left(1820\right)^{+}$} \\\hline
		\hline
		Fitted mass [GeV$/c^2$] & 1.6902 $\pm$ 0.0006 & 1.8236 $\pm$ 0.0003 & 1.6905 $\pm$ 0.0006 & 1.8234 $\pm$ 0.0003 \\
		Fitted $\Gamma$ [MeV] & 31.09 $\pm$ 1.9 & 23.0 $\pm$ 2.0 & 31.8 $\pm$ 1.8 & 24.2 $\pm$ 1.8 \\
        Input mass [GeV$/c^2$] & 1.6900 & 1.8230 & 1.6900 & 1.8230 \\
		Input $\Gamma$ [MeV] & 30 & 24 & 30 & 24 \\
		\hline
		\end{tabular}

\end{table*}
The obtained fit values are shown in Table \ref{tab:FitValuesFullTree}. In both cases, the fitted masses are in good agreement with the input values. 

\subsection{Impact and long-term perspective}
PANDA will be a strangeness factory where many different aspects of hyperon physics can be studied. Double- and triple strange hyperons are unknown territory both when it comes to production dynamics, spin observables and spectroscopy. Long-standing questions, such as relevant degrees of freedom and quark structure, can be investigated already during the first years with reduced detector setup and luminosity. Furthermore, the measurements in Phase One provide important milestones for the foreseen precision tests of CP conservation, that will be carried out when the design luminosity and the full PANDA setup are available in the subsequent Phase Two and Three. In the latter, copious amounts of weak, two-body hyperon decays will be recorded - several million exclusively reconstructed $\bar{\Lambda}\Lambda$ pairs every hour. This enables precise measurements of the decay asymmetry parameters. In the absence of CP violation, the asymmetry parameters of a hyperon have the same magnitude but the opposite sign of those of the antihyperon, \textit{e.g.} $\alpha=-\bar{\alpha}$. Differences in the decay asymmetry therefore indicate violation of CP symmetry. The $\bar{p}p \to \bar{Y}Y$ reaction provides a clean test of CP violation, since the initial state is a CP eigenstate and no mixing between the baryon and antibaryon is expected to occur. Since hyperons and antihyperons can be produced and detected at the same rate and in very large amounts, the prospects are excellent for ground-breaking symmetry tests that could help us to understand the matter-antimatter asymmetry of the Universe.
With the PS185 experiment at LEAR, the CP symmetry in $\bar p p \to \Lambda \bar \Lambda$ was studied and a sensitivity of $10^{-2}$ was reached~\cite{ps185hyp}.
This measurement was based upon $\sim 10^5$ reconstructed events which statistics can be obtained in Phase One of PANDA within only a few hours of beam time.
The perspective of such an asymmetry measurement in Phase Three with a ten times higher luminosity is, therefore, very promising.
In addition, Phase Three opens up the possibility to study also single-charm hyperons. A systematic comparison between the strange and the
charm sector will be an important step towards a coherent understanding of non-perturbative QCD at different scales.


\section{Charm and exotics}

The original constituent quark model (CQM) describes mesons and baryons. In CQM, mesons are described as quark-antiquark states ($q\bar{q}$) interacting through a potential. One of the motivations for this description was the non-observation of mesons with strangeness or charge larger than unity, neither had states been observed with other spin and parity combinations than those consistent with fermion-antifermion pairs. However, QCD allows for any colour-neutral combination of strongly interacting quarks and gluons and therefore, CQM-based models can be extended to incorporate the dynamics of glueballs, hybrids and multiquarks. These states are often referred to as \textit{QCD exotics}. 

Glueballs ($gg$ or $ggg$) are formed due to the self-coupling of the colour-charged gluons. This feature of the strong interaction is of particular interest since the glueball mass has no contribution from the Higgs mechanism. Instead, it is completely dynamically generated by the strong interaction. Most glueballs predicted by QCD or phenomenological models have the same quantum numbers as mesons and hence they can mix. As a consequence, it is a challenge to unambiguously determine the glueball fraction of an observed hadronic state. 

In addition to glueballs, there are meson- or baryon-like states for which QCD admits a gluonic component called hybrids, {\it e.g.} $q\bar{q}g$. Hybrids can, in addition to the spin-parity combinations allowed for regular hadrons, also have \textit{spin-exotic} quantum numbers. To establish the existence of hybrids experimentally, the decomposition of quantum numbers requires sophisticated partial-wave analysis (PWA) tools and large data samples. 

Also other colourless combinations of multiquark resonances are allowed within QCD. The study of multiquarks has experienced tremendous progress during the last decade. Examples of multiquark states are \textit{tetraquarks} ($qq\bar{q}\bar{q}$) or \textit{pentaquarks} ($qqqq\bar{q}$). However, many open questions remain, in particular about the internal structure of the observed states. Precision measurements of various resonance properties are needed, as well as ab-initio theoretical predictions, in order to reach deeper insights about the structure of multiquark states.

The search for exotic hadrons is being carried out at several energy scales, from the light $u$ and $d$ scale to the bottom quark scale. 
A fundamental question to be answered concerns the relevant degrees of freedom -- should excited light hadrons be described in terms of quarks and gluons, or are various dynamical effects, \textit{e.g.} at meson pair thresholds, more important? In the light quark sector, many resonances are broad and overlap in mass. This means that they mix if they have the same quantum numbers. The advantage of the light sector is that the production cross sections are generally large, allowing for large data samples to be collected within a short time. This is an advantage when determining spin and parity through partial-wave analyses. 

The physics of hidden-charm states, such as charmonium, is expected to be very different due to the higher mass of the charm quark ($m_{\rm c}\simeq$ 1.2 GeV/$c^2$ $>\Lambda_{\rm QCD}$). The strong coupling constant in this region is $\alpha_s \approx 0.3$, corresponding to an energy scale below the region in which perturbation theory starts to break down.
At these energies, quark and gluon degrees of freedom become relevant. The velocity of the charm quark is relatively small, $(v/c)^2\sim$0.3. Systems with charm can be partly described in a non-relativistic framework with relativistic effects added perturbatively, such as spin-spin and spin-orbit coupling~\cite{brambilla:2005}. One of the interesting questions is how large are the relativistic corrections actually. The structure of a separated energy scale ($m_{\rm c}\gg m_{\rm c}v/c\gg m_c (v/c)^2$) makes heavy-quark systems, such as charmonium, ideal probes to study the transition between perturbative and non-perturbative regimes~\cite{brambilla:2011,brambilla:2014}.

Meson-like systems composed of heavy and light constituent quarks, such as open-charm states, are complementary to that of hidden-charm meson-like states. Also here, various striking experimental observations have been made in the past~\cite{aubert:2003,besson:2003} pointing to the possible existence of narrow resonances that do not fit into the conventional heavy-light meson pattern. A recent example is the intriguing observation of LHCb, speculating the existence of an open-charm tetraquark with a mass around 2.9~GeV/$c^2$~\cite{Aaij2020}. Besides spectroscopy aspects, ground-state open-charm states decay weakly, providing access to, {\it e.g.}, semi-leptonic form factors. The field of open-charm spectroscopy and electro-weak processes will become accessible in the later stages of PANDA, beyond that of Phase One. Its success depends on the completion of the vertex reconstruction capabilities of PANDA and higher luminosities for excellent statistical significance. Differential cross section measurements will be accessible in Phase One, which allows for unique studies of the production mechanism of pairs of open-charm mesons and baryons in antiproton-proton collisions. Such measurements have the potential to study the intrinsic charm content of the nucleon and, thereby, shed light on the recently predicted nonvanishing asymmetric charm-anticharm sea from lattice QCD~\cite{sufian2020}. 

In the following, we discuss the Phase One perspectives of the meson-like spectroscopy programme of PANDA at various mass scales, starting from the light-quark sector to the hidden-charm region.

 \subsection{Light exotics} 
 
 \subsubsection{State of the art}
 
Lattice QCD calculations have resulted in detailed predictions for the glueball mass spectrum in the quenched approximation and more exploratory results in unquenched simulations~\cite{GlueballLattice}. There is consensus that the ground-state is a scalar ($J^{PC}=0^{++}$) in the mass range of about $1600\,\mathrm{MeV}/c^2$
which leads to mixing with nearby $q\bar{q}$ states~\cite{GlueballReview}.
Mixing scenarios include \textit{e.g.} the observed $f_0(1370)$, $f_0(1500)$ and $f_0(1710)$. 
Detailed experimental studies of their decay patterns, carried out mainly in antiproton annihilation experiments at CERN (Crystal Barrel and OBELIX at LEAR)
~\cite{Bertin:1997kh,Bertin:1998hu,Nichitiu:2002cj,Bargiotti:2003ev,Amsler:1992rx,Amsler:1994pz,Amsler:1994rv,Amsler:1994ah,Amsler:1995bz,Amsler:1995bf,Amsler:1995gf,Abele:1996nn,Abele:1996fr,Abele:1998qd,Abele:2001js,Abele:2001pv,Amsler:2002qq,Amsler:2006du}
and at Fermilab (E760 and E835)~\cite{Armstrong:1993fh, Uman:2006xb}, confirm this picture. 
A pseudoscalar glueball is predicted by lattice QCD above $2\,\mathrm{GeV}/c^2$. The much lighter $\eta(1440)$ has been suggested as a candidate, though it is unclear whether this is one single resonance or two ($\eta(1405)$ and $\eta(1475)$)\cite{PDG}. The possible existence of a $\eta(1275)$ complicates the picture further~\cite{pseudoglue}. 

\begin{sloppypar}
The lightest tensor ($J^{PC} = 2^{++}$) glueball is predicted in the mass range from $2$ to
$2.5\,\mathrm{GeV}/c^2$~\cite{GlueballReview}.
The possible mixing of two nonets
(${}^3P_2$ and ${}^3F_2$) results in five expected isoscalar
states. The JETSET collaboration at LEAR has reported a tensor 
component in the mass range around $2.2\,\mathrm{GeV}/c^2$ in the $\bar{p}p \to \phi\phi$ reaction \cite{Evangelista:1998zg}. However, due to the limited size of the data sample, no firm conclusions could be drawn. 
\end{sloppypar}

In the vicinity of meson-pair production thresholds, narrow meson-like excitations can appear. Prominent examples in the light quark sector are the $a_0(980)$ and the $f_0(980)$ scalar mesons. These states are strongly attracted by the $K\bar{K}$ threshold and are believed to have a large $K\bar{K}$ component. The narrow vector meson $\phi(2170)$, discovered by BaBar \cite{Aubert:2007ym}, is particularly interesting in this context. It does not fit into the $q\bar{q}$ model, it is comparatively narrow ($\approx 83\,\mathrm{MeV}$) and the mass is close to the $\phi f_0(980)$ threshold. It is debated whether the $\phi(2170)$ is an $s\bar{s}$ tetraquark or hybrid state \cite{Y2174th}. Close to the $K^*\bar{K}$ threshold, the COMPASS collaboration discovered a relatively narrow ($\Gamma\approx 153\,\mathrm{MeV}$) axial-vector meson, the $a_1(1420)$ \cite{Adolph:2015pws}. It has been interpreted as the isospin partner of the established $f_1(1420)$ \cite{Bertin:1997zu}. The latter can be attributed a molecular-type $K\bar{K}\pi$ component~\cite{Longacre:1990uc}, opening up for a possibility that also the $a_1(1420)$ is a molecular-type state. The first coupled-channel calculation related to a potential axial-vector molecule state originates from~\cite{Lutz:2003fm}. There are further interpretations proposed such a triangle singularity from rescattering of the $a_1(1260)$~\cite{Aceti:2016}. There has been significant progress in recent years in lattice scattering and coupled channel calculations including in the light meson sector~\cite{woss2019,dudek2016}.

Several experiments have reported large intensities in the spin-exotic $1^{-+}$ wave, referred to as $\pi_1(1400)$, $\pi_1(1600)$ and $\pi_1(2015)$ \cite{hybridrev}. Whereas the resonant nature of the $\pi_1(1400)$ and the $\pi_1(2015)$ is disputed, the $\pi_1(1600)$ is currently the strongest light hybrid candidate, recently re-addressed in COMPASS data~\cite{alekseev:2010,adolph:2015,aghasyan:2018,rodas:2019}. This implies the existence of so far undiscovered nonet partners.

\subsubsection{Potential for Phase One}

In the search for exotic hadrons, the gluon-rich environment and the access to all $\bar{q}q$-like quantum numbers in formation, give PANDA a unique advantage compared to $e^+e^-$ experiments. Furthermore, states with non-$q\bar{q}$ quantum numbers can be accessed in production.

The reaction $\bar{p}p\to\phi\phi$ is considered suitable for tensor glueball searches, since the production \textit{via} intermediate conventional $q\bar{q}$ states is OZI
suppressed in contrast to production \textit{via} an intermediate glueball. Already during Phase One of PANDA, we will collect data samples of this reaction that are two orders of magnitude larger than achieved by previous experiments. A potential tensor
component would reveal itself in energy
scans and amplitude analyses.

The $f_1(1420)$ can be identified through the decay to $K\bar{K}\pi$ and studied at a centre-of-mass energy of about $2.25\,\mathrm{GeV}$ in $\bar{p}p \to \pi^+\pi^- + K\bar{K}\pi$ and $\bar{p}p \to \pi^0/\eta + K\bar{K}\pi$. In the latter cases, the amplitude analysis is simpler since only one recoil ($\pi^0$ or $\eta$) is involved. The $a_1(1420)$ can be
accessed in $3\pi$ combinations from the $\bar{p}p\to \pi^+\pi^-\pi^+\pi^-$ reaction. The COMPASS analysis shows that very large samples are required \cite{Adolph:2015pws}. Here, PANDA will profit from the large expected production cross sections in $\bar{p}p$ annihilations. The cross sections for pion modes are in the order of mb, while reactions involving kaons are in the order of $100\,\mu\mathrm{b}$. The observed intensity of $f_1(1420)$ in
$\bar{p}p\to K^+ K^0 \pi^-\pi^+\pi^-$ is about $1\%$ \cite{bertin}. This makes the prospects excellent for studying the $f_1(1420)$ as well as searching for the $a_1(1420)$ in $\bar{p}p$ annihilations already during Phase One of PANDA.

Furthermore, insights on the nature of the $\phi(2170)$ will be obtained by studying other production mechanisms and hitherto unmeasured decay patterns. At PANDA, the $\phi(2170)$ will be accessible in reactions involving $\pi^0$, $\eta$, or $\pi^+\pi^-$ recoils at centre-of-mass
energies of about $2.6\,\mathrm{GeV}$. In a similar way, searches for hybrid candidates such as $\pi_1(1400)$, $\pi_1(1600)$ and $\pi_1(2015)$ can be performed.

\subsection{Charmonium-like exotics}

\subsubsection{State of the art}

In 2003, the discovery of a signal in the $J/\psi \pi^+\pi^-$ channel near the $D^0\bar D^{* 0}$ threshold completely changed our understanding of the charmonium spectra~\cite{X3872_belle}. Up to this point, the quark model originally published in 1978~\cite{Eichten:1978tg} had been very successful in describing all observed states. However, the new signal, established the state denoted $\chi_{c1}(3872)$ or $X(3872)$, turned out to have properties at odds with the quark model. After 2003, many more states in the charmonium and bottomonium mass range were discovered. While all states below the lowest $S$-wave open charm threshold behave in accordance with the quark model, the states above fit neither in mass nor in other properties. This family of exotic states is now referred to as the $XYZ$ states. Arguably the most prominent states, besides the aforementioned $\chi_{c1}(3872)$, are the vector-meson states $Y(4260)$~\cite{Y4260_babar} and $Y(4360)$~\cite{Y4360_babar} as well as the charged states $Z(4430)$~\cite{Z4430}, $Z_c(3900)$~\cite{Z3900_besiii}, $Z_c(4020)$~\cite{Z4020_besiii}, $Z_{cs}(3985)$~\cite{Zcs3985} in the charmonium sector and the charged states $Z_b(10610)$ and $Z_b(10650)$~\cite{Zb106x0_lhcb} in the bottomonium sector. The most prevalent interpretations of these states are hybrid mesons (quark states with an active gluon degree of freedom), compact tetraquarks (bound systems of diquarks and anti-diquarks), hadro-quarkonia (a compact heavy quarkonium surrounded by a light quark cloud) and hadronic molecules (bound systems of two mesons; when located very near the relevant $S$-wave threshold these can be very extended). Recent reviews of various calculations can be found in Refs.~\cite{ccbarrev1,rev_XYZ,ccbarrev3,Brambilla:2019esw}. In particular the $Z$ states -- charged states decaying into final states that contain both a heavy quark and its antiquark --- have received a lot of attention since they must contain at least four quarks~\cite{Olsen:2017bmm}.
In addition, lattice calculations predict a supermultiplet of hybrid mesons including exotic quantum numbers with a similar pattern in both charmonium and bottomonium~\cite{liu2012}.

As of today there is no consensus which one of the mentioned models explains the properties of the $XYZ$ states best. Clearly more experimental information is needed to make progress. The two most pressing issues are:

\begin{itemize}

\item Where are the spin partner states of the observed $XYZ$ states? Their location contains valuable information about the most prominent component of the states, since different assumptions lead to different effects of spin symmetry violation~\cite{Cleven:2015era}. PANDA is well prepared to hunt for those spin partner states, since the production mechanism is not constrained to certain quantum numbers.

\item What is the line shape of the near threshold states? This allows one to especially investigate the role of the two-meson component in a given state, since a strongly coupled continuum necessarily leaves an imprint in the line shapes~\cite{ccbarrev3}. Moreover, a virtual state cannot have a prominent compact component \cite{matuschek}.

\end{itemize}

\noindent PANDA can provide a significant contribution to answer these questions, in particular the second one, already in Phase One. Precision measurements of line-shape parameters of resonances provide crucial information that sheds light on their internal structure. The determination of these parameters for narrow states is particularly challenging and requires a facility with sufficient resolution to reach the necessary sensitivity.

In the following, we illustrate this by discussing the capability of PANDA to perform resonance energy scans, using the famous $\chi_{c1}(3872)$ state with $J^{PC}=1^{++}$ as a benchmark. The $\chi_{c1}(3872)$ has a small natural width; until recently the 90\% C.L. upper limit was estimated to be 1.2~MeV~\cite{BelleXwidthUL2011}. A new measurement from the LHCb data are compatible with an absolute Breit-Wigner decay width of $\Gamma = 1.39 \pm 0.24 \pm 0.10$ MeV for the $\chi_{c1}(3872)$~\cite{LHCb2020}. However, a Flatt\'e-like line shape model where the state is described by a resonance pole with a Full-Width-at-Half-Maximum of about 220~keV is equally probable. The result from LHCb emphasises the need for precision line-shape measurements with significantly better mass resolution than offered by experiments that rely on the detector resolution, typically around a few MeV. Only experiments like PANDA, where these resonances are accessible in formation, offer a direct and thus model-independent measurement of the line-shape.

The analysis presented in the following is meant as a demonstration of the precision capabilities of PANDA, but the technique can be applied to extract key properties of other resonances as well.

\begin{table*}[h!]
\small
\caption{Summary of parameter settings in the simulation study~\cite{xscan}. All parameters are defined in the text.}
\vspace{1ex}
\label{tab:parasum}
\centering
\begin{tabular}{c|c}
Input parameter & Input value\\\hline
${\cal B}(X\rightarrow J/\psi\rho^0)$       & 5\,\% ~\cite{Z3900_besiii,BelleXJ2Pi,CDFJrho}\\
${\cal B}(J/\psi\rightarrow e^+e^-)$      & 5.971\,\%  ~\cite{PDG}\\
${\cal B}(J/\psi\rightarrow \mu^+\mu^-)$     & 5.961\,\% ~ \cite{PDG}\\
${\cal B}(\rho^0\rightarrow \pi^+\pi^-)$    &  100\,\% ~\cite{PDG}\\\hline
\multirow{2}*{$\sigma_{\bar{p}p\to X,\max}$}    &  50\,nb~\cite{BelleXJ2Pi,LHCbX}\\
                                  &  [20, 30, 75, 100, 150]\,nb\\ 
$\sigma_{B,\mbox{\scriptsize DPM}}$  & 46\,mb~\cite{Flamino1984} \\
$\sigma_{B,\mbox{\scriptsize NR}}$      & 1.2\,nb~ \cite{ChenNRBG}\\\hline 
Total scan time $t_{\rm scan}$   & 80\,d\\
No. of scan points $N_{\rm scan}$ & 40  \\\hline
{Breit-Wigner width $\Gamma_X$}    & $[50,70,100,130,180,250,500]$\,keV  \\
{Line-shape parameter $E_{\rm f}$} & $-[10.0, 9.5, 9.0, 8.8, 8.3, 8.0, 7.5, 7.0]$\,MeV \\

\bottomrule
\end{tabular}
\end{table*}

\subsubsection{Potential for Phase One}
\label{sec:scan}

 PANDA offers a unique possibility to reach sub-MeV resolution exploiting the cooled antiproton beam from the HESR. This has been demonstrated by a feasibility study of the $\chi_{c1}(3872)$ line-shape measurement, to be carried out in a future energy scan designed to precisely measure absolute decay widths and line shapes \cite{xscan}. The $\chi_{c1}(3872)$, as well as all other non-exotic $J^{PC}$ combinations, can be created in formation in $\bar{p}p$ annihilation. 

 The details of the PANDA feasibility study can be found in Ref.~\cite{xscan}. In this paper, we focus on the conditions expected for Phase One. This implies an HESR beam momentum spread (beam energy resolution) of d$p/p=5 \cdot 10^{-5}$ (d$E_{\rm CMS}=83.9$\,keV) and an integrated luminosity of $\cal{L}=$ 1170\,(day $\cdot$ nb)$^{-1}$.

The reaction of interest is the direct formation $\bar{p}p \rightarrow \chi_{c1}(3872)$, where the $\chi_{c1}(3872)$ is identified by the two leptonic $J/\psi$ decay channels $\chi_{c1}(3872) \rightarrow J/\psi \rho^0 \rightarrow e^+ e^- \pi^+ \pi^-$ and $\chi_{c1}(3872) \rightarrow J/\psi \rho^0 \rightarrow \mu^+ \mu^- \pi^+ \pi^-$. The reconstruction efficiencies are $12.2\,\%$ and $15.2\,\%$, respectively, as determined with Monte Carlo simulations including a realistic GEANT detector implementation. The physics parameters as summarised in Tab.\,\ref{tab:parasum} have been used as input. 

In our study, we quantify i) the sensitivity of an absolute measurement of the natural decay width $\Gamma_{0}$ ii) the capability to distinguish two scenarios: a loosely bound $(D^0$-$\bar{D}^{\ast 0})$ molecular state and a virtual scattering state. 

Both scenarios have been studied under the assumption that PANDA will collect data in 40 energy points during 2x40 days of beam time, {\it i.e.} two days per energy point, with the Phase One operation conditions. This is considered a reasonable amount of time to allocate for this kind of measurements, especially since data for other purposes, \textit{e.g.} hyperon-antihyperon pair production, can be collected in parallel.

The parameter $\Gamma_0$ is determined by fitting a Voigt function, \textit{i.e.} a convolution of a Breit-Wigner with a natural decay width $\Gamma_0$ and a Gaussian with a standard deviation $\sigma_{\rm Beam}$, accounting for the beam momentum distribution. 

The molecular line shape differs significantly from that of a less sophisticated Breit-Wigner-like resonance shape. It depends on the given decay channel (here $J/\psi\pi^+\pi^-$) and on the dynamic Flatt\'{e} parameter $E_{\rm f}$~\cite{HanhartLS,KalashLS} (or the equivalent inverse scattering length, $\gamma$ in~\cite{BraatenLS}), that model a bound or virtual state. 

\begin{figure*}[tp!]
    \begin{center}
     \includegraphics[clip,trim= 0 0 10 0, width=0.32\linewidth, angle=0]{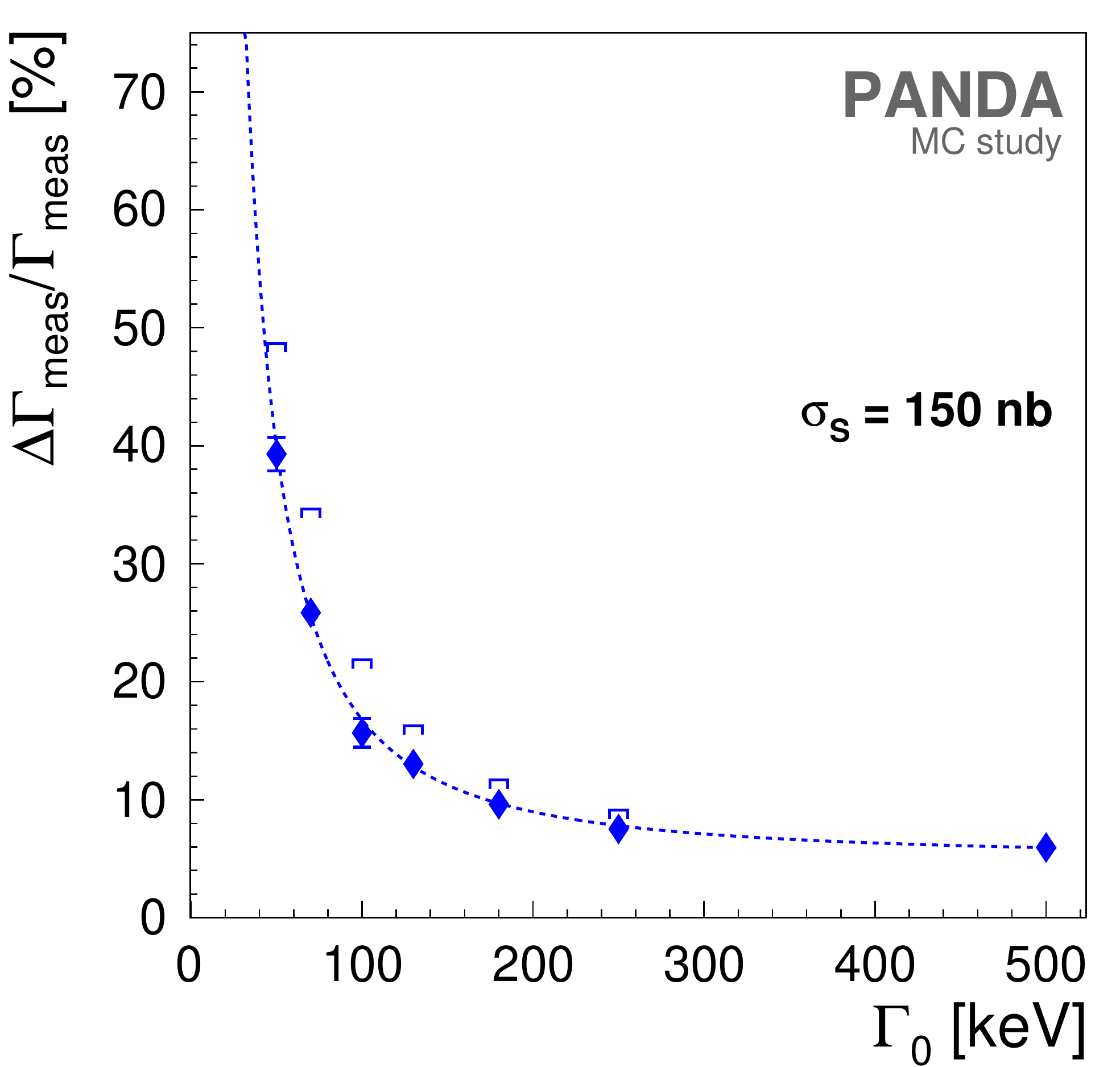}
     \includegraphics[clip,trim= 0 0 10 0, width=0.32\linewidth, angle=0]{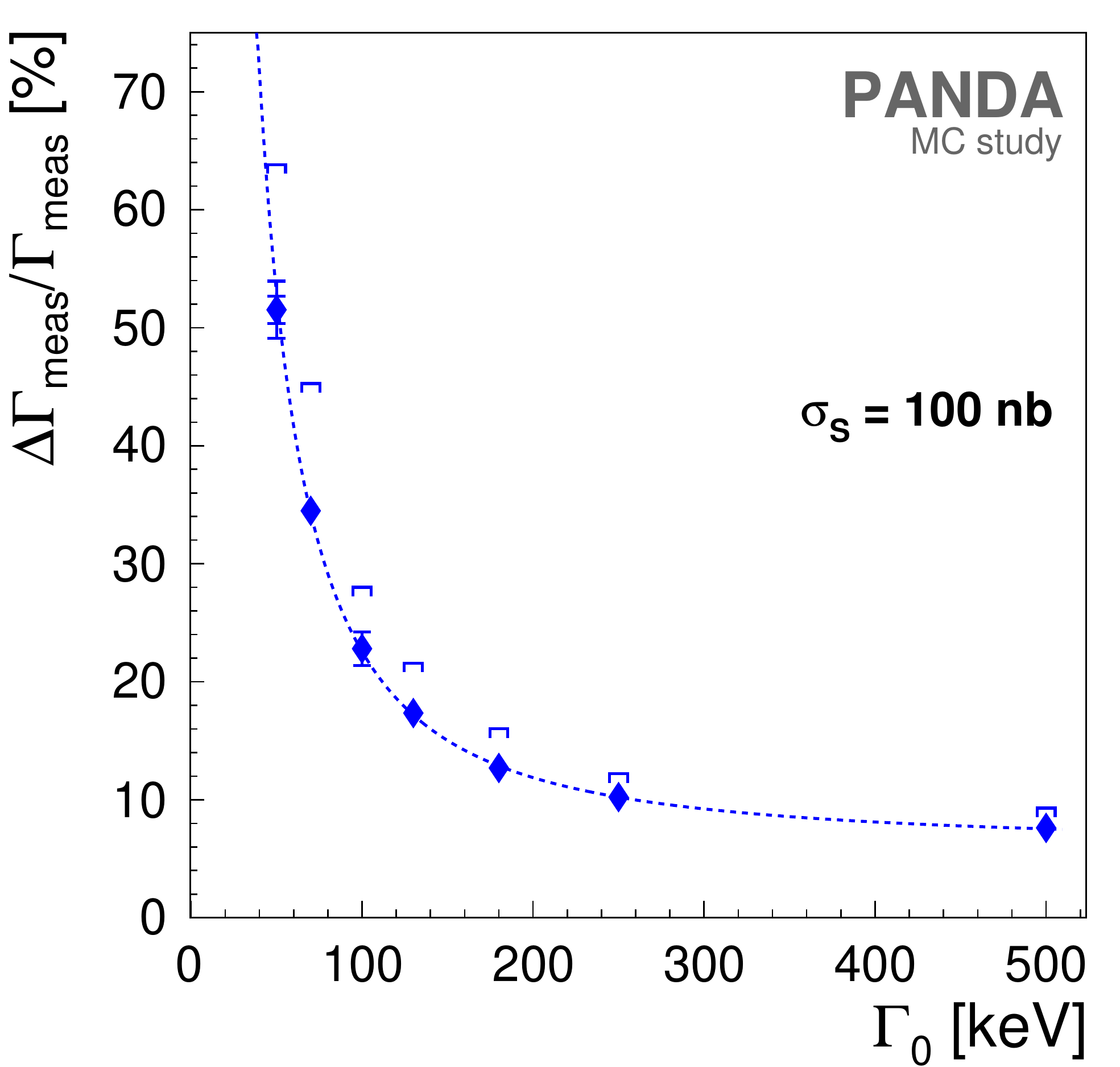}
     \includegraphics[clip,trim= 0 0 10 0, width=0.32\linewidth, angle=0]{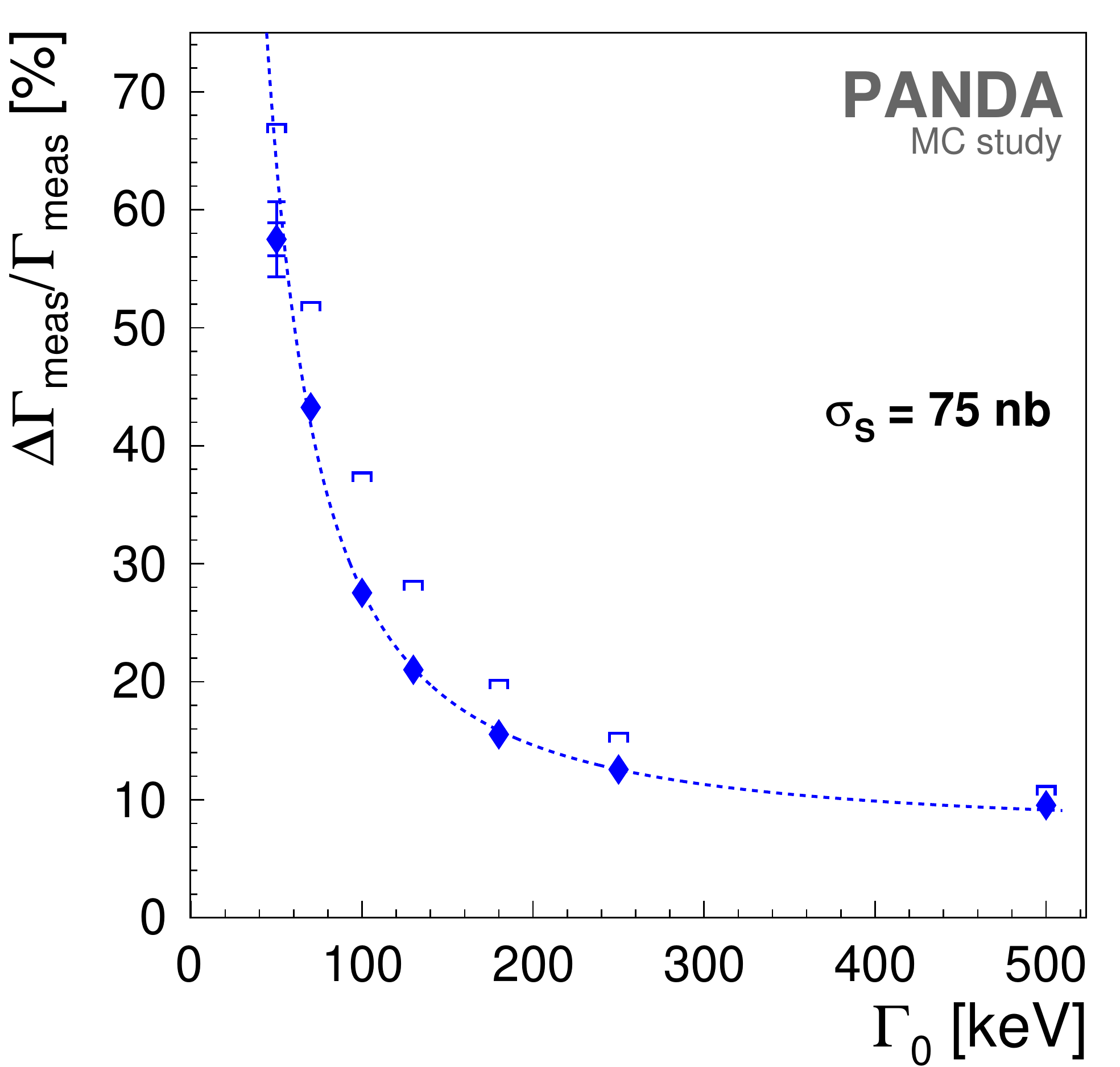}
     \includegraphics[clip,trim= 0 0 10 0, width=0.32\linewidth, angle=0]{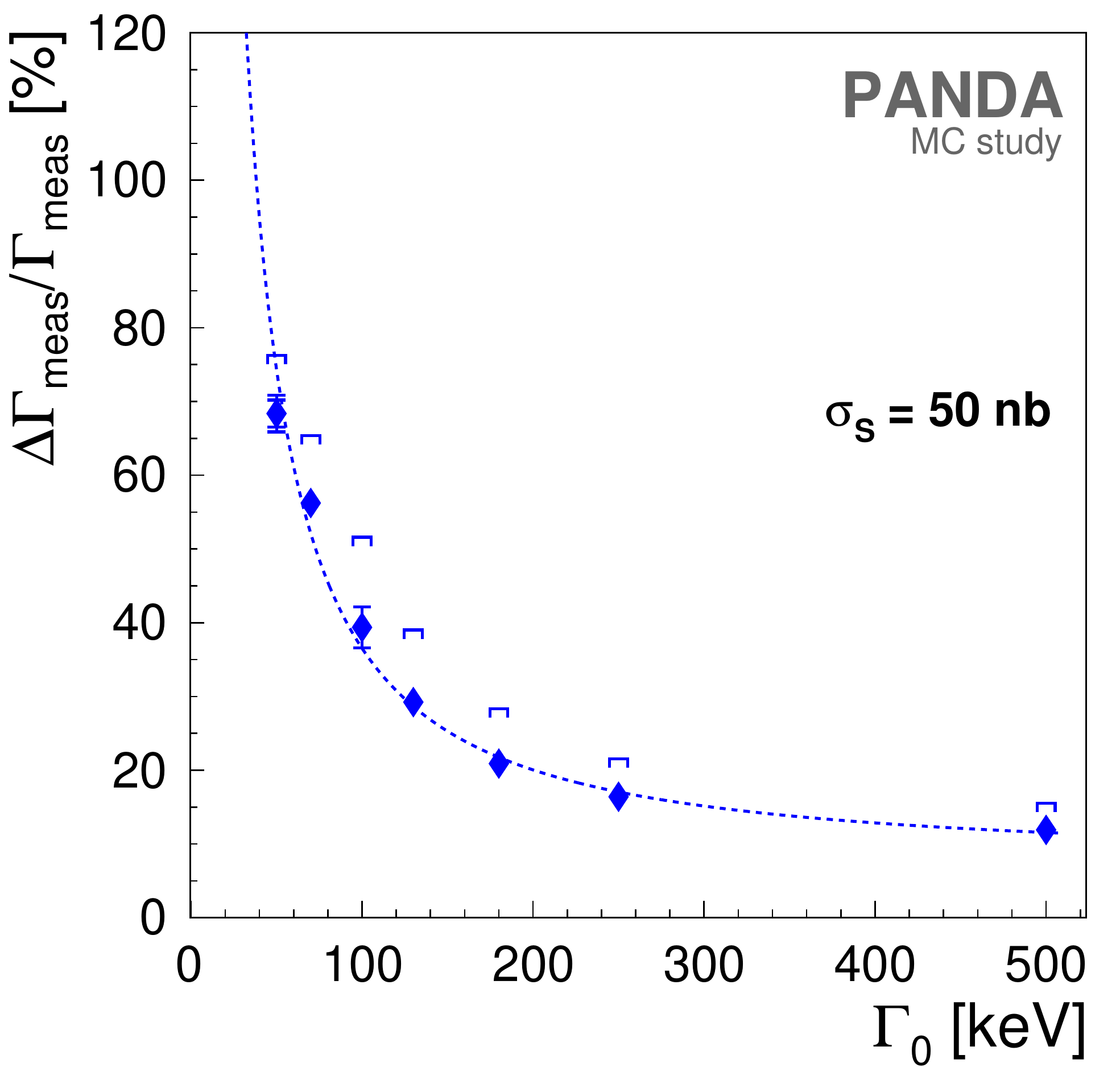}
     \includegraphics[clip,trim= 0 0 10 0, width=0.32\linewidth, angle=0]{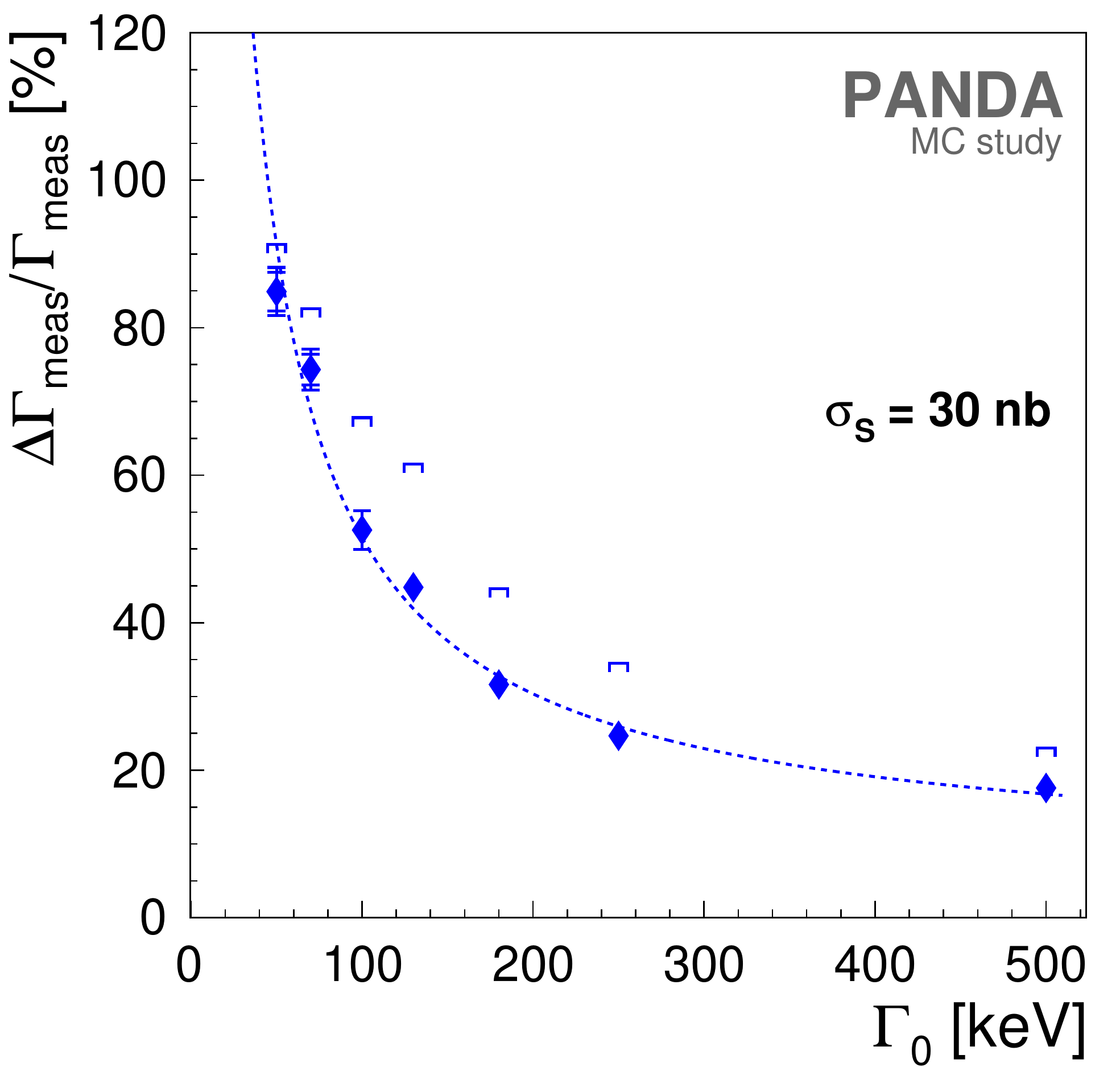}
     \includegraphics[clip,trim= 0 0 10 0, width=0.32\linewidth, angle=0]{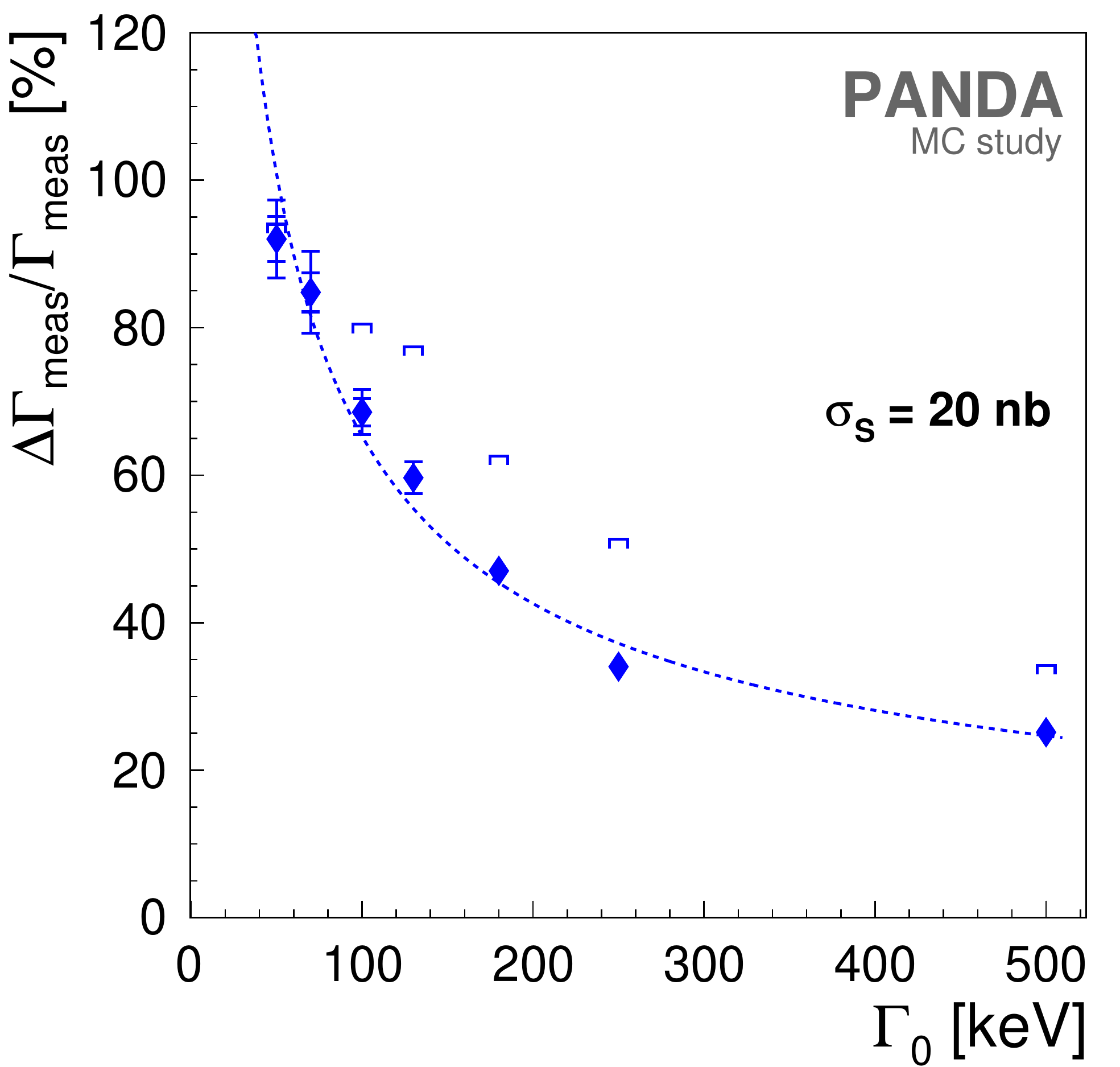}
  \caption{Sensitivity to the absolute Breit-Wigner width, parameterised in terms of the relative uncertainties $\Delta\Gamma_{\rm meas} / \Gamma_{\rm meas}$, shown as a function of the input decay width $\Gamma_{\rm 0}$ of a narrow resonance for six different input signal cross-sections $\sigma_{\rm S}$. All results are extracted for the Phase One HESR running mode. The inner error bars represent the statistical uncertainties and the outer the systematic ones. The bracket markers indicate the corresponding numbers for the case of DPM \cite{Capella1994} and non-resonant background upscaling according to~\cite{OHelene}, ignoring statistical and systematic errors.}
      \label{fig:SensitivityResultsBW}                                
     \end{center}
\end{figure*}
%
%
\begin{figure*}[h!]
    \begin{center}
     \vspace{-0.2cm}
     \includegraphics[clip,trim= 0 0 10 0, width=0.32\linewidth, angle=0]{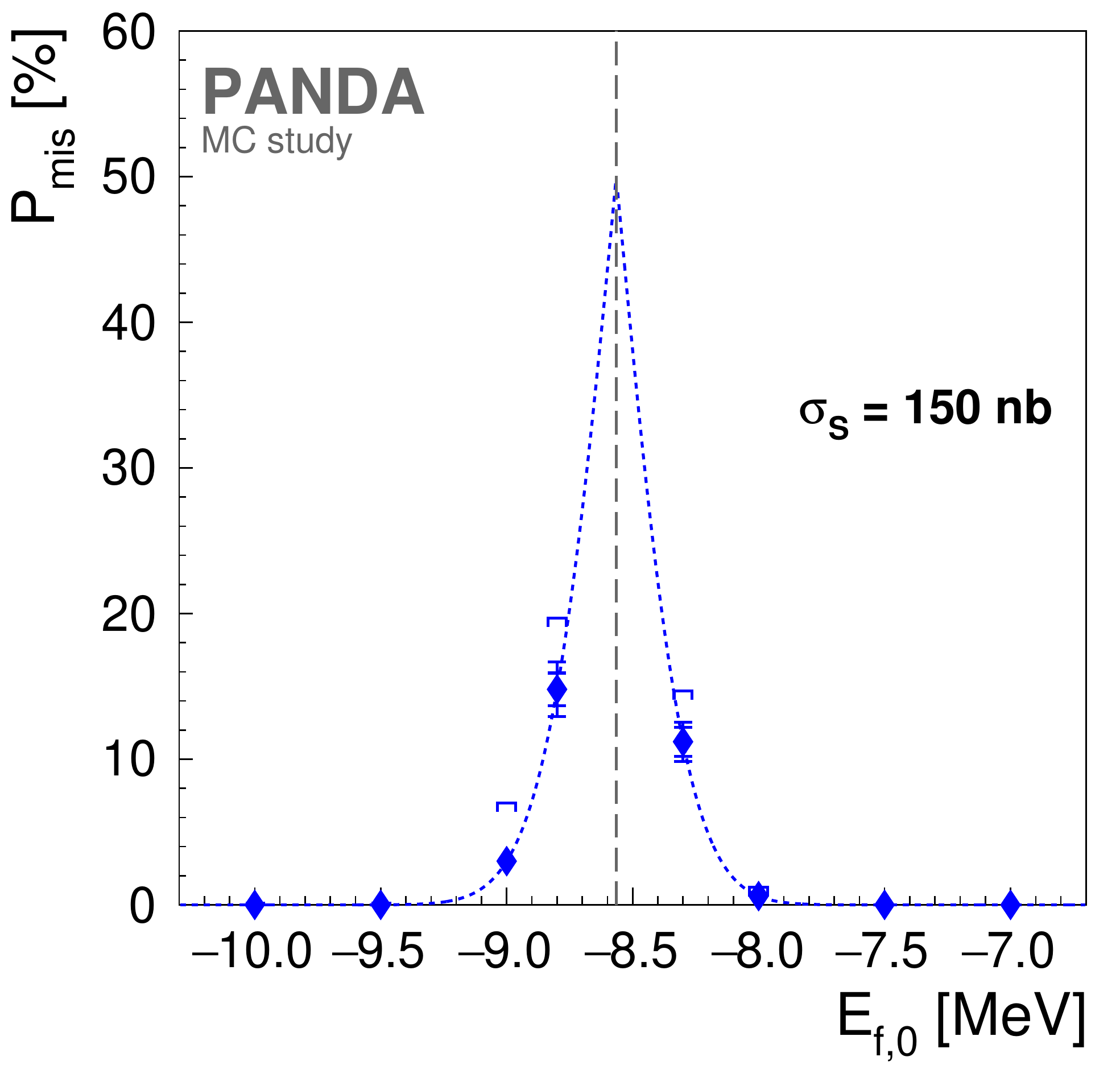}
     \includegraphics[clip,trim= 0 0 10 0, width=0.32\linewidth, angle=0]{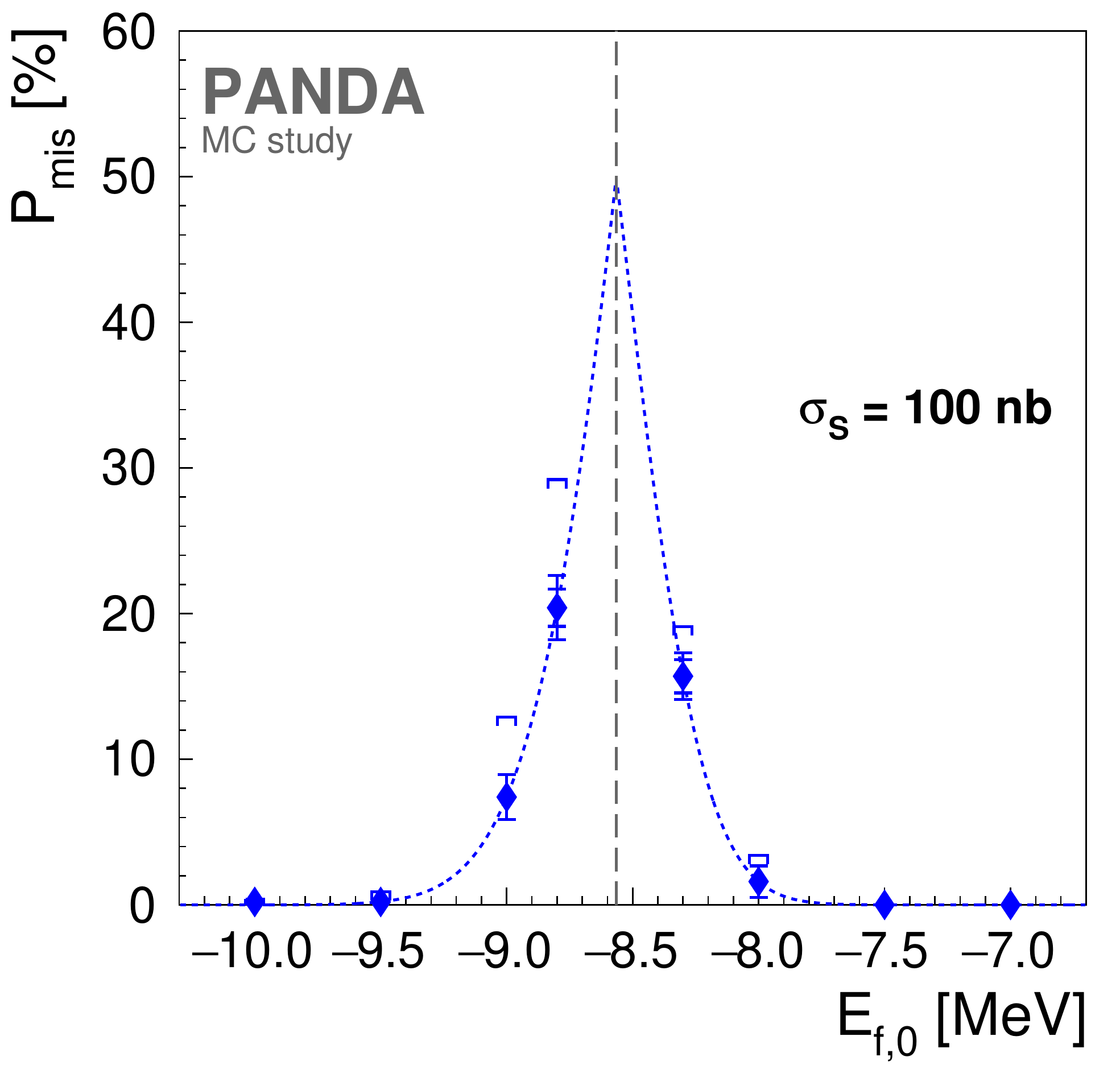}
     \includegraphics[clip,trim= 0 0 10 0, width=0.32\linewidth, angle=0]{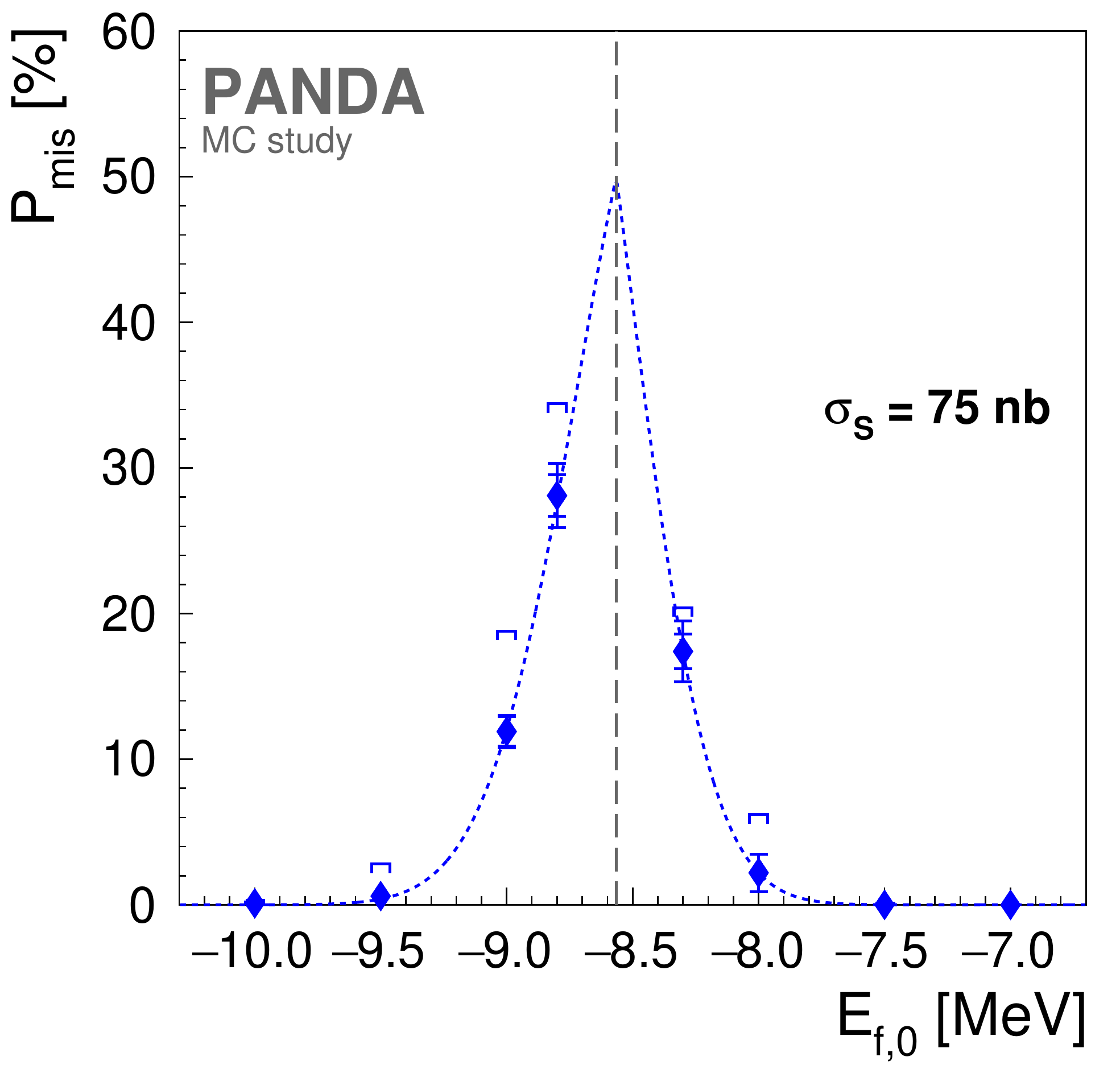}
     \includegraphics[clip,trim= 0 0 10 0, width=0.32\linewidth, angle=0]{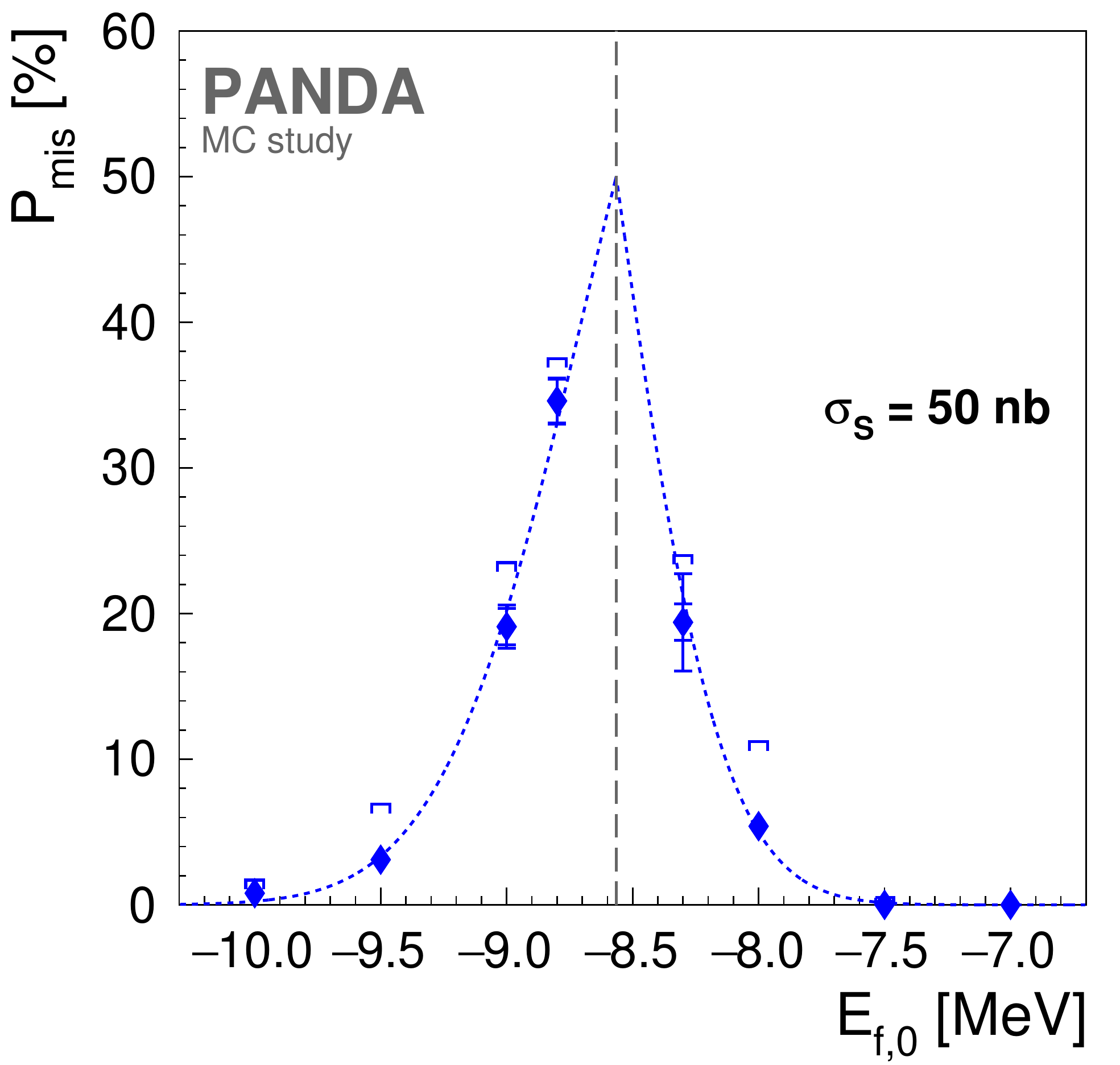}
     \includegraphics[clip,trim= 0 0 10 0, width=0.32\linewidth, angle=0]{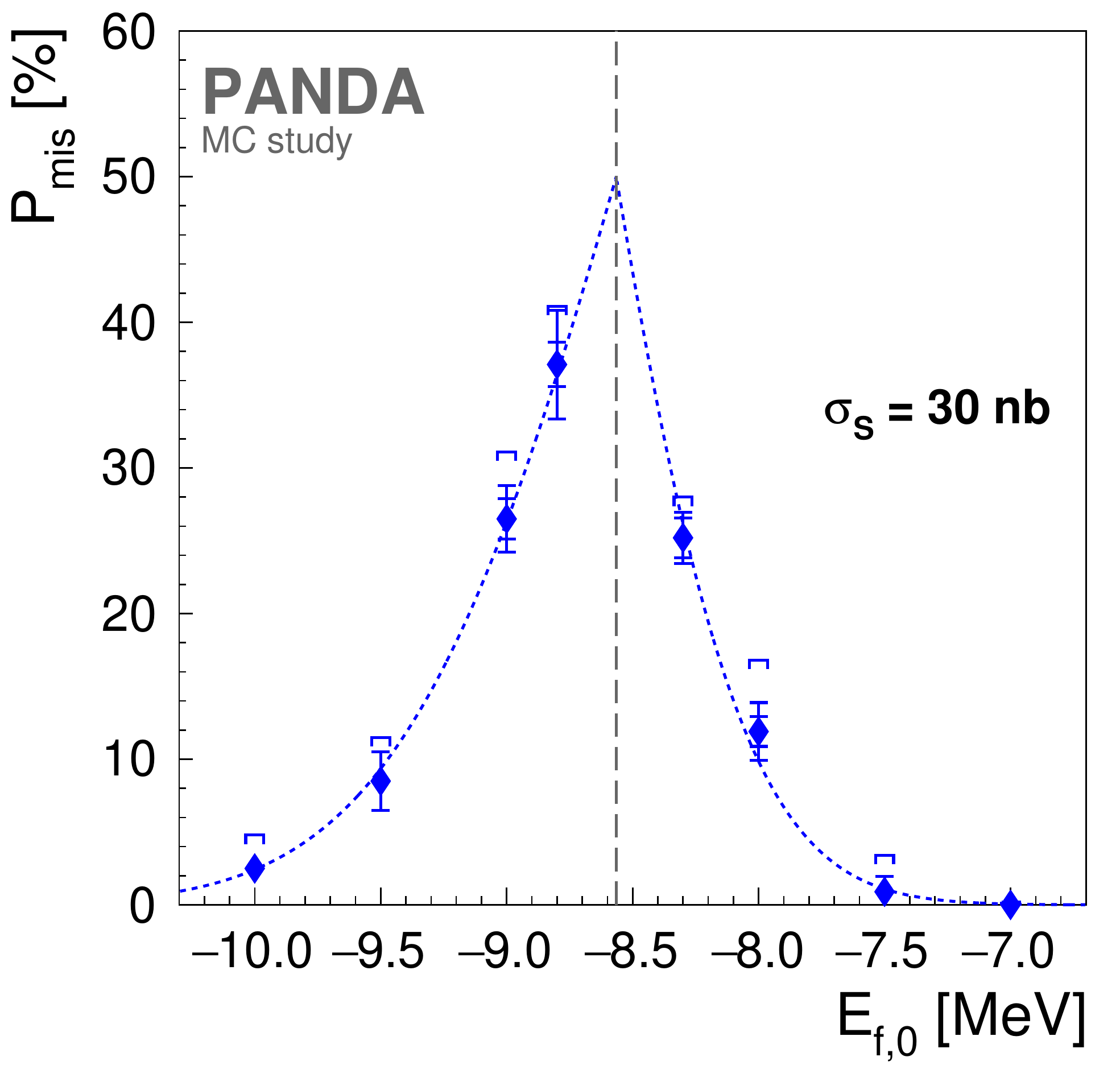}
     \includegraphics[clip,trim= 0 0 10 0, width=0.32\linewidth, angle=0]{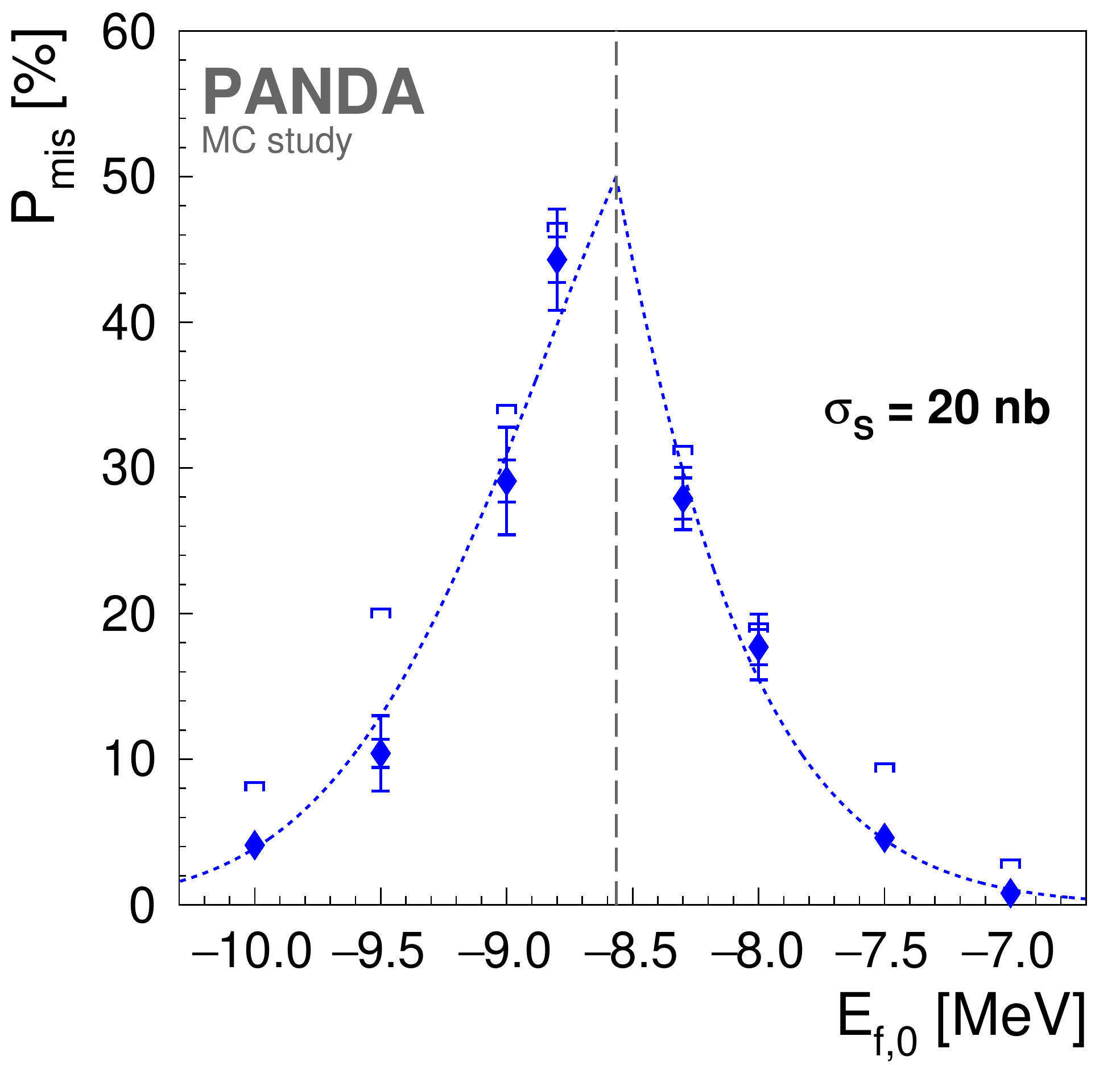}

     \caption{Sensitivity to the $\bar{D}^*D$ molecule scenario, parameterised in terms of the 
       mis-identification probability $P_{\rm mis}$, shown as a function of the input Flatt\'{e} parameter $E_{\rm f,0}$ of the $\chi_{c1}(3872)$ for six different input signal cross-section $\sigma_{\rm S}$. All results are extracted for the Phase One HESR running mode. The inner error bars represent the statistical uncertainty and the outer the systematic ones. The bracket markers indicate the corresponding numbers for the case of DPM \cite{Capella1994} and non-resonant background upscaling according to~\cite{OHelene}, ignoring statistical and systematic uncertainties.
     }
      \label{fig:SensitivityResultsEf}                                 
     \end{center}
\end{figure*}

For each of the six different input signal cross-sections, $\sigma_{\rm S}=(150, 100, 75, 50, 30, 20)\,$nb, the full procedure of simulation, 
PDF generation and Breit-Wigner/Molecule line shape fitting has been carried out, employing a maximum-likelihood method. 

The resulting sensitivities in terms of the relative uncertainty $\Delta\Gamma_{\rm meas} / \Gamma_{\rm meas}$ of the measured decay width are summarised for the Breit-Wigner case in Fig.\,\ref{fig:SensitivityResultsBW}. The corresponding sensitivity for the molecule case is parameterised in terms of the misidentification probability $P_{\rm mis}= N_{\rm mis-id}/N_{\rm MC}$. The $P_{\rm mis}$ as a function of the input Flatt\'{e} parameter $E_{\rm f,0}$ is shown in Fig.\,\ref{fig:SensitivityResultsEf}.

The available computing resources result in limited samples of DPM~\cite{Capella1994} background. This, in combination with an efficient background suppression of the order $\epsilon_{B,\rm gen}\approx 1\cdot10^{-10}$, results in a very small number of surviving background events which introduces an uncertainty. The impact is estimated by scaling up the number of background events determined from the 90\% confidence level upper limit, according to~\cite{OHelene}. The uncertainty due to non-resonant background from $p\bar{p}\rightarrow J/\psi\rho^{0}$ was determined in a similar way. The effect on the sensitivity is represented by bracket markers in Figs.\,\ref{fig:SensitivityResultsBW} and \ref{fig:SensitivityResultsEf}. 

\begin{figure*}[tp!]
    \begin{center}
     \vspace{-0.2cm}
     \includegraphics[clip,trim= 0 0 10 0, width=0.4\linewidth, angle=0]{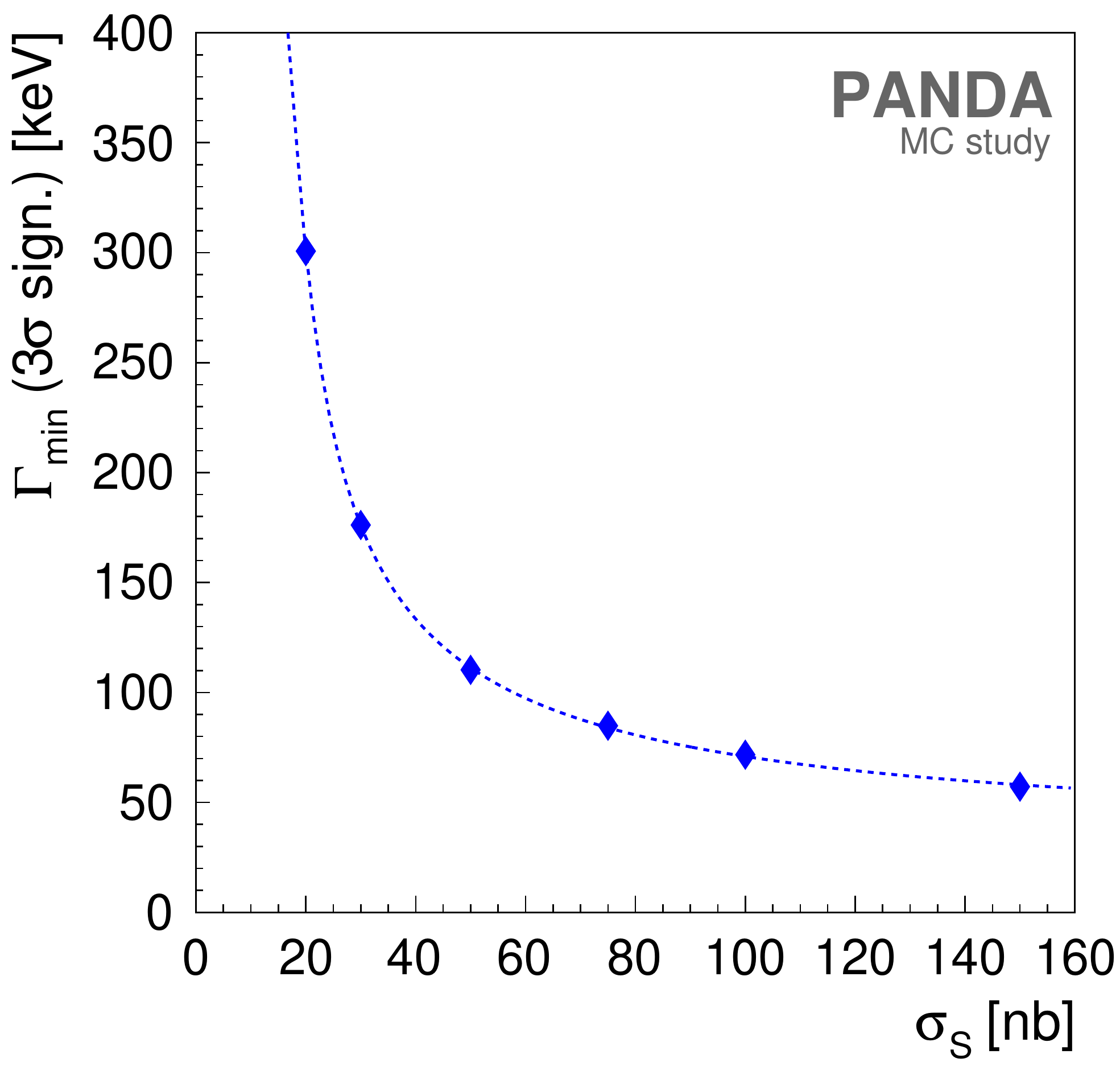}
     \hspace*{0.5cm}
     \includegraphics[clip,trim= 0 0 10 0, width=0.4\linewidth, angle=0]{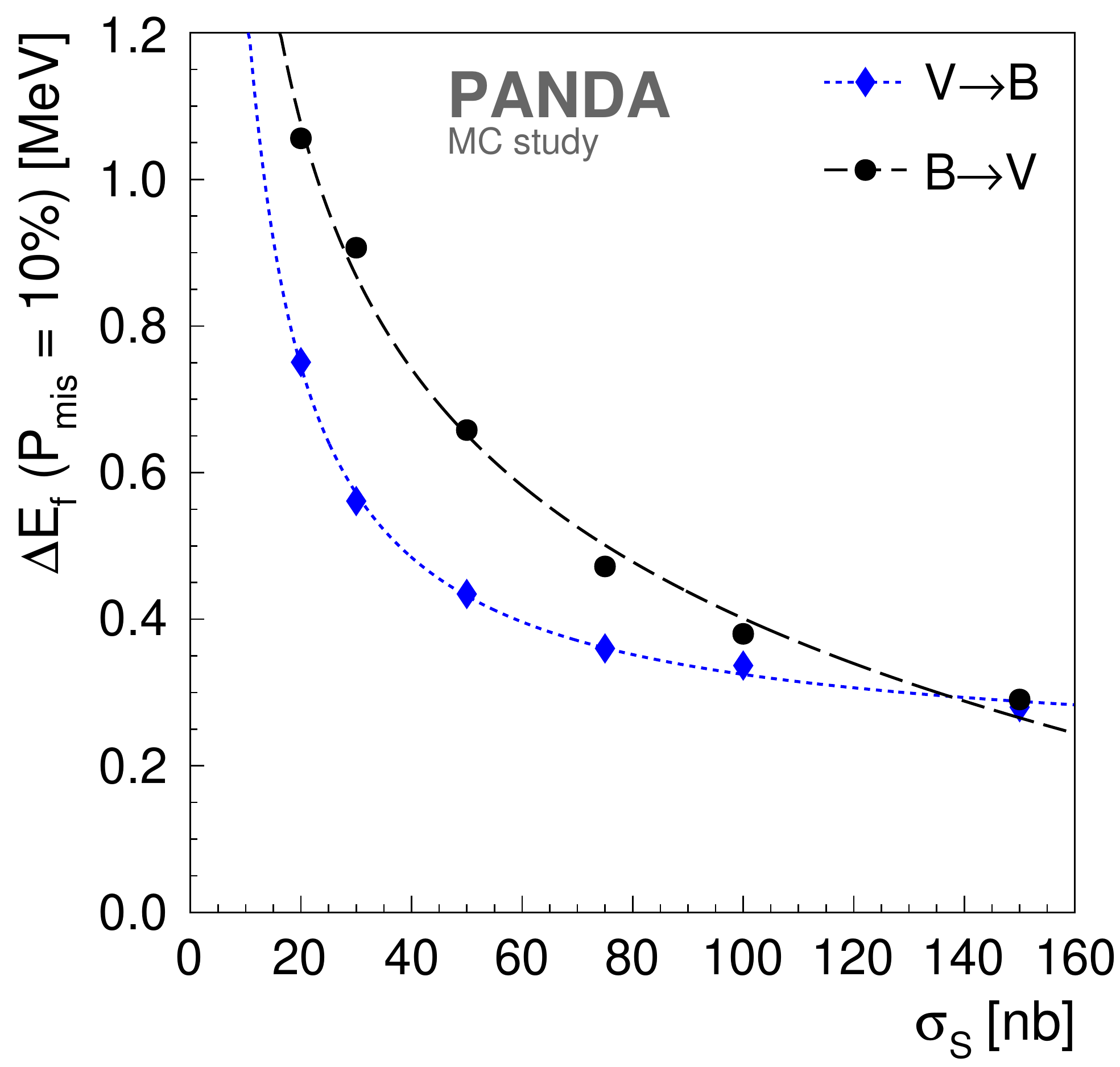}

     \caption{Left: Sensitivity in terms of $\Gamma_{min}$ for a 33\,\% relative error ($3\sigma$) BW width measurement.
       Right: Sensitivity in terms of the Flatt\'e parameter difference $\Delta E_f$ for a misidentification of $P_{\rm mis}=10\,\%$ for the molecular line-shape measurement. The black circles represent a bound molecular state misidentified as a virtual state ($P_{\rm mis, B\rightarrow V}$) and the blue diamonds a virtual state misidentified as a bound molecular state ($P_{\rm mis, V\rightarrow B}$).}
      \label{fig:SensitivityResultsBW_Ef}                                 
     \end{center}
\end{figure*}

A more compact representation of the results extracted from Figs.~\ref{fig:SensitivityResultsBW} and \ref{fig:SensitivityResultsEf} is shown in Fig.\,\ref{fig:SensitivityResultsBW_Ef} for the Breit-Wigner scenario (left panel) and the molecule scenario (right panel). In the BW case, the minimum $\Gamma_0$ is defined by the minimum width, for which a $3\sigma$ sensitivity is achieved in an absolute decay width measurement. This corresponds to a relative uncertainty $\Delta\Gamma_{\rm meas}/\Gamma_{\rm meas}$ of $33\,\%$. In the left panel of Fig.\,\ref{fig:SensitivityResultsBW_Ef}, the 3$\sigma$ sensitivity is shown as a function of the input $\sigma_{\bar{p}p\to X,\max}$. Trendlines for inter- and extrapolation are added using an empirical analytical function. 

In the molecule case, the capability of distinguishing a bound state from a virtual state is quantified in terms of the Flatt\'{e} parameter difference $\Delta E_{\rm f} := |E_{\rm f,0} - E_{\rm f,th}|$, where $E_{\rm f,0}$ is the input Flatt\'{e} parameter and $E_{\rm f,th}$ is the threshold energy separating a bound from a virtual state. The difference can be extracted from Fig.\,\ref{fig:SensitivityResultsEf} at $P_{\rm mis}=10\,\%$, assuming $E_{\rm f,th} = -8.5651$\,MeV according to Ref.~\cite{HanhartLS,KalashLS}. The results are shown as a function of the input cross section $\sigma_{\bar{p}p\to X,\max}$ in the right panel of Fig.\,\ref{fig:SensitivityResultsBW_Ef}. As expected, the larger the cross section, the better the performance in resolving small $\Delta E_{\rm f}$.

%
%

The achievable sensitivities have been calculated for one out of the six input cross sections, $\sigma_{\rm S} = 50$\,nb, in line with the experimental upper limit on $p\bar{p} \rightarrow \chi_{c1}(3872)$ production provided by the LHCb experiment. For values of the natural decay width larger than $\Gamma_0 = 110$\,keV a $3\sigma$ relative error $\Delta\Gamma_{\rm meas}/\Gamma_{\rm meas}$ better than 33\,\%, is achieved already in Phase One with 80 days of dedicated beam time for one resonance scan measurement. The nature of the state -- bound or virtual -- can be correctly determined with a probability of 90\,\% provided for $\Delta E_{\rm f} \approx 700\,$keV. The presented work serves as an example, but the same approach will be applied to narrow resonances in general, achieving sub-MeV resolutions.


\subsection{Impact and long-term perspectives}

The planned Phase One line-shape measurement of the $\chi_{c1}(3872)$ and other states with $J^{PC} \neq 1^{--}$ can reveal the intriguing nature of hadronic states.
Once the nature of these states has been understood, it might lead to further insights on the relevant degrees-of-freedom that give rise to
hadronisation at different energy scales. 

In addition, PANDA has excellent discovery potential for hitherto unknown, exotic meson-like states thanks to the gluon-rich environment provided by $\bar{p}p$ annihilations as well as the access to all $\bar{q}q$-like quantum numbers in formation. In particular, this opens up for extensive searches for spin partners of the $XYZ$ states. Discoveries and measurements of the properties of spin partners provide valuable insights on the prominent components, since different assumptions lead to different effects of spin symmetry violation~\cite{Cleven:2015era}. 

In later phases of PANDA, when the design luminosity is reached, studies of hadrons with open charm will commence.
The structure and dynamics of these systems, composed of heavy and light constituent quarks, are complementary to that of hidden-charm meson-like states. The decay of the lowest lying states occurs primarily via weak processes, providing experimental access to the semi-leptonic form factors and the CKM parameters. 
Moreover, spectroscopy studies of the excited states can provide new insights in the non-perturbative QCD domain that are not accessible in the hidden-charm sector. 
This opens the possibility to search for exotic open-charm states. Hence, PANDA can build upon the BABAR and CLEO discoveries of the narrow exotic candidates $D_{s0}^*(2317)$~\cite{aubert:2003} and $D_{s1}(2460)$~\cite{besson:2003}, respectively. PANDA has the potential to measure the width of the $D_{s0}^*(2317)$ with a resolution in the order of 0.1~MeV via an energy scan near the threshold of the associated $D_s^{\pm}D_{s0}^*(2317)^{\mp}$ production~\cite{mertens:2012} and to search for other higher order excitations of open-charm states. This is particularly important since the width is sensitive to a possible molecular component of the state \cite{ccbarrev3,matuschek,lin2013,guo2018}.   


 
\section{Hadrons in nuclei}

Hadron reactions with nuclear targets provide a great opportunity to study how nuclear forces emerge from QCD. In particular, these reactions offer an angle to the onset of colour transparency at intermediate energies, the short-distance structure of the nuclear medium, and the effects of the nuclear potential on hadron properties. Two important aspects make antiproton probes unique in this regard:

\begin{itemize}
    \item The kinematic threshold for the production of heavy mesons (e.g. charmonia, $D$, $D^*$) and antibaryons is accessible at small beam momenta.
    \item The existence of two-body annihilation channels at large momentum transfer.
\end{itemize}

\noindent Close to threshold, the produced particles are rather slow in the laboratory frame. Since the coherence lengths are small compared to the internucleon distance, these particles interact with the nuclear residue as ordinary hadrons. The probability for such multiple interactions is quantified by the \textit{nuclear transparency} $T(A)$ and is given by the ratio of the cross section of an exclusive nuclear process with the corresponding elementary (nucleon) reaction. The antiproton beam gives access to hadron channels that are difficult to study with other probes at low momenta, for example $J/\Psi$. 

Slow particles are influenced by the nuclear mean field potentials. Antiprotons are particularly suitable for implanting low-momentum antibaryons or mesons into the nuclear environment, where resulting effects of the nuclear potential on their masses and decay widths can be studied. Nuclear potentials are crucial to gain valuable insights into neutron stars \cite{lattimer}. 

At higher beam momenta, the factorisation theorem mentioned in Section \ref{sec:nucleonstructure} becomes valid, splitting the reaction into a hard, pQCD calculable part and a soft part described by GPDs. This relies on the assumption that soft gluonic exchanges between the incoming and outgoing quark configurations are suppressed, which in turn is only possible if these configurations are colour neutral and have transverse sizes substantially smaller than the normal hadron size. While well-established at large momentum transfer, it is still an open question at which scale this phenomenon, known as \textit{colour transparency} (CT), sets in. 

Interactions at large momentum transfers also probe the short-distance ($\leq 1.2$ fm) structure of the nuclear medium itself. In this region, effects from non-perturbative QCD discussed in sections 4 to 6 come into play in the dynamics of the nuclear repulsive core, a rather unexplored territory \cite{fomin}, which is expected to have an effect on cold, dense nuclear matter such as neutron stars. 

\subsection{Antihyperons in nuclei}
\label{sec:antihyp}
\subsubsection{State of the art}

Nuclei made of protons and neutrons have been studied for more than a century. Hypernuclei, where one of the nucleons is replaced by a hyperon, and hyperatoms, where a hyperon is attached to a nucleus in an atomic orbit, have been explored since more than six decades. As a result, valuable information about the nuclear potentials of $\Lambda$ and $\Sigma^-$ hyperons has been obtained \cite{felicello}.

It was recently pointed out in Ref.~\cite{PhysRevC.82.024602} that in-medium interactions of antibaryons may cause compressional effects and may thus provide additional information on the nuclear equation of state \cite{GAITANOS2015181}. The data for antibaryons in nuclei are however rather scarce. So far, only antiprotons have been subjected to experimental studies. The antiproton optical potential has been addressed in studies of elastic $\bar{p} A$ scattering at KEK \cite{Nakamura:1984xw} and LEAR \cite{Garreta:1984zau,Garreta:1984rs}. The fits to the angular distributions of the scattered antiprotons, indicate that the potential has a shallow attractive real part $\mbox{Re}(V_{\rm opt}) = -(0-70)$ MeV and a deep imaginary part $\mbox{Im}(V_{\rm opt}) = -(70-150)$ MeV in the centre of a nucleus. This is in contrast to results from the analysis of X-ray transitions in antiprotonic atoms and of radiochemical data. Here, the real part turned out to be much deeper, $\mbox{Re}(V_{\rm opt}) = -110$ MeV, whereas the imaginary part was found to be $\mbox{Im}(V_{\rm opt}) = -160$ MeV \cite{Friedman2005}. However, the calculations of the $\bar{p} A$ elastic scattering as well as those of antiprotonic atoms, are sensitive to the $\bar{p}$ potential at the nuclear periphery. The production of $\bar{p}$ in $p A$ and $AA$ collisions, on the other hand, is sensitive to the antiproton potential deeply inside the nuclei and seems to favour $\mbox{Re}(V_{\rm opt})=-(100-250)$ MeV
at normal nuclear density as predicted by microscopic transport calculations \cite{Teis1994,Spieles1996,Sibirtsev1998}. Antiproton absorption cross sections on nuclei as well as the $\pi^+$ and proton momentum
spectra produced in $\bar p$ annihilation nuclei at LEAR calculated within the Giessen Boltzmann-Uehling-Uhlenbeck (GiBUU) model \cite{PhysRevC.80.021601} are consistent with $\mbox{Re}(V_{\rm opt}) \simeq -150$ MeV,
\textit{i.e.} about a factor of four weaker than expected from naive G-parity relations.

In Ref. \cite{GAITANOS2011193} it has been suggested that this discrepancy is a consequence of the missing energy dependence of the proton-nucleus optical potential in conventional relativistic mean-field models. The energy- and momentum dependence required for such an effect can be recovered by extending the relativistic hadrodynamic Lagrangian by non-linear derivative interactions \cite{GAITANOS2011193,GAITANOS20099,GAITANOS2013133}, hence mimicking many-body forces \cite{0004-637X-808-1-8}. Since hyperons and antihyperons play an important role in the interpretation of high-energy heavy-ion collisions and dense hadronic systems, it needs to be investigated how these concepts carry over to the strangeness sector. However, antihyperons annihilate quickly in nuclei and conventional spectroscopic studies are therefore challenging or even unfeasible. Instead, quantitative information about the potentials can be obtained from exclusive antihyperon-hyperon production in $\bar{p}A$ annihilations close to threshold. However, so far no such experimental data exist on nuclear potentials of antihyperons.

\subsubsection{Potential for Phase One}

In the absence of feasible spectroscopic methods, schematic calculations performed in Refs. \cite{josef1,josef2,alicia} indicate that the transverse momentum asymmetry
\begin{equation}
\alpha_{T}=\frac{p_{T}(Y) - p_{T}(\bar{Y})}{p_{T}(Y) + p_{T}(\bar{Y})},
\label{eq:hyp01}
\end{equation}
where $p_T({Y}/\bar{Y})$) is the transverse momentum of the hyperon/antihyperon, is sensitive to the depth of the antihyperon potential. Other observables of interest are polarisation and coplanarity.

As concluded in Section \ref{simhyperprod}, a unique feature of antiproton interactions within the PANDA energy range is the large production cross sections of hyperon-antihyperon pairs. However, due to the strong absorption of antibaryons in nuclei, the exclusive production rate of antihyperon-hyperon pairs is expected to be smaller in antiproton-nucleus collisions compared to antiproton-proton interactions.

Realistic calculations for the Phase One feasibility have been performed using the GiBUU transport model \cite{Buss20121}. Here we show recent results obtained with GiBUU, release 2017, which incorporates, inter alia, updates in the kaon dynamics and an improved parametrisations of the
hyperon-nucleon (S = -1) collision channels at low hyperon momenta
with respect to the previously used release 1.5 \cite{alicia,singhhyp}. Non-linear derivative interactions were not included. A simple scaling factor $\xi_{\overline{p}}$ = 0.22 was applied for the antiproton potential to ensure a Schr\"odinger equivalent antiproton potential of about 150\,MeV at saturation density \cite{PhysRevC.80.021601}. Since no experimental information exists so far for antihyperons in nuclei, G-parity symmetry was adopted as a starting point. The calculations were carried out for different values of the antihyperon scaling factor $\xi_{\overline{Y}}$. The calculations were performed for the following cases:
\begin{itemize}
    \item $\bar{\Lambda}\Lambda$ pair production in $\bar{p}$+$^{20}$Ne interactions at $p_{beam} = 1.52$ GeV/$c$.
    \item $\bar{\Lambda}\Lambda$ pair production in $\bar{p}$+$^{20}$Ne interactions at $p_{beam} = 1.64$ GeV/$c$.
    \item $\bar{\Lambda}\Sigma^-$ pair production in $\bar{p}+^{20}$Ne interactions at $p_{beam} = 1.64$ GeV/$c$.
    \item $\bar{\Xi}^+\Xi^-$ pair production in $\bar{p}+^{20}$Ne interactions at $p_{beam} = 2.90$ GeV/$c$.
\end{itemize}
A beam momentum of 1.64 GeV/$c$ is also used for the study of the $\overline{p}p \rightarrow \overline{\Lambda}\Lambda$ which will serve as a point of reference. At the lower beam momentum of 1.52 GeV/$c$, the production of $\Sigma$ is strongly suppressed, hence reducing experimental ambiguities.

\begin{figure}[tb]
\includegraphics[width=0.49\textwidth]{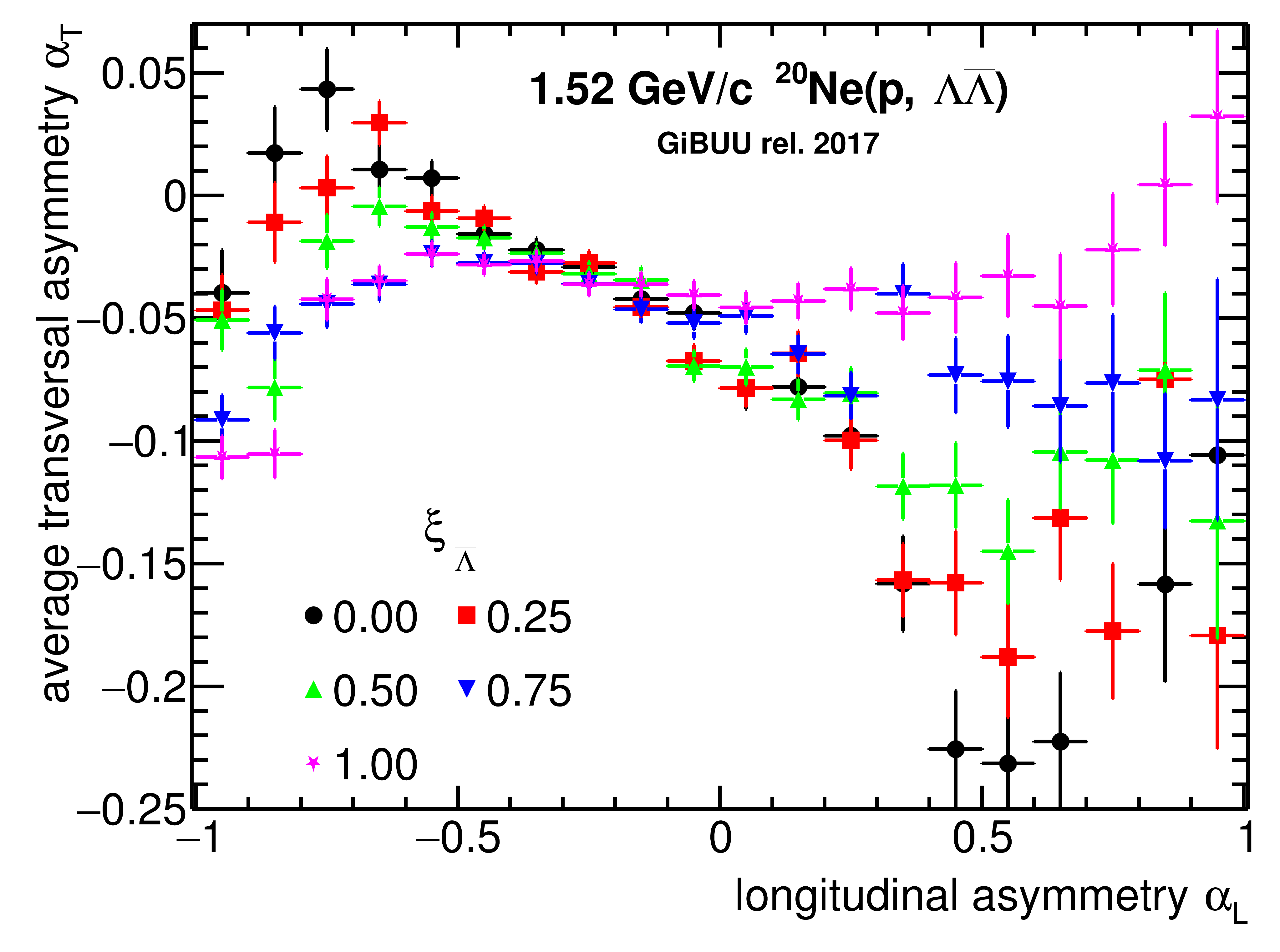}
\includegraphics[width=0.49\textwidth]{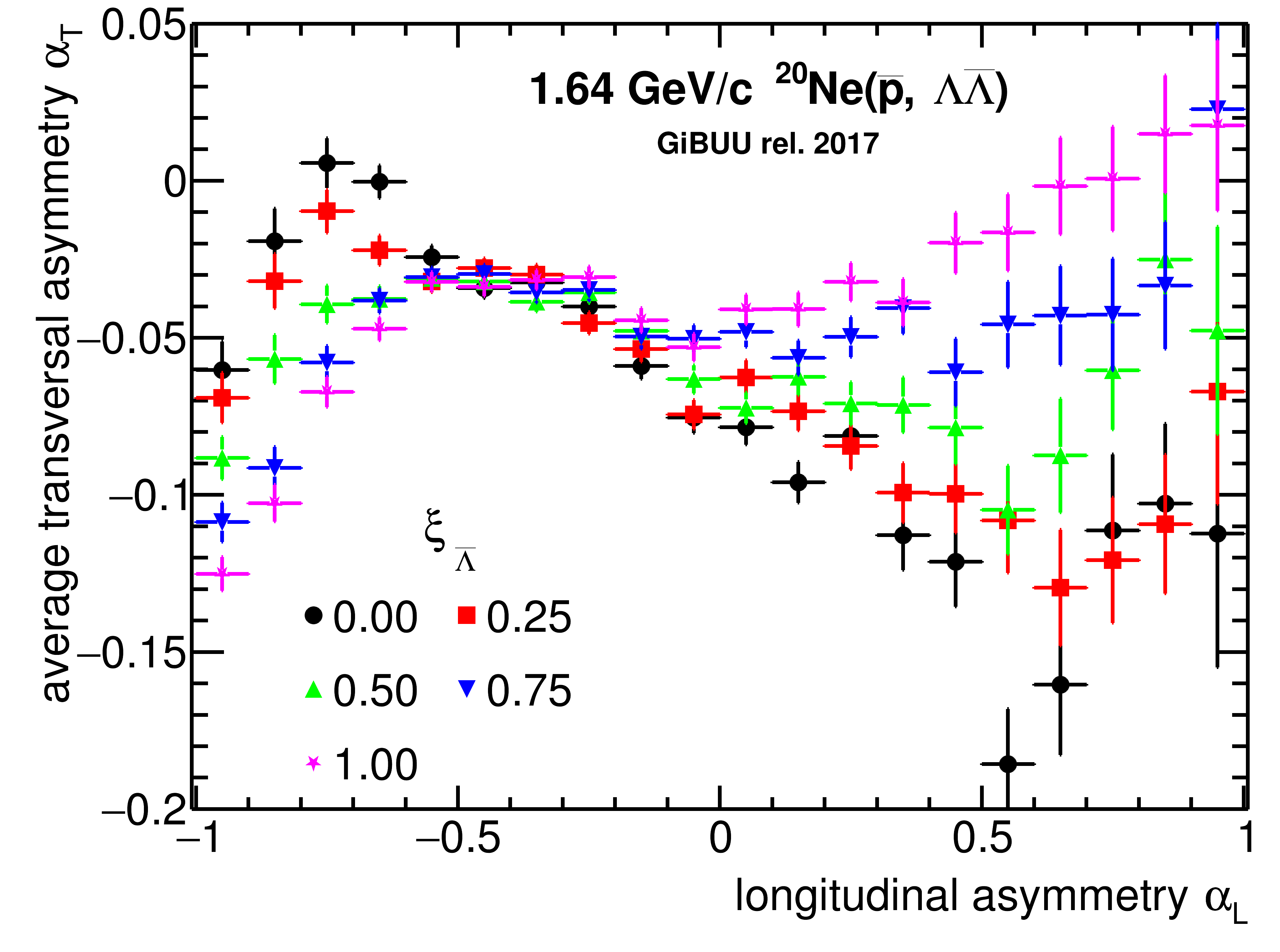}
\caption{Average transverse momentum asymmetry  $\alpha_{T}$ (Eq. \ref{eq:hyp01}) as a function of the longitudinal momentum asymmetry for $\Lambda\overline{\Lambda}$-pairs produced
exclusively in 1.52\,GeV/$c$ (upper panel) and 1.64\,GeV/$c$ (lower panel)
$\bar{p}$+$^{20}$Ne interactions. The different symbols show the GiBUU predictions for different scaling factors $\xi_{\overline{\Lambda}}$ of the $\overline{\Lambda}$-potential. The vertical error bars correspond to the statistical precision that can be obtained for 12 hours of data taking with the conditions specified in the text.}
\label{fig:hyp01}
\end{figure}

\begin{figure}[tb]
\includegraphics[width=0.49\textwidth]{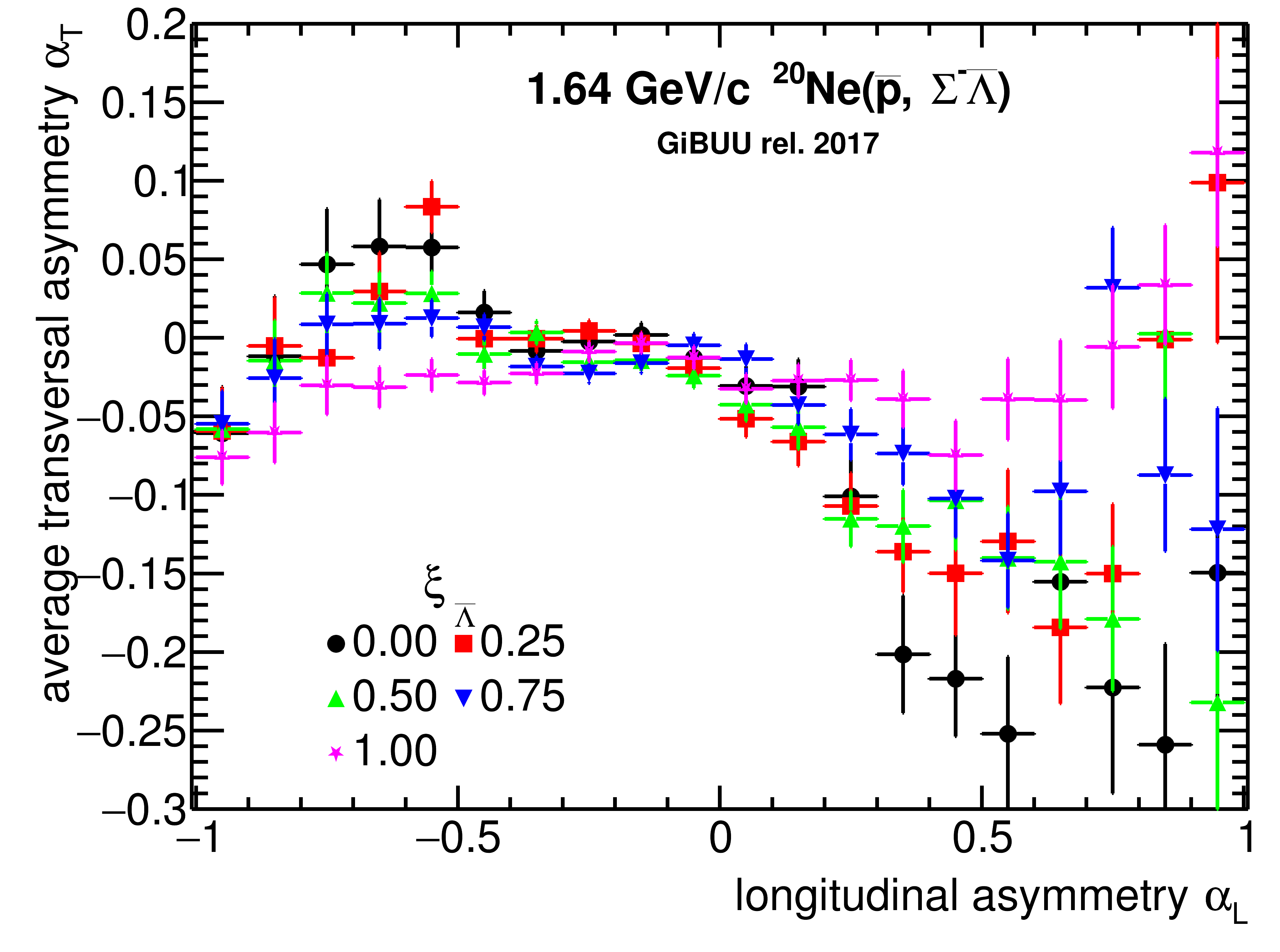}
\includegraphics[width=0.49\textwidth]{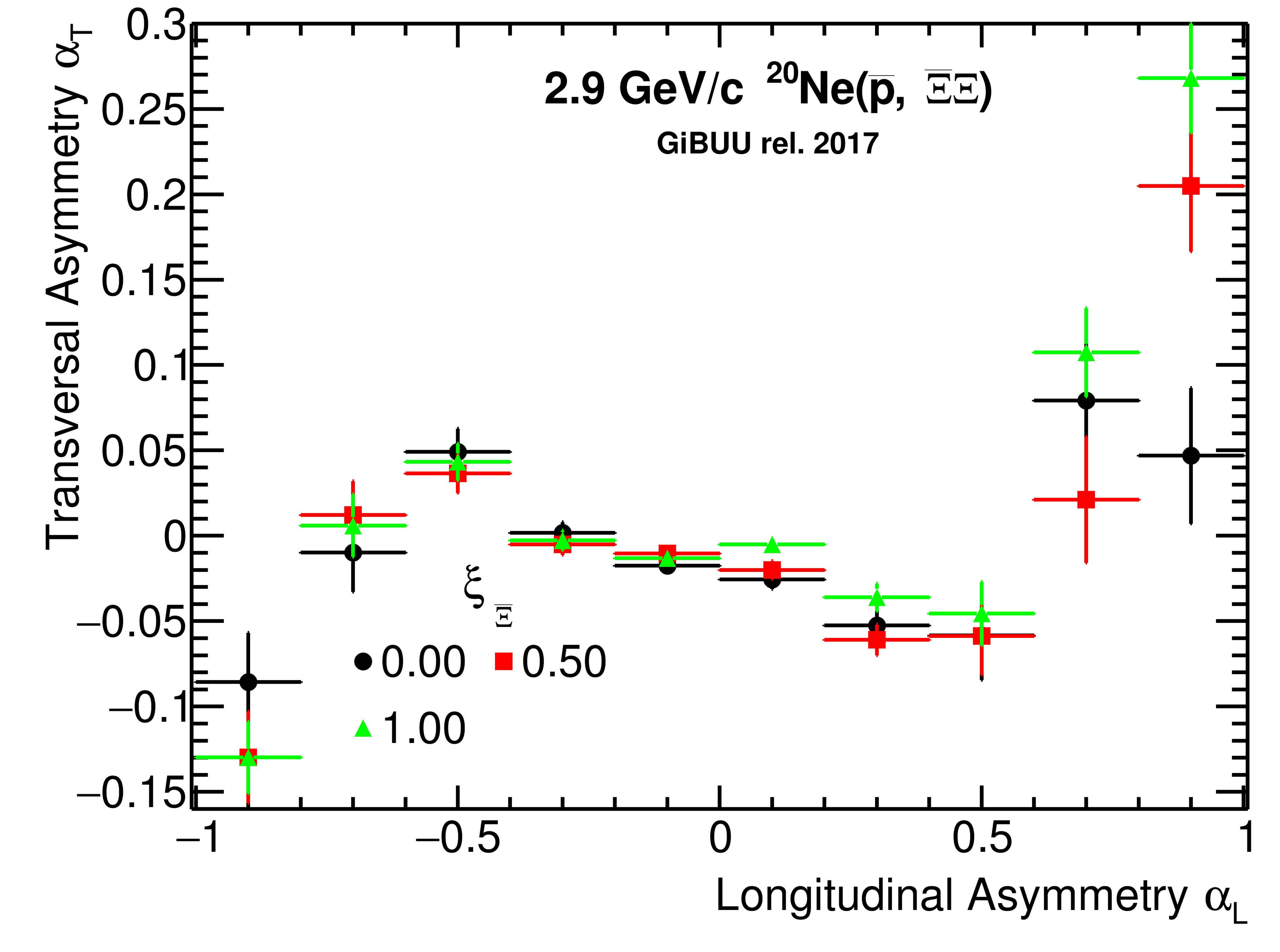}
\caption{Average transverse momentum asymmetry as a function of the
longitudinal momentum asymmetry for $\Sigma^-\overline{\Lambda}$ pairs (upper panel) and $\Xi^-\overline{\Xi}^+$ pairs (lower panel) produced
exclusively in 1.64\,GeV/$c$ $\overline{p}$+$^{20}$Ne and 2.90\,GeV/$c$ $\overline{p}$+$^{20}$Ne interactions, respectively \cite{singhhyp}. The different symbols show the GiBUU predictions for different scaling factors for the antihyperon potentials. The vertical error bars correspond to the statistical precision that can be obtained for two months of data taking with the conditions specified in the text.}
\label{fig:hyp02}
\end{figure}

\noindent The resulting distributions of transverse asymmetry $\alpha_{T}$ as a function of the longitudinal asymmetry $\alpha_{L}$, defined in the same way but with $T \to L$, are shown in Figs.~\ref{fig:hyp01} ($\bar{\Lambda}\Lambda$) and \ref{fig:hyp02} ($\bar{\Lambda}\Sigma^-$ and $\bar{\Xi}^+\Xi^-$). For $\bar{\Lambda}\Lambda$, we observe a remarkable sensitivity of $\alpha_{T}$ to the potential at negative values of $\alpha_{L}$ (Fig.~\ref{fig:hyp01}), and it is clear that secondary effects do not wipe out the dependence. The large $\alpha_{T}$ sensitivity as well as
the negative shift in $\alpha_{T}$ are linked to the substantial $\overline{\Lambda}$ transverse momentum smearing due to secondary scattering.

In order to estimate the expected event rate we assume an interaction rate of 10$^{6}$\,s$^{-1}$, 20\% beam loss in the HESR due to the complex target and
a reconstruction efficiency of 10\%, which is slightly smaller than that of the elementary $\bar{p}p \to \bar{\Lambda}\Lambda$ presented in Tab.~\ref{tab:PEMFF}.
With these assumptions we can obtain 2 (1) $\bar{\Lambda}\Lambda$ per second for $p_{beam}$ = $1.64 (1.52)$ GeV/$c$. One day of data taking with 90\% effective run time at 1.64 GeV/$c$ will yield 15$\cdot$10$^4$ events, corresponding to a sample size two times as large as the one presented in Fig.~\ref{fig:hyp01}. One week of data taking would also enable measurements of polarisation and coplanarity.

For the results presented in the right panel of Fig.~\ref{fig:hyp02}, about 12000 $\Xi^-\overline{\Xi}^+$ pairs were generated for each value of the scaling factor $\xi_{\overline{\Xi}^+}$. With the Phase One luminosity and a $\Xi^-\overline{\Xi}^+$ reconstruction efficiency of 5\% (slightly smaller than that of the elementary $\bar{p}p \to \bar{\Xi}^+\Xi^-$ presented in Tab.~\ref{tab:hyperons}), this requires a running time of about two months.

The studies proposed here will benefit from measurements of the reference reaction $\overline{p}p \rightarrow Y\overline{Y}$. However, as discussed in Section \ref{sec:hyperprod}, such measurements already constitute an important part of the hyperon production programme and can, thanks to the predicted large production rate, be completed in a very short time. The results from our calculations illustrate that even with rather conservative assumptions about luminosity, PANDA can provide unique and relevant information on the behaviour of antihyperons in nuclei already during Phase One.




\subsubsection{Impact and long-term perspectives}

Already in Section~\ref{sec:hyperprod}, it was concluded that PANDA will be a strangeness factory. In combination with nuclear targets, this opens up unique possibilities for pioneering studies of the nuclear antihyperon potentials already during Phase One. In the future, when the luminosity is increased, a unique programme for double- and possibly triple strange hyperatom- and hypernuclear studies will follow~\cite{singhhyp}.


\subsection{Meson-nucleus reactions}

\subsubsection{State of the art}

Colour Transparency (CT) has mainly been studied in the high-energy regime, \textit{e.g.} at Fermilab and HERA \cite{strikman}. At intermediate energies, some evidence was found by the CLAS collaboration for an onset of CT in exclusive meson production with electron probes at momentum transfers of a few GeV \cite{clasctpi, clasctrho}. 

Two-body hadron-nucleus reactions are also sensitive to short-range nucleon--nucleon correlations \cite{farrar1988mf}. These have been studied experimentally for example in two-nucleon knockout reactions with proton beams at BNL \cite{tang,piasetzky} and with electron beams at JLab \cite{subedi,shneor}. It was found that inside ground-state nuclei, the short-range nucleon-nucleon interaction can give rise to correlated nucleon pairs with large relative momenta but small centre-of-mass momenta, called short-range correlated (SRC) pairs.

\subsubsection{Potential for Phase One}

Despite describing different physics phenomena, CT and SRC can be studied with similar probes and momentum regimes and with similar methods. Reactions with antiproton probes have the advantage that they give access to mesons that are unlikely to be produced with electron beams, for example kaons. 

To establish the onset of CT in the intermediate energy regime indicated by CLAS, studies of \textit{e.g.} exclusive meson production in $\bar{p}p$ and $\bar{p}A$ have been suggested \cite{brodsky,farrar}. At large momentum transfer, a $q\bar{q}$ pair is more probable to be in a small-size configuration than a $qqq$ triplet due to combinatorics. Therefore, two-meson annihilation channels provide a very promising search-ground for such studies. It is noteworthy that the main feature of the nuclear target, \textit{i.e.} the possibility of initial- and final state interactions with spectator nucleons, can be explored already for the deuteron. The wave function of the deuteron is relatively well-known which allows for more robust theoretical predictions. The simplest opportunity to study CT is the $d(\bar p, \pi^- \pi^0)p$ process at large momentum transfer in the elementary $\bar p n \to \pi^- \pi^0$ reaction \cite{Larionov:2019xdn}. 
The ``golden'' channel for nuclear transparency is considered to be $A(\bar p, J/\psi)(A-1)^*$. During Phase One, it will be difficult to study charmonium production for heavy nuclei due to the limited luminosity, but studies of the integrated cross section with a deuteron target may be started. Calculations of exclusive charmonium production $d(\bar p, J/\psi)n$ \cite{Larionov:2019mwa} predict a quite large cross section of $\sim 5$ nb at the quasi-free peak
($p_{\rm lab}=4.07$ GeV/$c$).

The same two-body antiproton reactions can be used to study the decay of a short-range correlation after removal of one nucleon, for example $\bar p + A \to h_1+h_2 + \, N_{back} \, +(A-2)^*$, where $N_{back}$ refers to a backward-going nucleon \cite{piasetzky}. In these reactions, it is possible to test the validity of factorisation of the cross section into the elementary cross section, the decay function, and the absorption factor using different final states. Such tests in combination with analogous studies at JLab would contribute to detailed understanding of the dynamics of interactions with short-range correlations and high density fluctuations in nuclei.

In SRC studies, antiprotons give access to correlated $pp$ and $pn$ pairs without the necessity of identifying and determining the momentum of an outgoing neutron. Instead, a struck neutron can be identified by reconstructing processes like $\bar{p}n \to \pi^-\pi^0$ or $\bar{p}n \to \pi^+\pi^-\pi^-$ in the PANDA detector. The wave function of the SRC pair may include the contribution of non-hadronic degrees of freedom. The simplest case is again provided by the deuteron wave function which may include the $\Delta-\Delta$ component predicted by meson-exchange model calculations \cite{Haidenbauer:1993pw} as well as in the quark model \cite{Glozman:1993pm}. The presence of the $\Delta^{++}-\Delta^-$ configuration in the deuteron may be tested in the exclusive reaction $\bar p d \to \pi^- \pi^- \Delta^{++}$ \cite{Larionov:2018lpk}. In the PANDA momentum range, the signal process due to the antiproton annihilation on the valence $\Delta^-$ dominates over two-step background processes. This is valid in a broad kinematic range of the produced $\Delta^{++}$ also for $\Delta-\Delta$ probabilities in the deuteron as low as $\sim 0.3\%$.

\subsubsection{Impact and long-term perspective}

At large beam momenta, PANDA can contribute with studies of colour transparency and short-range correlated nucleon-nucleon pairs, and offers access to final states which are difficult or unfeasible to study with electron or proton beams. 

The larger luminosities of the later stages of PANDA will allow for more extensive studies of charmonium production $A(\bar p, J/\psi)(A-1)^*$ reactions, both for deuterium target and beyond. Exclusive studies of differential cross sections and $J/\psi$ and $\psi^\prime(2S)$ transparency ratios shed further light on colour transparency, as discussed in detail in Refs. \cite{farrar,Larionov:2013axa}. 

The $J/\psi N$ absorption cross section is of particular interest for studies of Quark-Gluon Plasma in heavy-ion collisions \cite{Matsui:1986dk}. 


\section{Summary and conclusions}

The Standard Model of particle physics gives an accurate description of phenomena that occur at very high energies exploiting the
basic interactions among quarks and gluons.
However, describing why and how these quarks and gluons form hadrons remains an open question. The most prominent examples are the building blocks of visible matter, \textit{i.e.} protons and the neutrons. Furthermore, it is a challenge to describe quantitatively how the effective forces between these composite objects emerge from first principles: how do protons and neutrons form atomic nuclei, and how do these form the macroscopic objects of our universe, for example neutron stars?

A central theme in strong interaction phenomena is the non-Abelian nature of QCD, \textit{i.e.} the self-coupling of the force carriers. Self-coupling is present in all non-Abelian theories such as gravity, but hadrons are so far the only objects for which these effects can be studied in a controlled way in the laboratory. 

The PANDA experiment will provide an advanced and multi-faceted facility for studies of different aspects of the strong interaction. PANDA will utilise a beam of antiprotons: a unique and highly versatile probe for hadron physics. The beam energy provided by the HESR storage ring is optimised to shed light on the very regime where quarks form hadrons. Combined with a near 4$\pi$ multipurpose detector, PANDA will offer the broadest hadron physics programme of any existing or planned experiment in the world. 
 
The PANDA physics programme will benefit from the recent theoretical developments (lattice QCD, effective field theory, Dyson-Schwinger approach, AdS/CFT, etc.), but also provide guidance from data to the construction of new theoretical and phenomenological tools, as well as refinements of the existing ones. The close collaboration between theory and experiment will hence be mutually beneficial and has potential to give new insights in the dynamics of non-linear interacting systems on a quantum scale.

In this paper, we have discussed the potential of PANDA during the first phase of data collection, Phase One, when the luminosity will be $\approx$20 times lower than the FAIR design value and the experimental setup will be slightly reduced. The four main physics domains of PANDA -- nucleon structure, strangeness physics, charm and exotics, and hadrons in nuclei -- have been discussed within the context of Phase One. Highlights have been outlined and the potential for PANDA to push the frontiers beyond state of the art was demonstrated for selected examples. PANDA will be a key experiment that can investigate the nucleon structure, perform line-shape measurements
of non-vector charmonium-like states, study multistrange hyperons at a large scale and antihyperons in nuclei. It offers complementary approaches to topics like
time-like form factors, light hadron spectroscopy and colour transparency. In later phases of the PANDA experiment, the full setup and the design luminosity enable an even wider programme that also includes open-charm production, triple-strange hyperon physics, hyperatom and hypernuclear physics and searches for physics beyond the Standard Model \textit{e.g.} through hyperon decays.

\section*{Acknowledgement}

We acknowledge the support of the Theory Advisory Group (ThAG) of PANDA and we value the various discussions that took place with the ThAG sharpening the physics programme of PANDA. We appreciate the comments and feedback we received from Gunnar Bali, Nora Brambilla, Stan Brodsky, Alexei Larionov, Horst Lenske, Stefan Leupold, Ulf Mei\ss{}ner, Simone Pacetti, Mark Strikman, and Lech Szymanowski.

We acknowledge financial support from
the Bhabha Atomic Research Centre (BARC) and the Indian Institute of Technology Bombay, India;
the Bundesministerium f\"ur Bildung und Forschung (BMBF), Germany;
the Carl-Zeiss-Stiftung 21-0563-2.8/122/1 and 21-0563-2.8/131/1, Mainz, Germany;
the CNRS/IN2P3 and the Universit\'{e} Paris-Sud, France;
the Czech Ministry (MEYS) grants LM2015049, CZ.02.1.01/0.0/0.0/16 and 013/0001677,
the Deutsche Forschungsgemeinschaft (DFG), Germany;
the Deutscher Akademischer Austauschdienst (DAAD), Germany;
the European Union's Horizon 2020 research and innovation programme under grant agreement No 824093.
the Forschungszentrum J\"ulich, Germany;
the Gesellschaft f\"ur Schwerionenforschung GmbH (GSI), Darmstadt, Germany;
the Helmholtz-Gemeinschaft Deutscher Forschungszentren (HGF), Germany;
the INTAS, European Commission funding;
the Institute of High Energy Physics (IHEP) and the Chinese Academy of Sciences, Beijing, China;
the Istituto Nazionale di Fisica Nucleare (INFN), Italy;
the Ministerio de Educacion y Ciencia (MEC) under grant FPA2006-12120-C03-02;
the Polish Ministry of Science and Higher Education (MNiSW) grant No. 2593/7, PR UE/2012/2, and the National Science Centre (NCN) DEC-2013/09/N/ST2/02180, Poland;
the State Atomic Energy Corporation Rosatom, National Research Center Kurchatov Institute, Russia;
the Schweizerischer Nationalfonds zur F\"orderung der Wissenschaftlichen Forschung (SNF), Swiss;
the Science and Technology Facilities Council (STFC), British funding agency, Great Britain;
the Scientific and Technological Research Council of Turkey (TUBITAK) under the Grant No. 119F094
the Stefan Meyer Institut f\"ur Subatomare Physik and the \"Osterreichische Akademie der Wissenschaften, Wien, Austria;
the Swedish Research Council and the Knut and Alice Wallenberg Foundation, Sweden.


\clearpage
\vspace{2cm}
\begin{thebibliography}{99}
\bibitem{fritsch} H.~Fritzsch, M.~Gell-Mann and H.~Leutwyler, Phys. Lett. B {\bf 47}, 365 (1973).

\bibitem{wilczek} H.~D. Politzer, Phys. Rev. Lett. {\bf 30}, 1346 (1973); D.~J. Gross and F.~Wilczek, Phys. Rev. Lett. {\bf 30}, 1343 (1973); S. Weinberg, Phys. Rev. Lett. {\bf 31}, 494 (1973).


\bibitem{PDG} P.A.~Zyla \textit{et al.} (Particle Data Group), Prog. Theor. Exp. Phys. \textbf{2020}, 083C01 (2020).


\bibitem{pspin1} C.~A.~Aidala \textit{et al.,} Rev. Mod. Phys. \textbf{85}, 655 (2013).

\bibitem{pspin2} C.~Alexandrou \textit{et al.,} Phys. Rev. Lett. \textbf{119}, 142002 (2017).
\bibitem{pmass} Y-B.~Yang \textit{et al.,} Phys. Rev. Lett. \textbf{121}, 212001 (2018).

\bibitem{glueballs1} H.~Fritzsch and M.~Gell-Mann, \textit{Current algebra: Quarks and what else?} Proc. of the XVI Int. Conf. on High Energy Physics, Chicago, 1072, Vol. 2, p. 135, eConf C 720906V2 135 (1972); H.~Fritzsch and P.~Minkowski, \textit{Psi Resonances, Gluons and the Zweig Rule}, Nuovo Cim. A \textbf{30}, 393 (1975).

\bibitem{hybrids1} T.~Barnes, PhD thesis, Caltech (1977); R.L.~Jae and K.~Johnson, Phys. Lett. {\bf 60}, 201 (1976); R.~Giles and S.-H H.~Tye, Phys. Rev. Lett. {\bf 37}, 1175 (1976); D.~Horn and J.~Mandula, Phys. Rev. D {\bf 17}, 898 (1978).

\bibitem{ccbarrev1} A.~Esposito, A.~Pilloni and A.~D.~Polosa, Phys.\ Rept.\  {\bf 668}, 1 (2016).

\bibitem{rev_XYZ} R.~F.~Lebed, R.~E.~Mitchell and E.~S.~Swanson, Prog. Part. Nucl. Phys. \textbf{93}, 143 (2017).

\bibitem{Brambilla:2019esw}
N.~Brambilla, S.~Eidelman, C.~Hanhart, A.~Nefediev, C.~P.~Shen, C.~E.~Thomas, A.~Vairo and C.~Z.~Yuan,
Phys. Rept. \textbf{873}, 1 (2020).


\bibitem{brodsky1980} S.~J.~Brodsky, P.~Hoyer, C.~Peterson, and N.~Sakai, Phys. Lett. \textbf{93B}, 451 (1980).
\bibitem{harris1996} B.~W.~Harris, J.~Smith, and R.~Vogt, Nucl. Phys. B \textbf{461}, 181 (1996).
\bibitem{franz2000} M.~Franz, M.~V.~Polyakov, and K.~Goeke, Phys. Rev. D \textbf{62}, 074024 (2000).

\bibitem{ccbarrev3} F.~K.~Guo, C.~Hanhart, U.~G.~Meißner, Q.~Wang, Q.~Zhao and B.~S.~Zou, Rev.\ Mod.\ Phys.\  {\bf 90},  015004 (2018).

\bibitem{colortrans} P.~Jain, B.~Pire and J.P.~Ralston, Physics Reports {\bf 271}, 67 (1996), and references therein.

\bibitem{bielich:2008} J.~Schaffner-Bielich, Nucl. Phys. A \textbf{804}, 309 (2008).

\bibitem{lutz:2015} Matthias F.~M.~Lutz {\it et al.}, Nucl. Phys A \textbf{948}, 93 (2016).

\bibitem{Sakharov:1967dj}A.~D.~Sakharov, Pisma Zh. Eksp. Teor. Phys. Fiz. \textbf{5} 32 (1967).

\bibitem{Amsler19} C.~Amsler, arXiv:1908.08455 (2019).

\bibitem{Durante19} M.~Durante {\it et al.}, Physica Scripta \textbf{94}, 033001 (2019).

\bibitem{hesrtdr} R.~Maier {\it et al.}, HESR Technical Design Report V. 3.1.2 (2008).

\bibitem{Agakishievetal:2009} G.~Agakishiev {\it et al.} (HADES collaboration), Eur. Phys. J. A \textbf{41}, 243 (2009).

\bibitem{Blomqvist:1998} K.~I.~Blomqvist {\it et al.}, Nucl. Instrum. Methods Phys. Res., Sect. A \textbf{403}, 263 (1998).

\bibitem{Magnet-TDR} W.~Erni {\it et al.} (PANDA Collaboration), \textit{The PANDA Solenoid and Dipole Spectrometer Magnets}, Technical Design Report (2005).

\bibitem{Target-TDR} W.~Erni {\it et al.} (PANDA Collaboration), \textit{The PANDA Internal Targets}, Technical Design Report (2012).

\bibitem{MVD-TDR} W.~Erni {\it et al.} (PANDA Collaboration), \textit{The PANDA Micro Vertex Detector}, Technical Design Report (2012).

\bibitem{STT-TDR} W.~Erni {\it et al.} (PANDA Collaboration), \textit{The PANDA Straw Tube Tracker}, Technical Design Report, Eur. Phys. J. A \textbf{49} 25 (2013).

\bibitem{DIRC-TDR} B.~Singh {\it et al.} (PANDA Collaboration), \textit{The PANDA Barrel DIRC Detector}, Technical Design Report, J. Phys. G: Nucl. Part. Phys. \textbf{46}, 045001 (2019).

\bibitem{TOF-TDR} B.~Singh {\it et al.} (PANDA Collaboration), \textit{The PANDA Barrel TOF}, Technical Design Report (2016).

\bibitem{EMC-TDR} W.~Erni {\it et al.} (PANDA Collaboration), \textit{The PANDA Electromagnetic Calorimeter}, Technical Design Report (2008).

\bibitem{Muon-TDR} W.~Erni {\it et al.} (PANDA Collaboration), \textit{The PANDA Muon System}, Technical Design Report (2012).

\bibitem{FTS-TDR} B.~Singh {\it et al.} (PANDA Collaboration), \textit{The PANDA Forward Tracker}, Technical Design Report (2018).

\bibitem{FTOF-TDR} B.~Singh {\it et al.} (PANDA Collaboration), \textit{The PANDA Forward Time-of-Flight detector (FToF wall)}, Technical Design Report (2018).

\bibitem{FSC-TDR} B.~Singh {\it et al.} (PANDA Collaboration), \textit{The PANDA Forward Spectrometer Calorimeter}, Technical Design Report (2016).

\bibitem{LMD-TDR} B.~Singh {\it et al.} (PANDA Collaboration), \textit{The PANDA Luminosity Detector}, Technical Design Report (2018).

\bibitem{PANDAROOT} S.~Spataro {\it et al.}, \textit{The PandaRoot framework for simulation,reconstruction and analysis}, J. Phys.: Conf. Ser. \textbf{331}, 032031 (2011).

\bibitem{FAIRROOT} M.~Al-Turany {\it et al.}, \textit{The FairRoot Framework}, J. Phys.: Conf. Ser. \textbf{396}, 022001 (2012).

\bibitem{ROOT} R.~Brun and F.~Rademakers,
\textit{ROOT - An Object Oriented Data Analysis Framework}, Nucl. Inst. \& Meth. in Phys. Res. A \textbf{389}, 81 (1997).

\bibitem{EvtGen} A.~Ryd \textit{et al.}, \textit{EvtGen: A Monte Carlo Generator for B-Physics}, EVTGEN-V00-11-07 (2005).

\bibitem{Capella1994} A.~Capella \textit{et al.,} Physics 1089 Reports, \textbf{236}(4), 225 (1994).

\bibitem{FTF} B.~Andersson \textit{et al.}, Nucl. Phys. B \textbf{281}, 289 (1987); B.~Nilsson-Almquist and E.~Stenlund, Comp. Phys. Commun. \textbf{43}, 387 (1987).


\bibitem{Collins1989} J. C. Collins, D. E. Soper, and G. F. Sterman, Adv. Ser. Direct. High Energy Phys. {\bf 5}, 1 (1989).

\bibitem{Ji:1996nm} X.~D.~Ji, Phys. Rev. D \textbf{55}, 7114 (1997).

\bibitem{Radyushkin:1996nd}A.~V.~Radyushkin, Phys. Lett. B \textbf{380}, 417 (1996).

\bibitem{Mueller:1998fv} D.~M\"uller, D.~Robaschik, B.~Geyer, F.~M.~Dittes and J.~Ho\v{r}ej\v{s}i, Fortsch. Phys. \textbf{42}, 101 (1994).

\bibitem{Collins:1996fb} J.~C.~Collins, L.~Frankfurt and M.~Strikman, Phys. Rev. D \textbf{56}, 2982 (1997).

\bibitem{Berger:2001xd} E.~R.~Berger, M.~Diehl and B.~Pire, Eur. Phys. J. C \textbf{23}, 675 (2002).

\bibitem{Radyushkin:1998rt} A.~V.~Radyushkin, Phys. Rev. D \textbf{58}, 114008 (1998).

\bibitem{Diehl:1998kh} M.~Diehl, T.~Feldmann, R.~Jakob and P.~Kroll, Eur. Phys. J. C \textbf{8}, 409 (1999).

\bibitem{Barone2010} V.~Barone, F.~Bradamante and A.~Martin, Prog. Part. Nucl. Phys. \textbf{65}, 267 (2010).

\bibitem{Frankfurt:1999fp} L.~L.~Frankfurt, P.~V.~Pobylitsa, M.~V.~Polyakov and M.~Strikman, Phys. Rev. D \textbf{60}, 014010 (1999).

\bibitem{Pire:2004ie} B.~Pire and L.~Szymanowski, Phys. Rev. D \textbf{71}, 111501 (2005).

\bibitem{Singh:2014pfv} B.~P.~Singh \textit{et al.} (PANDA Collaboration), Eur. Phys. J. A \textbf{51}, 107 (2015).

\bibitem{Singh:2016qjg} B.~P.~Singh \textit{et al.} (PANDA Collaboration), Phys. Rev. D \textbf{95}, 032003 (2017). 

\bibitem{Freund:2002cq} A.~Freund, A.~V.~Radyushkin, A.~Schafer and C.~Weiss, Phys. Rev. Lett. \textbf{90}, 092001 (2003).

\bibitem{Kroll:2005ni} P.~Kroll and A.~Sch\"afer, Eur. Phys. J. A \textbf{26}, 89 (2005).

\bibitem{Kroll:2013kra} P.~Kroll and A.~Sch\"afer, Eur. Phys. J. A \textbf{50}, 1 (2014).
  

\bibitem{Hofstadter:1956qs} R.~Hofstadter, Rev. Mod. Phys. \textbf{28}, 214 (1956).

\bibitem{rosenbluth} M.~N.~Rosenbluth, Phys. Rev. \textbf{79}, 615 (1950).

\bibitem{Perdrisat:2006hj} C.~F.~Perdrisat, V.~Punjabi and M.~Vanderhaeghen, Prog. Part. Nucl. Phys. \textbf{59}, 694 (2007).

\bibitem{Puckett:2010ac} A.~J.~R.~Puckett \textit{et al.}, Phys. Rev. Lett. \textbf{104}, 242301 (2010); J.~Arrington, P.~G.~Blunden and W.~Melnitchouk, Prog. Part. Nucl. Phys. \textbf{66}, 782 (2011).

\bibitem{Akhiezer:1968ek} A.~I.~Akhiezer and M.~P.~Rekalo, Sov. Phys. Dokl. \textbf{13}, 572 (1968).

\bibitem{guichon:2003} P.~A.~Guichon, M.~Vanderhaeghen, Phys. Rev. Lett. \textbf{91}, 142303 (2003).

\bibitem{vectordom} D.~Schildknecht, Acta Phys. Polon. B \textbf{37}, 595 (2006).

\bibitem{hammer2020} H.~W.~Hammer and U.-G.~Mei\ss{}ner,
       Sci. Bull. \textbf{65}, 257 (2020).

\bibitem{Denig2013}  A.~Denig and G.~Salme, Prog. Part. Nucl. Phys. \textbf{68}, 113 (2013).
\bibitem{Pacetti2015}  S.~Pacetti, R.~Baldini Ferroli and E.~Tomasi-Gustafsson, Phys. Rept. \textbf{550}, 1 (2015).

\bibitem{Bardin:1994am} G.~Bardin \textit{et al.}, Nucl. Phys. B \textbf{411}, 3 (1994).

\bibitem{Lees:2013uta} J.~P.~Lees \textit{et al.} (BaBar Collaboration), Phys. Rev. D \textbf{88}, 072009 (2013).

\bibitem{Ablikim:2015vga} M.~Ablikim \textit{et al.} (BESIII Collaboration), Phys. Rev. D \textbf{91}, 112004 (2015).

\bibitem{Ablikim:2019njl} M.~Ablikim \textit{et al.} (BESIII Collaboration), Phys. Rev. D \textbf{99}, 092002 (2019).
  
\bibitem{Ablikim:2021kjh} M.~Ablikim \textit{et al.} (BESIII Collaboration), arXiv:2102.10337 (2021).

\bibitem{Akhmetshin:2015ifg} R.~R.~Akhmetshin \textit{et al.} (CMD-3 Collaboration), Phys. Lett. B \textbf{759}, 634 (2016).

\bibitem{phragmen} E.~C.~Tichmarsh, \textit{The theory of functions}, Oxford University Press (1939).

\bibitem{Ablikim:2019eau} M.~Ablikim \textit{et al.} (BESIII Collaboration), Phys. Rev. Lett. \textbf{124}, 042001 (2020).

\bibitem{Haidenbauer:2014kja} J.~Haidenbauer, X.~W.~Kang and U.~G.~Mei\ss{}ner, Nucl. Phys. A \textbf{929}, 102 (2014).

\bibitem{Lorentz2015} I. T. Lorentz, H.-W. Hammer, and U.-G. Mei\ss{}ner, Phys. Rev. D \textbf{92}, 034018 (2015).
\bibitem{Bianconi2016} A. Bianconi and E. Tomasi-Gustafsson, Phys. Rev. C \textbf{93}, 035201 (2016).

\bibitem{Singh:2016dtf} B.~Singh \textit{et al.} (PANDA Collaboration), Eur. Phys. J. A \textbf{52}, 325 (2016).

\bibitem{Zimmermann2020} G.~Barucca \textit{et al.} (PANDA Collaboration), Eur. Phys. J. A \textbf{57}, 30 (2021).

\bibitem{Zichichi:1962ni} A.~Zichichi, S.~M.~Berman, N.~Cabibbo and R.~Gatto, Nuovo Cim. \textbf{24}, 170 (1962).

\bibitem{Gakh:2005wa} G.~I.~Gakh and E.~Tomasi-Gustafsson, Nucl. Phys. A \textbf{761}, 120 (2005).

\bibitem{Dbeyssi:2011tv} A.~Dbeyssi, E.~Tomasi-Gustafsson, G.~I.~Gakh and M.~Konchatnyi, Nucl. Phys. A \textbf{894}, 20 (2012).




\bibitem{dubnickova96} A.~Z.~Dubnickova \textit{et al.,} Z. Phys. C \textbf{70} 473 (1996).

\bibitem{Adamuscin:2006bk} C.~Adamuscin, E.~A.~Kuraev, E.~Tomasi-Gustafsson and F.~E.~Maas, Phys. Rev. C \textbf{75}, 045205 (2007).

\bibitem{Guttmann:2012sq} J.~Guttmann and M.~Vanderhaeghen, Phys. Lett. B \textbf{719}, 136 (2013).

\bibitem{Boucher:2011} J. Boucher, PhD thesis, University of Paris-Sud XI and of the Johannes Gutenberg University, 2011.

\bibitem{FFtomasi} E.~Tomasi-Gustafsson and M.~P.~Rekalo, Phys. Lett. B \textbf{504}, 291 (2001).

\bibitem{Lees:2013b}  J.~P.~Lees \textit{et al.} (BaBar Collaboration), Phys. Rev. D \textbf{87}, 092005 (2013).



\bibitem{e835} M. Ambrogiani \textit{et al.}, (E835 Collaboration), Phys. Rev. D \textbf{60}, 032002 (1999); M. Andreotti \textit{et al.}, Phys. Lett. B \textbf{559}, 20 (2003).


\bibitem{fenice} A. Antonelli \textit{et al.}, Nucl. Phys. B \textbf{517}, 3 (1998).

\bibitem{e760} T. A. Armstrong \textit{et al.}, Phys. Rev. Lett. \textbf{70}, 1212 (1993).

\bibitem{dm1} B.~Delcourt \textit{et al.}, Phys. Lett. B \textbf{86}, 395 (1979).

\bibitem{dm2} D.~Bisello  \textit{et al.}, Nucl. Phys. B \textbf{224}, 379 (1983); Z. Phys. C \textbf{48}, 23 (1990).

\bibitem{cleo} T.~K.~Pedlar \textit{et al.}, [CLEO], Phys. Rev. Lett. \textbf{95}, 261803 (2005).

\bibitem{adone73} M. Castellano \textit{et al.}, Nuovo Cimento A \textbf{14}, 1 (1973).


\bibitem{Schafer:1994vr}
M.~Schafer, H.~C.~Donges and U.~Mosel, Phys. Lett. B \textbf{342}, 13 (1995).

\bibitem{Dieperink:1997jh}
A.~E.~L.~Dieperink and S.~I.~Nagorny, Phys. Lett. B \textbf{397}, 20 (1997).


\bibitem{Vidana2018}
I.~Vida\~{n}a,
Proc. R. Soc. A {\bf 474}, 20180145 (2018).


\bibitem{rafelski}J.~Rafelski and B.~Müller, Phys. Rev. Lett \textbf{48}, 1066 (1982).

\bibitem{bes3nature}M.~Ablikim \textit{et al.} (BESIII collaboration), Nature Phys. \textbf{15}, 631 (2019).

\bibitem{alicehyp}J.~Adam \textit{et al.} (ALICE collaboration), Nature Physics \textbf{13}, 535 (2017).

\bibitem{quarkgluon} M.~Kohno and W.~Weise, Phys. Lett. B \textbf{179}, 15 (1986); H.~R.~Rubinstein and H.~Snellman, Phys. Lett. B \textbf{165}, 187 (1985); S.~Furui and A.~Faessler, Nucl. Phys. A \textbf{486}, 669 (1987); M.~Burkardt and M.~Dillig Phys. Rev. C \textbf{37}, 1362 (1988); M.~A.~Alberg \textit{et al.}, Z. Phys. A \textbf{331}, 207 (1988).

\bibitem{kaonexchange} F.~Tabakin and R.~A.~Eisenstein, Phys. Rev. C \textbf{31}, 1857 (1985); M.~Kohno and W.~Weise, Phys. Lett. B. \textbf{179}, 15 (1985); P.~La France \textit{et al.}, Phys. Lett. B \textbf{214}, 317 (1988); R.~G.~E.~Timmermans \textit{et al.}, Phys. Rev. D \textbf{45}, 2288 (1992); J.~Haidenbauer \textit{et al.}, Phys. Rev. C \textbf{46}, 2516 (1992).

\bibitem{quarkgluonhadron} P.~G.~Ortega \textit{et al.}, Phys. Lett. B. \textbf{696} 352 (2011).

\bibitem{XiQG} P.~Kroll and W.~Schweiger, Nucl. Phys. A \textbf{474}, 608 (1987) ;P.~Kroll, B.~Quadder and W.~Schweiger, Nucl. Phys. B \textbf{316}, 373 (1989); H.~Genz, M.~Nowakowski and D.~Woitschitzsky, Phys. Lett. B \textbf{260}, 179 (1991).

\bibitem{XiMEX} J.~Haidenbauer, K.~Holinde and J.~Speth, Phys. Rev. C \textbf{47}, 2982 (1993).

\bibitem{kaidalov} A.~B.~Kaidalov and P.~E.~Z.~Volkovitsky, Phys. C \textbf{63} 51 (1994).

\bibitem{HaidenbauerHyp}J.~Haidenbauer and U.-G.~Mei\ss{}ner, Phys. Lett. B \textbf{761}, 456 (2016).

\bibitem{ps185hyp} P.~D.~Barnes \textit{et al.},
Nucl. Phys. A \textbf{526}, 575 (1991); P.~D.~Barnes \textit{et al.},
Phys. Rev. C \textbf{62}, 055203 (2000); E.~Klempt,~F.~Bradamante, A.~Martin, J.-M.~Richard, Phys. Rep. \textbf{368}, 119 (2002); K.~D.~Paschke \textit{et al.},
Phys. Rev. C \textbf{74}, 015206 (2006).

\bibitem{bes3hyp} M.~Ablikim \textit{et al.,} Phys. Rev. Lett. \textbf{123}, 122003 (2019).

\bibitem{PaschkeQuinn}K.~Paschke and B.~Quinn, Phys. Lett. B \textbf{495}, 49 (2000).

\bibitem{koch} W.~Koch \textit{Analysis of scattering and decay} ed. Nikolic M (New York-London-Paris:Gordon and Breach, 1968).

\bibitem{perotti}E.~Perotti \textit{et al.,} \textit{Extraction of Polarization Parameters in the $\mathrm{\bar{p} p} \rightarrow \bar{\Omega} \Omega$ Reaction}, J. Phys. Conf. Ser. \textbf{1024}, 012019 (2018). 

\bibitem{erikthesis} E.~Thom\'e, \textit{Multi-Strange and Charmed Antihyperon-Hyperon Physics for PANDA},
Ph. D. Thesis, Uppsala University, 2012.

\bibitem{LEAR} T.~Johansson 2003, Proceedings of \textit{8th Int. Conf. on Low Energy Antiproton Physics}, 95 (2003).

\bibitem{Musgrave1965}B.~Musgrave and G.~Petmezas, Nuovo Cim. \textbf{35}, 735 (1965);C.~Baltay \textit{et al.,} Phys. Rev. B \textbf{140}, 1027 (1965).

\bibitem{pandaphysics} The PANDA collaboration, Physics Performance Report (2009).

\bibitem{grapethesis} S.~Grape, \textit{Studies of PWO Crystals and Simulations of the $\overline{p}p \rightarrow \overline{\Lambda}\Lambda,\overline{\Lambda}\Sigma^0$ Reactions for the PANDA experiment} Ph.D. Thesis, Uppsala University, 2009.

\bibitem{walter} W.~Ikegami~Andersson, \textit{Exploring the Merits and Challenges of Hyperon Physics with PANDA at FAIR}, PhD Thesis, Uppsala University (2020).

\bibitem{pandahyperons} G.~Barucca \textit{et al.} (PANDA collaboration), \textit{The potential of hyperon-antihyperon studies with PANDA at FAIR}, in print Eur. Phys. J A, arXiv[hep-ex]:2009.11582 (2020). 

\bibitem{gabriela}G.~Perez~Andrade, \textit{Production of the $\Sigma^0$ hyperons in the PANDA experiment at FAIR}, Master Thesis, Uppsala University (2019).

\bibitem{sigma6}H.~Becker \textit{et al.,} Nucl. Phys. B \textbf{141}, 48 (1978).


\bibitem{quarkmodel} M.~Gell-Mann, Phys. Lett. \textbf{8}, 214 (1964); G.~Zweig, CERN Report 8183/TH.401 (1964);~\textit{ibid.} 8419/TH.412 (1964).
\bibitem{alexandrou} C.~Alexandrou \textit{et al.,} Phys. Rev. D \textbf{90}, 074501 (2014).
\bibitem{lat1} R.~A.~Briceno, H.-W.~Lin, and D.~R.~Bolton, Phys. Rev. D \textbf{86}, 094504 (2012).
\bibitem{lat2} H.~Na and S.~Gottlieb, PoS LATTICE2008, 119 (2008); \textit{ibid.}, PoSLAT2007, 124 (2007).
\bibitem{lat3} L.~Liu, H.-W.~Lin, K.~Orginos, and A.~Walker-Loud, Phys. Rev. D \textbf{81}, 094505 (2010).
\bibitem{lat4} Y.~Namekawa \textit{et al.} (PACS-CS Collaboration), Phys. Rev. D \textbf{87}, 094512 (2013).
\bibitem{doublecharm} R.~Aaij \textit{et al.} (LHCb Collaboration), Phys. Rev. Lett. \textbf{119}, 112001 (2017).

\bibitem{borsanyi2015} Sz. Borsanyi \textit{et al.}, Science \textbf{347}, 1452 (2015).
\bibitem{leinweber2015} D. Leinweber \textit{et al.}, arXiv:1511.09146 (2015).
\bibitem{sun2020} M.~Sun \textit{et al.} ($\chi$QCD Collaboration), Phys. Rev. D \textbf{101}, 054511 (2020). 

\bibitem{Eichmann:2016yit} G.~Eichmann, H.~Sanchis-Alepuz, R.~Williams, R.~Alkofer and C.~S.~Fischer, Prog.\ Part.\ Nucl.\ Phys.\  {\bf 91}, 1 (2016).

\bibitem{Kolomeitsev:2003kt} E.~E.~Kolomeitsev and M.~F.~M.~Lutz, Phys.\ Lett.\ B {\bf 585}, 243 (2004).
\bibitem{Hyodo:2020czb} T.~Hyodo and M.~Niiyama, arXiv:2010.07592 [hep-ph].
\bibitem{Mai:2020ltx} M.~Mai, arXiv:2010.00056 [nucl-th].
\bibitem{Meissner:2020khl} U.-G.~Mei\ss{}ner, Symmetry \textbf{12}, 981 (2020).

\bibitem{Brodsky:2015} S.~J.~Brodsky, G.~F.~de~Teramond, H.~G.~Dosch and J.~Erlich, Phys. Rep.~\textbf{584}, 1 (2015).
\bibitem{Deur:2015} A.~Deur, S.~J.~Brodsky, G.~F.~de~Teramond, Phys. Lett. B \textbf{750}, 528 (2015).
\bibitem{Brodsky:2016} S.~J.~Brodsky, G.~F.~de~T{\'e}ramond, H.~G.~Dosch, C.~Lorc{\'e}, Phys. Lett. B \textbf{759}, 171 (2016).
  

\bibitem{Klempt} E.~Klempt and J.~M.~Richard, Rep. Prog. Phys. {\bf 82} 1095 (2010).
\bibitem{Crede} V.~Crede and W.~Roberts, Rep. Prog. Phys. {\bf 76} 076301 (2013).
\bibitem{dslight}G.~Eichmann, C.~S.~Fischer and H.~Sanchis-Alepuz, Phys. Rev. D \textbf{94}, 094033 (2016).

\bibitem{dsstrange}C.~S.~Fischer and G.~Eichmann, \textit{Overview of multiquark states}, PoS \textbf{Hadron2017} (2018). 

\bibitem{claslambda1405}K.~Moriya \textit{et al.} (CLAS collaboration), Phys. Rev. Lett. \textbf{112}, 082004 (2014).
\bibitem{dalitz:1959} R.~H.~Dalitz and S.~F.~Tuan, Phys. Rev. Lett. \textbf{2}, 425 (1959); Annals Phys. \textbf{10}, 307 (1960).
\bibitem{adelaide}J.~M.~M.~Hall \textit{et al.,} Phys. Rev. Lett. \textbf{114}, 132002 (2015).
\bibitem{babaromega} B.~Aubert \textit{et al.} (BaBar Collaboration), Phys. Rev. Lett. \textbf{97}, 112001 (2006).
\bibitem{Lange2001}D.~J.~Lange, Nucl. Instrum. Meth. A \textbf{462}, 152 (2001).
\bibitem{puetz:2021} G.~Barucca \textit{et al.} (PANDA collaboration), \textit{Study of Excited $\Xi$ Baryons with the PANDA Detector}, in print Eur. Phys. J A, arXiv:2012.01776 (2021). 
\bibitem{Flamino1984}V.~Flaminio \textit{et al.,}, Report CERN–HERA–84–01 (1984).



\bibitem{Aaij2020}
R.~Aaij \textit{et al.} (LHCb Collaboration), arXiv:2009.00026 (2020). 

\bibitem{brambilla:2005} N.~Brambilla, A.~Pineda, J.~Soto, and A.~Vairo, Rev. Mod. Phys. \textbf{77}, 1423 (2005).
\bibitem{brambilla:2011} N.~Brambilla \textit{et al.}, Eur. Phys. J. \textbf{71}, 1534 (2011).
\bibitem{brambilla:2014} N.~Brambilla \textit{et al.}, Eur. Phys. J. \textbf{74}, 2981 (2014).

\bibitem{sufian2020} R.~S.~Sufian, T.~Liu, A.~Alexandru, S.~J.~Brodsky, G.~F.~de~T{\'e}ramond, Phys. Lett. \textbf{808}, 135633 (2020).

\bibitem{aubert:2003} B.~Aubert \textit{et al.} (BaBar Collaboration), Phys. Rev. Lett. \textbf{90}, 242001 (2003).

\bibitem{besson:2003} D.~Besson \textit{et al.} (CLEO Collaboration), Phys. Rev. D \textbf{68}, 032002 (2003).


\bibitem{GlueballLattice}
C. Morningstar and M. J. Peardon, AIP Conf. Proc. {\bf 688}, 220 (2004); C. J. Morningstar and M. J. Peardon, Phys. Rev. D {\bf 60}, 034509 (1999); 
Y. Chen {\it et al.}, Phys. Rev. D {\bf 73}, 014516 (2006); E. Gregory {\it et al.}, arXiv:1208.1858 (2021).

\bibitem{GlueballReview}
F.~E.~Close, Rept. Prog. Phys. \textbf{51}, 833 (1988); S.~Godfrey and J.~Napolitano, Rev. Mod. Phys. \textbf{71}, 1411 (1999);
C.~Amsler and N.~A.~Tornqvist, Phys. Rept. \textbf{389}, 61 (2004);
E.~Klempt and A.~Zaitsev, Phys. Rept. \textbf{454}, 1 (2007); V.~Crede and C.~Meyer, Prog. Part. Nucl. Phys. \textbf{63}, 74 (2009).


\bibitem{Bertin:1997kh} 
  A.~Bertin {\it et al.} (OBELIX Collaboration),
  Phys.\ Lett.\ B {\bf 408}, 476 (1997).

\bibitem{Bertin:1998hu} 
  A.~Bertin {\it et al.} (OBELIX Collaboration),
  Phys.\ Rev.\ D {\bf 57}, 55 (1998).

\bibitem{Nichitiu:2002cj} 
  F.~Nichitiu {\it et al.} (OBELIX Collaboration),
  Phys.\ Lett.\ B {\bf 545}, 261 (2002).

\bibitem{Bargiotti:2003ev} 
  M.~Bargiotti {\it et al.} (OBELIX Collaboration),
  Eur.\ Phys.\ J.\ C {\bf 26}, 371 (2003).

\bibitem{Amsler:1992rx} 
  C.~Amsler {\it et al.} (Crystal Barrel Collaboration),
  Phys.\ Lett.\ B {\bf 291}, 347 (1992).

\bibitem{Amsler:1994pz} 
  C.~Amsler {\it et al.} (Crystal Barrel Collaboration),
  Phys.\ Lett.\ B {\bf 333}, 277 (1994).

\bibitem{Amsler:1994rv} 
  C.~Amsler {\it et al.} (Crystal Barrel Collaboration),
  Phys.\ Lett.\ B {\bf 322}, 431 (1994).

\bibitem{Amsler:1994ah} 
  C.~Amsler {\it et al.} (Crystal Barrel Collaboration),
  Phys.\ Lett.\ B {\bf 340}, 259 (1994).

\bibitem{Amsler:1995bz} 
  C.~Amsler {\it et al.} (Crystal Barrel Collaboration),
  Phys.\ Lett.\ B {\bf 353}, 571 (1995).

\bibitem{Amsler:1995bf} 
  C.~Amsler {\it et al.} (Crystal Barrel Collaboration),
  Phys.\ Lett.\ B {\bf 355}, 425 (1995).

\bibitem{Amsler:1995gf} 
  C.~Amsler {\it et al.} (Crystal Barrel Collaboration),
  Phys.\ Lett.\ B {\bf 342}, 433 (1995).

\bibitem{Abele:1996nn} 
  A.~Abele {\it et al.} (Crystal Barrel Collaboration),
  Phys.\ Lett.\ B {\bf 385}, 425 (1996).

\bibitem{Abele:1996fr} 
  A.~Abele {\it et al.} (Crystal Barrel Collaboration),
  Phys.\ Lett.\ B {\bf 380}, 453 (1996).

\bibitem{Abele:1998qd} 
  A.~Abele {\it et al.},
  Phys.\ Rev.\ D {\bf 57}, 3860 (1998).

\bibitem{Abele:2001js} 
  A.~Abele {\it et al.} (Crystal Barrel Collaboration),
  Eur.\ Phys.\ J.\ C {\bf 19}, 667 (2001).

\bibitem{Abele:2001pv} 
  A.~Abele {\it et al.} (Crystal Barrel Collaboration),
  Eur.\ Phys.\ J.\ C {\bf 21}, 261 (2001).

\bibitem{Amsler:2002qq} 
  C.~Amsler {\it et al.} (Crystal Barrel Collaboration),
  Eur.\ Phys.\ J.\ C {\bf 23}, 29 (2002).

\bibitem{Amsler:2006du} 
  C.~Amsler {\it et al.} (Crystal Barrel Collaboration),
  Phys.\ Lett.\ B {\bf 639}, 165 (2006).

\bibitem{Armstrong:1993fh} 
  T.~A.~Armstrong {\it et al.} (E760 Collaboration),
  Phys.\ Lett.\ B {\bf 307}, 394 (1993).

\bibitem{Uman:2006xb} 
  I.~Uman, D.~Joffe, Z.~Metreveli, K.~K.~Seth, A.~Tomaradze and P.~K.~Zweber,
  Phys.\ Rev.\ D {\bf 73}, 052009 (2006).

\bibitem{pseudoglue} A.~Masoni \textit{et al.}, J. Phys. G \textbf{32}, R 293 (2006).





\bibitem{Evangelista:1998zg}
  C.~Evangelista {\it et al.} (JETSET Collaboration),
  Phys.\ Rev.\ D {\bf 57}, 5370 (1998).
  
  
\bibitem{Aubert:2007ym}
  B.~Aubert {\it et al.} (BaBar Collaboration),
  Phys.\ Rev.\ D {\bf 74}, 091103 (2006);
  B.~Aubert {\it et al.} (BaBar Collaboration),
  Phys.\ Rev.\ D {\bf 77}, 092002 (2008);
  C.~P.~Shen {\it et al.} (Belle Collaboration),
  Phys.\ Rev.\ D {\bf 80}, 031101 (2009).  
    
\bibitem{Y2174th}
  G.~Ding, M.~Yan, Phys. Lett. B 650 (2007), 390-400;
  Z.~Wang, Nucl. Phys. A {\bf 791}, 106 (2007).
  
\bibitem{Adolph:2015pws}
  C.~Adolph {\it et al.} (COMPASS Collaboration),
  Phys.\ Rev.\ Lett.\  {\bf 115}, 082001 (2015).
  
  \bibitem{Bertin:1997zu}
  A.~Bertin {\it et al.} (OBELIX Collaboration),
  Phys.\ Lett.\ B {\bf 400}, 226 (1997).
  
  \bibitem{Longacre:1990uc}
  R.~S.~Longacre,
  Phys.\ Rev.\ D {\bf 42}, 874 (1990);
   N.~.A.~T\"ornqvist, Z. Phys. C {\bf 61}, 525 (1994).
  
  \bibitem{Lutz:2003fm} M.~F.~M. Lutz and E.~E.~Kolomeitsev, Nucl. Phys. A \textbf{730}, 392 (2004). 
   
   \bibitem{Aceti:2016}
   F.~Aceti, L.~R.~Dai, and E.~Oset, Phys. Rev D \textbf{94}, 096015 (2016).
  
  \bibitem{woss2019} A.~J.~Woss (Hadron Spectrum Collaboration) \textit{et al.}, arXiv:2009.10034; Phys. Rev. D \textbf{100}, 054506 (2019).
 
  \bibitem{dudek2016} J.~J.~Dudek, R.~G.~Edwards, and D.~J.~Wilson (Hadron Spectrum Collaboration), Phys. Rev. D \textbf{93}, 094506 (2016).
   
   \bibitem{hybridrev} C.~A.~Meyer and E.~S.~Swanson, Prog. Part. Nucl. Phys. \textbf{82}, 21 (2015).

\bibitem{alekseev:2010} M.~G.~Alekseev \textit{et al.} (COMPASS Collaboration), Phys. Rev. Lett. \textbf{104}, 241803 (2010).
\bibitem{adolph:2015} C.~Adolph \textit{et al.} (COMPASS Collaboration), Phys. Lett. B \textbf{740}, 303 (2015).
\bibitem{aghasyan:2018} M.~Aghasyan \textit{et al.}, Phys. Rev. D \textbf{98}, 092003 (2018).
\bibitem{rodas:2019} A.~Rodas \textit{et al.} [Joint Physics Analysis Center], Phys. Rev. Lett. \textbf{122}, 042002 (2019).

\bibitem{bertin} A.~Bertin \textit{et al.,} Phys. Lett. B \textbf{400}, 226 (1997).


\bibitem{X3872_belle} S.-K.~Choi \textit{et al.,} Phys. Rev. Lett. \textbf{91}, 262001 (2003).

\bibitem{Eichten:1978tg}
  E.~Eichten, K.~Gottfried, T.~Kinoshita, K.~D.~Lane and T.~M.~Yan, Phys.\ Rev.\ D {\bf 17}, 3090 (1978); Erratum: [Phys.\ Rev.\ D {\bf 21}, 313 (1980)].


\bibitem{Y4260_babar} B.~Aubert \textit{et al.} (BaBar Collaboration), Phys. Rev. Lett. \textbf{95}, 142001 (2005).

\bibitem{Y4360_babar} B.~Aubert \textit{et al.} (BaBar Collaboration), Phys. Rev. Lett. \textbf{98}, 212001 (2007); X.~L.~Wang \textit{et al.,} Phys. Rev. Lett. \textbf{99}, 142002 (2007).

\bibitem{Z4430} S.~Choi \textit{et al.} (Belle Collaboration), Phys. Rev. Lett. \textbf{100}, 142001 (2008).

\bibitem{Z3900_besiii} M.~Ablikim \textit{et al.} (BESIII Collaboration), Phys. Rev. Lett. \textbf{110}, 252001 (2013).

\bibitem{Z4020_besiii} M.~Ablikim \textit{et al.} (BESIII Collaboration), Phys. Rev. Lett. \textbf{111}, 242001 (2013).

\bibitem{Zcs3985} M.~Ablikim \textit{et al.} (BESIII Collaboration), arXiv:2011.07855 [hep-ex] (2020).

\bibitem{Zb106x0_lhcb} A.~Garmash \textit{et al.} (Belle Collaboration), Phys. Rev. Lett. \textbf{116}, 212001 (2016).  

\bibitem{Olsen:2017bmm} S.~L.~Olsen, T.~Skwarnicki and D.~Zieminska, Rev.\ Mod.\ Phys.\  {\bf 90}, 015003 (2018).

\bibitem{liu2012} L.~Liu \textit{et al.} (Hadron Spectrum Collaboration), Journal of High Energy Physics \textbf{1207}, 126 (2012); S.~M.~Ryan, D.~J.~Wilson, arXiv:2008.02656.

\bibitem{Cleven:2015era}
 M.~Cleven, F.~K.~Guo, C.~Hanhart, Q.~Wang and Q.~Zhao, Phys.\ Rev.\ D {\bf 92}, 014005 (2015).  
 
 \bibitem{matuschek} I.~Matuschek, Vadim Baru, Feng-Kun Guo and Christoph Hanhart, \textit{On the nature of near-threshold bound and virtual states}, [arXiv:2007:05329 [hep-ph] (2020).
  
\bibitem{BelleXwidthUL2011} S.-K. Choi \textit{et al.} (Belle Collaboration), Phys.\ Rev.\ D \textbf{84}, 052004 (2011).
  
\bibitem{LHCb2020} R.~Aaij \textit{et al.} (LHCb Collaboration), Phys.\ Rev.\ D \textbf{102}, 092005 (2020).

\bibitem{xscan} G.~Barucca \textit{et al.} (PANDA Collaboration), Eur. Phys. J. A \textbf{55}, 42 (2019).




\bibitem{HanhartLS} C.~Hanhart \textit{et al.,} Phys. Rev. D \textbf{76}, 034007 (2007).

\bibitem{KalashLS} Yu.~S.~Kalashnikova \textit{et al.,} Phys. Atom. Nucl. \textbf{73}, 1592 (2010).

\bibitem{BraatenLS} E.~Braaten \textit{et al.,} Phys. Rev. D \textbf{77}, 014029 (2008).

\bibitem{BelleXJ2Pi} C.-Z.~Yuan, \textit{Exotic Hadrons}, proceedings of the \textit{XXIX Physics in Collision Conference}, arXiv[hep-ex]: 0910.3138v2 (2009).

\bibitem{CDFJrho} A.~Abulencia \textit{et al.,} Phys. Rev. Lett. \textbf{96}, 102002 (2006).

\bibitem{LHCbX} R.~Aaij \textit{et al.} (LHCb Collaboration), Eur. Phys. J. C \textbf{73}, 4262 (2013); \textit{ibid.} LHCb-PAPER-2016-016 (2016).

\bibitem{ChenNRBG} G.~Y.~Chen, Phys. Rev. D \textbf{77} 097501 (2008).

\bibitem{OHelene} O.~Helene, Nucl. Instrum. Meth. Phys. Res. \textbf{212}, 319 (1983).

\bibitem{mertens:2012} M.~Mertens for the PANDA Collaboration, Hyperfine Interactions \textbf{209}, 111 (2012).

\bibitem{lin2013} L.~Liu, K.~Orginos, F.-K.~Guo, C.~Hanhart and U.-G.~Meißner, Phys. Rev. D \textbf{87}, No. 1, 014508 (2013).

\bibitem{guo2018} X.-Y.~Guo, Y.~Heo and M.~F.~M.~Lutz, Phys. Rev. D \textbf{98} No. 1, 014510 (2018).


\bibitem{lattimer} J.~M.~Lattimer and A.~W.~Steiner, Astrophys. J., \textbf{784}, 123 (2014).

\bibitem{fomin} N.~Fomin, D.~Higinbotham, M.~Sargsian and P.~Solvignon, Annu. Rev. Nucl. Part. Sci. \textbf{67} 129 (2017).

\bibitem{felicello} A.~Felicello and T.~Nagae, Rep. Prog. Phys. \textbf{78}, 096301 (2015).

\bibitem{PhysRevC.82.024602}
A.~B.~Larionov, I.~N.~Mishustin, L.~N.~Satarov, and W.~Greiner,
Phys. Rev. C {\bf 82}, 024602 (2010).

\bibitem{GAITANOS2015181}
T.~Gaitanos and M.~Kaskulov,
Nucl. Phys. A {\bf 940}, 181 (2015). 

\bibitem{Nakamura:1984xw} K. Nakamura \textit{et al.,} Phys. Rev. Lett. \textbf{52}, 731 (1984).

\bibitem{Garreta:1984zau} D.~Garreta \textit{et al.,} Phys. Lett. B \textbf{135}, 266 (1984).

\bibitem{Garreta:1984rs} D.~Garreta \textit{et al.,} Phys. Lett. B \textbf{149}, 64 (1984).

\bibitem{Friedman2005}
E. Friedman, A. Gal, J. Mares, Nucl. Phys. A {\bf 761}, 283 (2005).

\bibitem{Teis1994}
S. Teis, W. Cassing, T. Maruyama, U. Mosel, Phys. Rev. C {\bf 50}, 388 (1994).

\bibitem{Spieles1996} C. Spieles, M. Bleicher, A. Jahns, R. Mattiello, H. Sorge, H. Stöcker, W. Greiner,
Phys. Rev. C {\bf 53}, 2011 (1996).

\bibitem{Sibirtsev1998} A. Sibirtsev, W. Cassing, G.I. Lykasov, M.V. Rzjanin, Nucl. Phys. A {\bf 632}, 131 (1998).

\bibitem{PhysRevC.80.021601}
A.B. Larionov, I.A. Pshenichnov, I.N. Mishustin, and W. Greiner,
Phys. Rev. C {\bf 80}, 021601 (2009).





\bibitem{GAITANOS2011193}
T. Gaitanos and M. Kaskulov and H. Lenske,
Phys. Lett. B {\bf 703}, 193 (2011).

\bibitem{GAITANOS20099}
T. Gaitanos, M. Kaskulov and U. Mosel,
Nucl. Phys. A {\bf 828}, 9 (2009). 

\bibitem{GAITANOS2013133}
T. Gaitanos and M. Kaskulov, Nucl. Phys. A {\bf 899}, 133 (2013). 

\bibitem{0004-637X-808-1-8}
R. O. Gomes and V. Dexheimer and S. Schramm and C. A. Z. Vasconcellos,
The Astrophysical Journal {\bf 808}, 1 (2015).

\bibitem{josef1} J.~Pochodzalla, Phys. Lett. B \textbf{669}, 306 (2008).

\bibitem{josef2} J.~Pochodzalla, Hyperfine Interact. \textbf{194}, 255 (2009).

\bibitem{alicia} A.~Sanchez Lorente, S.~Bleser, M.~Steinen and J.~Pochodzalla, Phys. Lett. B \textbf{749}, 421 (2015).

\bibitem{Buss20121} O.~Buss \textit{et al.,} Phys. Rep. \textbf{512}, 1 (2012).

\bibitem{singhhyp} B.~Singh \textit{et al.,} Nucl. Phys. A \textbf{954}, 323 (2016).

\bibitem{strikman} D.~Dutta, K.~Hafidi and M.~Strikman, Prog. Part. Nucl. Phys. Rep. \textbf{69}, 1 (2013).

\bibitem{clasctpi} B.~Clasie \textit{et al.}
Phys. Rev. Lett. \textbf{99}, 242502 (2007).

\bibitem{clasctrho} L.~El-Fassi \textit{et al.}
Phys. Lett. B \textbf{712}, 326 (2007).

\bibitem{farrar1988mf} G.~R.~Farrar, H.~Liu, L.~L.~Frankfurt and M.~I.~Strikman, Phys. Rev. Lett. \textbf{62}, 1095 (1989).


\bibitem{tang} A.~Tang \textit{et al.}
Phys. Rev. Lett. \textbf{90}, 042301 (2003).

\bibitem{piasetzky} E.~Piasetzky \textit{et al.}
Phys. Rev. Lett. \textbf{97}, 162504 (2006).

\bibitem{subedi} R.~Subedi \textit{et al.,} Science \textbf{320}, 1476 (2008).

\bibitem{shneor} R.~Shneor \textit{et al.}
Phys. Rev. Lett. \textbf{99}, 072501 (2007).

\bibitem{brodsky} S.~J.~Brodsky and A.~H.~Mueller,
Phys. Lett. B \textbf{206}, 685 (1988).

\bibitem{farrar} G.~R.~Farrar, L.~L.~Frankfurt, M.~I.~Strikman and H.~Liu,
Nucl. Phys. B \textbf{345}, 125 (1990).

\bibitem{Larionov:2019xdn} A.~B.~Larionov and M.~Strikman, Eur. Phys. J. A \textbf{56}, 21 (2020).

\bibitem{Larionov:2019mwa} A.~B.~Larionov, A.~Gillitzer and M.~Strikman, Eur. Phys. J. A \textbf{55}, 154 (2019).

\bibitem{Haidenbauer:1993pw} J.~Haidenbauer, K. Holinde and M.~B.~Johnson, Phys. Rev. C \textbf{48}, 2190 (1993). 

\bibitem{Glozman:1993pm} L.~{\relax Ya}~Glozman, V.~G.~Neudachin and I.~T.~Obukhovsky, Phys. Rev. C \textbf{48}, 389 (1993).

\bibitem{Larionov:2018lpk} A.~B.~Larionov, A.~Gillitzer, J.~Haidenbauer and M.~Strikman, Phys. Rev.C \textbf{98}, 054611 (2018).

\bibitem{Larionov:2013axa} A.~B.~Larionov, M.~Bleicher, A.~Gillitzer and M.~Strikman, Phys. Rev. C \textbf{87}, 054608 (2013).

\bibitem{Matsui:1986dk} T.~Matsui and H.~Satz, Phys. Lett. B \textbf{178}, 416 (1986).














\end{thebibliography}

\end{document}